\documentclass[a4paper,11pt]{article}
\pdfoutput=1
\usepackage{jinstpub}
\makeatletter
\gdef\@fpheader{Accepted for publication in JINST}
\makeatother

\usepackage{booktabs}
\usepackage{multirow}
\usepackage{xcolor}
\usepackage{subcaption}
\usepackage{tabularx}
\usepackage{float}
\usepackage{siunitx}
\usepackage{enumitem}
\usepackage{lineno}

\graphicspath{{figures_rev/}{figures/}{../figures/}}

\newcommand{\sigt}{\sigma_t}
\newcommand{\Npe}{N_\mathrm{pe}}
\newcommand{\taud}{\tau_d}
\newcommand{\taur}{\tau_r}
\newcommand{\tauwls}{\tau_\mathrm{WLS}}
\newcommand{\fdet}{f_\mathrm{det}}
\newcommand{\Fdet}{F_\mathrm{det}}
\newcommand{\fscint}{f_\mathrm{scint}}
\newcommand{\fwls}{f_\mathrm{WLS}}
\newcommand{\sigelec}{\sigma_\mathrm{elec}}
\newcommand{\sigtransit}{\sigma_\mathrm{transit}}
\newcommand{\sigphoton}{\sigma_\mathrm{photon}}

\title{Comprehensive study of timing resolution in plastic scintillator detectors
with wavelength-shifting fiber and silicon photomultiplier readout}

\author{H.~Che$^{a,}$\footnote{\href{mailto:c.haohui@wustl.edu}{\texttt{c.haohui@wustl.edu}}} and G.~Yang$^{b,}$\footnote{ \href{mailto:gyang1@bnl.gov}{\texttt{gyang1@bnl.gov}}} \\[0.3em]

\small $^{a}$Washington University in St.\ Louis, St.\ Louis, MO 63130, USA\\[0.2em]
\small $^{b}$Brookhaven National Laboratory, Upton, NY 11973, USA
}

\abstract{We present a comprehensive study of the timing resolution achievable in plastic
scintillator detectors read out through wavelength-shifting (WLS) fibers coupled to silicon
photomultipliers (SiPMs), combining a semi-analytical framework, toy Monte Carlo validation,
and full Geant4 optical photon simulation. The analytical model traces the complete photon
detection chain: scintillation emission, WLS fiber re-emission, optical transit time
dispersion, SiPM single-photon time resolution, and electronics quantization. It expresses
the timing resolution $\sigt$ as a function of the detected photoelectron yield $\Npe$,
scintillator decay constants ($\taur$, $\taud$), WLS re-emission time ($\tauwls$), fiber
numerical aperture, detector geometry, and readout electronics parameters. The analytical
predictions are validated at two levels. First, toy Monte Carlo simulations ($2\times 10^5$
events per parameter point across 80 grid points spanning 8 fiber types and $\Npe$ from 5
to 200) achieve analytical-to-MC agreement of $1.0000 \pm 0.0018$. Second, full Geant4
optical photon simulations track the entire scintillation, wavelength-shifting, and
fiber transport chain in realistic detector geometries, confirming the analytical timing
predictions to within 10--20\% for $\Npe \gtrsim 10$ and providing first-principles photoelectron yield calibration.
A comprehensive parameter scan covering 7 scintillator materials, 8 WLS fiber types,
5 SiPM models, 5 electronics configurations, 3 readout topologies, and 3 boundary
conditions produces quantitative design maps and lookup tables for detector optimization.}


\begin{document}

\maketitle
\flushbottom

\section{Introduction}
\label{sec:intro}

Precise timing measurement in plastic scintillator detectors is of critical importance
across a broad range of particle physics experiments. In neutrino experiments such as
the Super Fine-Grained Detector (SuperFGD)~\cite{SuperFGD_2020} of the T2K near detector
upgrade, sub-nanosecond timing enables particle identification through time-of-flight
and rejection of out-of-bunch backgrounds. In large-area muon detection systems,
calorimeters, and cosmic-ray observatories, the achievable timing resolution directly
impacts reconstruction performance, triggering efficiency, and physics reach.

The readout chain for a modern plastic scintillator detector is typically formed with plastic scintillators \cite{Eljen_scint} that convert ionization energy to photons, wavelength-shifting (WLS) fibers~\cite{Kuraray_fiber,SaintGobain_fiber} that re-emit the scintillation light at a different color for better transport to photodetectors,  silicon photomultipliers (SiPMs) \cite{Hamamatsu_MPPC} that detect re-emitted photons with sub-nanosecond resolution, and finally front-end electronics that digitize signals recorded by SiPMs. Each part of the readout chain contributes to the system's overall timing resolution with distinct mechanisms.

Published measurements from experiments such as SuperFGD~\cite{SuperFGD_timing},
MINERvA~\cite{MINERvA_timing}, MINOS~\cite{MINOS_timing}, and the Scintillator
Bar Detector (SciBar)~\cite{SciBar_timing} provide valuable reference points.
However, a generalized model that connects the measurable component parameters to the achievable resolution is needed to predict the time resolution of new scintillator detector configurations.

In this work, we construct a semi-analytical model expressed in
terms of experimentally measurable material and detector parameters. The model enables both direct
comparison with published data and prediction for new configurations. We validate the
analytical predictions at two levels: high-statistics toy Monte Carlo simulations
demonstrate agreement at the $0.2\%$ level across 80 parameter-space grid points.
A full Geant4 optical photon simulation tracks scintillation, WLS absorption
and re-emission, fiber propagation, and SiPM detection in realistic three-dimensional
detector geometries.

\section{Analytical framework}
\label{sec:model}

The timing resolution of a scintillator detector with WLS fiber and SiPM readout
is determined by a chain of stochastic processes. We formalize this chain with eight functions (F1-F8): F1, F2, F5, and F7 describe individual physical
processes, while F3, F4, F6, and F8 combine them, by convolution, order statistics, and
quadrature sum, into the final timing resolution.

\subsection{F1: Scintillator emission profile}
\label{sec:F1}

The scintillation photon emission time following energy deposition is described by
a bi-exponential probability density function (PDF) ~\cite{Moszynski1979} of one rise and one decay:
\begin{equation}
\fscint(t; \taur, \taud) = \frac{1}{\taud - \taur}
\left( e^{-t/\taud} - e^{-t/\taur} \right), \quad t \geq 0,
\label{eq:fscint}
\end{equation}
where $\taur$ is the scintillation rise time and $\taud$ the scintillation decay time.
This distribution is properly normalized, $\int_0^\infty \fscint(t)\,dt = 1$, and has
mean $\langle t \rangle = \taud + \taur$ with variance
$\mathrm{Var}[t] = \taud^2 + \taur^2$.

A crucial mathematical property of equation~(\ref{eq:fscint}) is that $\fscint$ can be written as
the convolution of two exponential distributions:
\begin{equation}
\fscint = \mathrm{Exp}(\taur) \otimes \mathrm{Exp}(\taud),
\label{eq:fscint_conv}
\end{equation}
where $\mathrm{Exp}(\tau)$ denotes an exponential distribution with rate parameter
$1/\tau$. This identity, which follows directly from the Laplace transform, implies
that a random variate $T \sim \fscint$ can be sampled as $T = X + Y$ where
$X \sim \mathrm{Exp}(\taur)$ and $Y \sim \mathrm{Exp}(\taud)$ are independent.
This provides an exact and efficient sampling algorithm for Monte Carlo studies, and each scintillation photon time requires only two exponential random variates.

The peak of $\fscint$ occurs at $t_\mathrm{peak} = \taur \taud / (\taud - \taur)
\cdot \ln(\taud/\taur)$. The rising edge width, parameterized as
$\sqrt{\taur \cdot \taud}$, sets the fundamental timing scale.

\subsection{F2: WLS fiber re-emission}
\label{sec:F2}

The WLS fiber absorbs a scintillation photon and, after a characteristic delay $\tauwls$,
re-emits a photon at a longer wavelength. The re-emission time profile is modeled as a
single exponential:
\begin{equation}
\fwls(t; \tauwls) = \frac{1}{\tauwls} e^{-t/\tauwls}, \quad t \geq 0.
\label{eq:fwls}
\end{equation}
The single exponential describes the first-order radiative decay of the excited WLS
dye~\cite{Knoll2010}. A re-emitted photon can
be absorbed again by the dye before it leaves the fiber. However, because the absorption and emission
bands of the WLS dye overlap only weakly \cite{Kuraray_fiber}, such re-absorption is rare, and its
effect on the time profile is neglected here. The full extent of its effect on our prediction is analyzed in section \ref{sec:g4_effects}

\subsection{F3: Combined photon time PDF}
\label{sec:F3}

The total time from scintillation to WLS re-emission is the convolution of $\fscint$ and $\fwls$:
\begin{equation}
f_\mathrm{scint+WLS}(t) = (\fscint \otimes \fwls)(t).
\label{eq:conv}
\end{equation}
For the bi-exponential scintillator emission convolved with a single exponential WLS
re-emission, this yields a closed-form triple-exponential expression:
\begin{equation}
f_\mathrm{scint+WLS}(t) = \frac{1}{\taud - \taur} \left[
\frac{\taud}{\taud - \tauwls} e^{-t/\taud}
- \frac{\taur}{\taur - \tauwls} e^{-t/\taur}
+ \left( \frac{\taur}{\taur - \tauwls} - \frac{\taud}{\taud - \tauwls} \right)
e^{-t/\tauwls}
\right],
\label{eq:triple}
\end{equation}
valid when $\taur \neq \tauwls \neq \taud$. Degenerate cases are handled by introducing
infinitesimal perturbations to the time constants.

The physical significance of this convolution is that the WLS fiber acts as a temporal
low-pass filter. A WLS fiber with $\tauwls$ much larger than $\taud$ broadens the
photon arrival time distribution substantially, degrading timing resolution by suppressing
the information-rich sharp leading edge.

\subsection{F4: Single-photon detection PDF}
\label{sec:F4}

The full single-photon detection time PDF includes the SiPM single-photon time
resolution (SPTR):
\begin{equation}
\fdet(t) = f_\mathrm{scint+WLS}(t) \otimes G(0, \sigma_\mathrm{SPTR}),
\label{eq:fdet}
\end{equation}
where $G(0, \sigma_\mathrm{SPTR})$ is a Gaussian with standard deviation
$\sigma_\mathrm{SPTR} = \mathrm{SPTR}_\mathrm{FWHM} / 2\sqrt{2\ln 2}$,
using the standard Gaussian relation $\mathrm{FWHM} = 2\sqrt{2\ln 2}\,\sigma \approx 2.355\,\sigma$.
A Gaussian is the standard parameterization of the SiPM single-photon time
response~\cite{Gundacker_SPTR,Acerbi2019}. While the measured response also has a small
non-Gaussian tail~\cite{Acerbi2019}, it is inconsequential when compared to the overall timing resolution especially considering the SPTR itself is already a small timing contributor.
This convolution is computed numerically on a fine time grid ($dt = 10$~ps,
range 0--80~ns, extended to 120~ns for mirror configurations) using direct convolution.

\subsection{F5: Photoelectron yield}
\label{sec:F5}

The mean number of photoelectrons detected at position $x$ from the readout end is:
\begin{equation}
\Npe(x) = Y_\mathrm{scint} \cdot \frac{dE}{dx} \cdot d
\cdot \varepsilon_\mathrm{cap}
\cdot \mathrm{PDE}
\cdot e^{-x/\Lambda_\mathrm{att}}
\cdot C_\mathrm{geom},
\label{eq:Npe}
\end{equation}
where $Y_\mathrm{scint}$ is the scintillator light yield (photons/MeV),
$dE/dx$ is the energy loss rate (MeV/cm),
$d$ is the scintillator thickness (cm),
$\varepsilon_\mathrm{cap}$ is the single-direction fiber trapping fraction,
PDE is the SiPM photon detection efficiency,
$\Lambda_\mathrm{att}$ is the WLS fiber attenuation length,
and $C_\mathrm{geom}$ is a geometry-dependent correction factor.

The actual number of photoelectrons in each event is drawn from a Poisson distribution:
$N \sim \mathrm{Poisson}(\Npe(x))$.

\subsection{F6: Order statistic variance}
\label{sec:F6}

For a leading-edge (LE) discriminator triggering on the $k$-th photoelectron, the
timestamp is the $k$-th order statistic $T_{(k:N)}$~\cite{order_stats} of $N$
independent samples from $\fdet$. The foundational analysis of statistical timing
limits in scintillation counters dates to Post and Schiff~\cite{Post1950}. The variance of this order statistic is computed
numerically using:
\begin{equation}
\mathrm{Var}[T_{(k:N)}] = \int_0^\infty t^2 \, g_{k:N}(t) \, dt
- \left[ \int_0^\infty t \, g_{k:N}(t) \, dt \right]^2,
\label{eq:orderstat}
\end{equation}
where the PDF of the $k$-th order statistic from $N$ samples of $\fdet$ is:
\begin{equation}
g_{k:N}(t) = \frac{N!}{(k-1)!(N-k)!} [\Fdet(t)]^{k-1} [1 - \Fdet(t)]^{N-k} \fdet(t),
\label{eq:orderpdf}
\end{equation}
with $\Fdet(t) = \int_0^t \fdet(t')\,dt'$ being the CDF of the detection time PDF.
The CDF is computed using the cumulative trapezoidal rule for sub-percent accuracy.

The photon statistics contribution to the timing resolution is then:
\begin{equation}
\sigphoton^2 = \mathrm{Var}[T_{(k:\Npe)}].
\label{eq:sigphoton}
\end{equation}

\subsection{F7: Transit time spread}
\label{sec:F7}

Photons propagating through the WLS fiber at position $x$ from the readout end arrive
with a transit time spread due to the range of guided mode angles:
\begin{equation}
\sigtransit(x) = \frac{x \cdot \mathrm{NA}^2}{2 \sqrt{12} \cdot c \cdot n_\mathrm{core}},
\label{eq:transit}
\end{equation}
where NA is the fiber numerical aperture, $n_\mathrm{core}$ is the core refractive index,
and $c$ is the speed of light. This formula follows from a uniform distribution of guided
mode angles between the axial (shortest) and critical-angle (longest) propagation paths.
The fastest photon travels axially at $v_\mathrm{max} = c/n_\mathrm{core}$. The slowest
guided photon propagates at the critical angle with effective speed
$v_\mathrm{min} = (c/n_\mathrm{core})\sqrt{1 - \mathrm{NA}^2/n_\mathrm{core}^2}
\approx (c/n_\mathrm{core})(1 - \mathrm{NA}^2/2n_\mathrm{core}^2)$.
The transit time spread at distance $x$ is
$\Delta t = x/v_\mathrm{min} - x/v_\mathrm{max} \approx x \cdot \mathrm{NA}^2 / (2 c \cdot n_\mathrm{core})$.
For a uniform distribution over this range, the standard deviation is
$\sigma = \Delta t / \sqrt{12}$, yielding the factor $2\sqrt{12}$ in the denominator.

\subsection{F8: Total timing resolution}
\label{sec:F8}

The total timing resolution is the quadrature sum of all contributions:
\begin{equation}
\sigt^2 = \sigphoton^2(\Npe, \fdet, k) + \sigtransit^2(x, \mathrm{NA}, n_\mathrm{core})
+ \sigelec^2,
\label{eq:master}
\end{equation}
where $\sigelec = \Delta t_\mathrm{TDC} / \sqrt{12}$ is the electronics quantization
contribution from a TDC with bin size $\Delta t_\mathrm{TDC}$.

\subsubsection{Readout topology and far-end conditions}
\label{sec:topologies}

\begin{figure}[htbp]
\centering
\includegraphics[width=\textwidth]{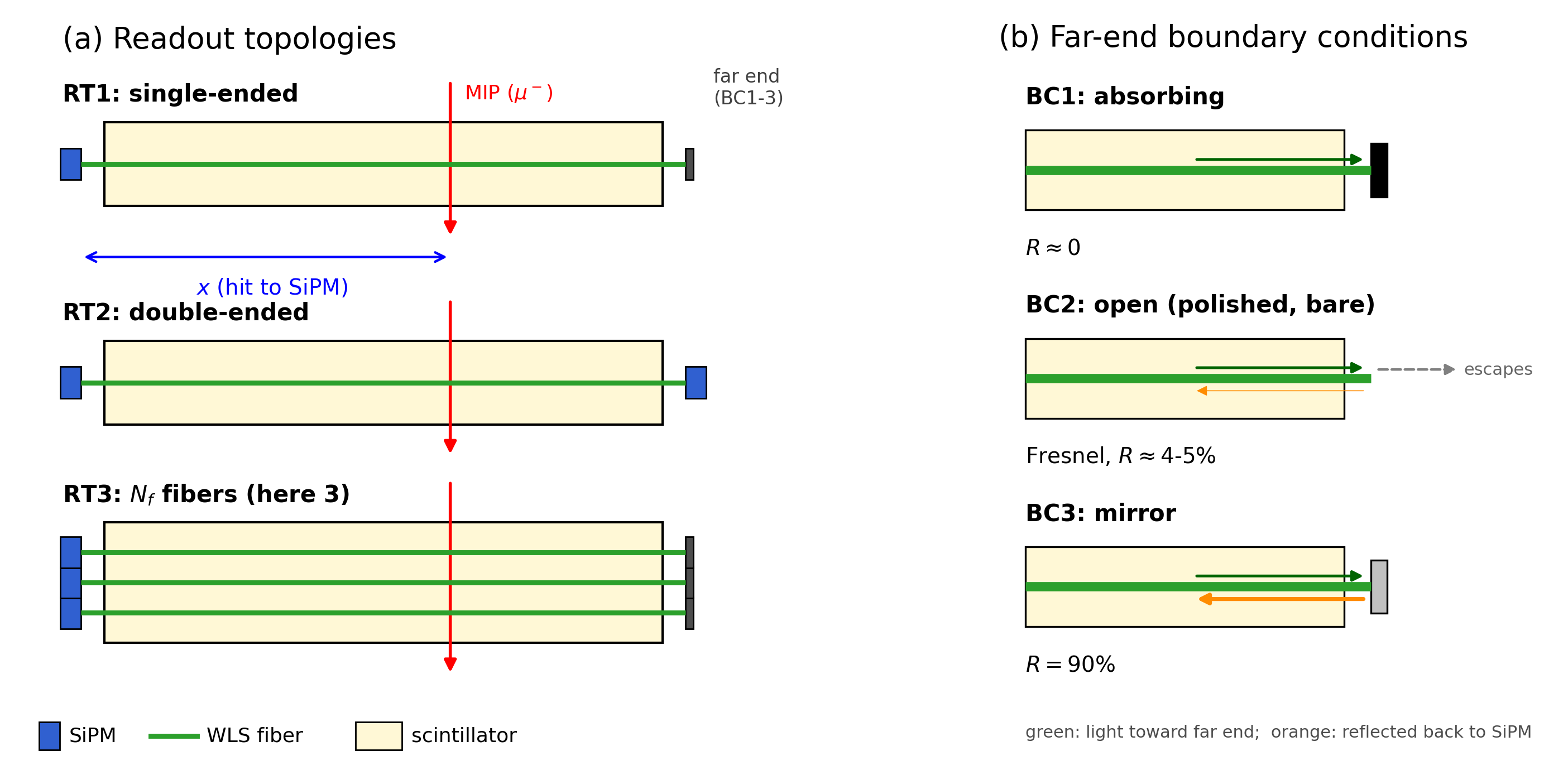}
\caption{Schematic of the configurations studied. (a) Readout topologies: single-ended
(RT1), double-ended (RT2), and $N_f$ independently read out fibers (RT3); the hit position $x$
is measured from the SiPM. (b) Far-end boundary conditions: absorbing (BC1), open with Fresnel
reflection (BC2), and mirror (BC3).}
\label{fig:schematic}
\end{figure}

We consider three readout topologies: Single-ended (RT1) topology with one SiPM at one end of the fiber, Double-ended (RT2) topology with SiPMs at both ends of the fiber, and Multi-fibered (RT3) topology with $N_f$ fibers independently read out by SiPMs. The timing resolution varies with different readout topologies, with RT1 having the simplest one that is governed directly by equation (\ref{eq:master}). For RT2 and RT3, additional complexity is introduced by their more complicated topologies. RT2, with its double-ended readout, has an overall timing resolution that is the weighted mean time of measurements on SiPMs on both ends achieving: 
  \begin{equation}
  \sigt^\mathrm{RT2} = \frac{1}{\sqrt{1/\sigt^{L,2} + 1/\sigt^{R,2}}}.
  \label{eq:RT2}
  \end{equation}
Equation (\ref{eq:RT2}) reduces to $\sigt^\mathrm{RT2} \approx \sigt/\sqrt{2}$ for symmetric readout. \\
For RT3, the overall timing resolution is $\sigt^\mathrm{RT3} = \sigt / \sqrt{N_f}$ for uncorrelated channels.\\
\\
Another important aspect of the detector's physical design that affects timing resolution is the properties at the far-end (the end without SiPMs) of the fiber/scintillator. Here we consider three different far-end boundary conditions: Absorbing (BC1), Open/Fresnel (BC2), and Mirror (BC3); the three readout topologies and the three boundary conditions are sketched in figure~\ref{fig:schematic}. An absorbing material at the far-end absorbs all photons reached there ($R =0\%$). Open/Fresnel type describes a bare, polished fiber end, i.e.\ far-ends that reflect a small fraction ($R \approx 4\%$; the normal-incidence Fresnel reflectance calculated from the standard reflectance formula~\cite{Hecht2017} for the $n = 1.49$ PMMA--air interface is $\approx 3.87\%$ ) of photons back towards the SiPM. For BC3 the far end is mirrored with $R = 90\%$.

\section{Analytical parametric studies}
\label{sec:analytical}

\subsection{Single-photon time profile and leading-edge information}
\label{sec:profiles}

\begin{figure}[htbp]
\centering
\includegraphics[width=\textwidth]{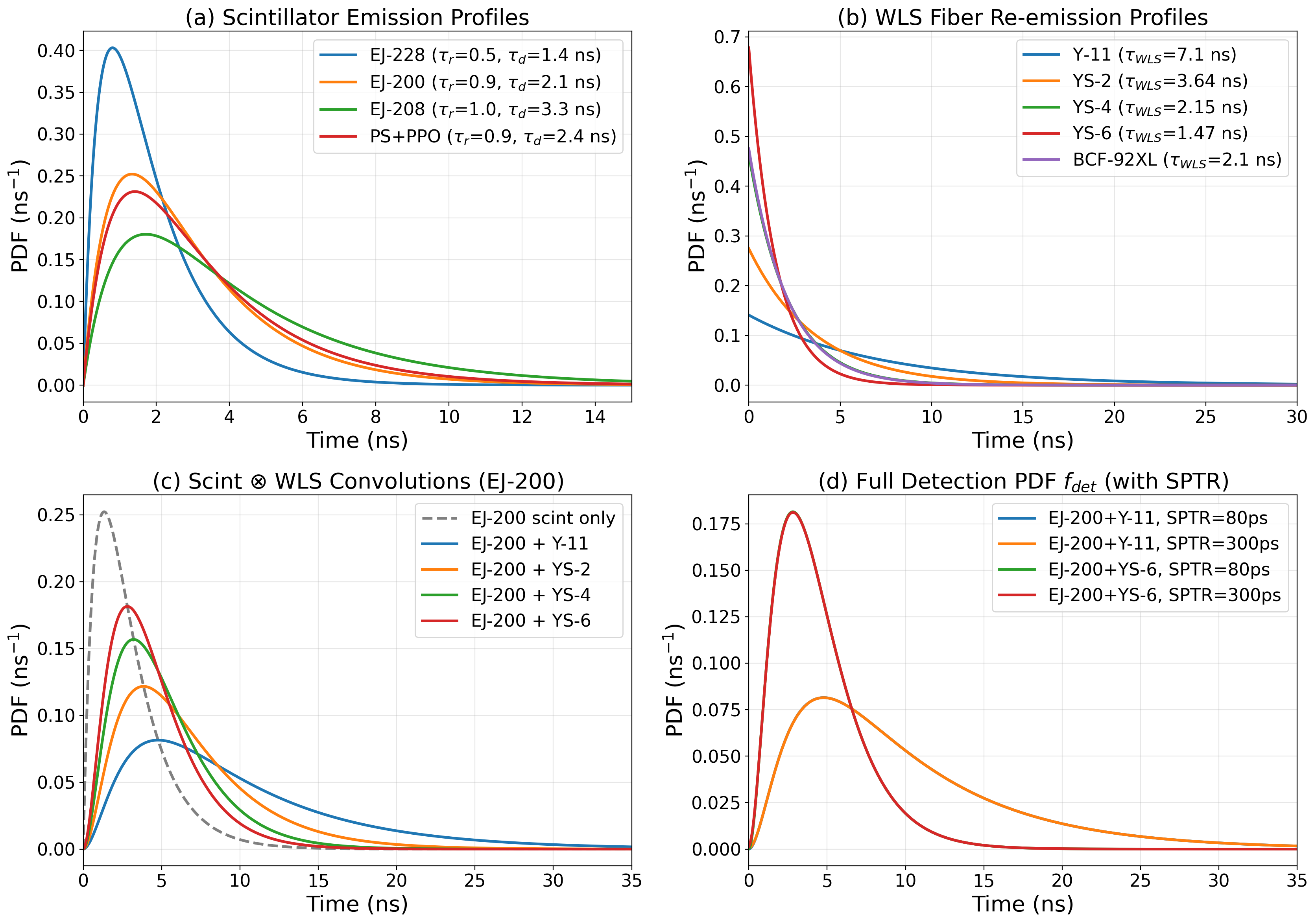}
\caption{Building blocks of the single-photon detection time PDF. (a) Scintillator
emission profiles $\fscint(t)$ for four materials. (b) WLS fiber re-emission profiles
$\fwls(t)$. (c) Convolved scintillator+WLS PDFs. (d) Full detection PDFs $\fdet(t)$
including SiPM SPTR convolution. The WLS fiber acts as a temporal low-pass filter.
Y-11 ($\tauwls = 7.1$~ns) broadens the distribution dramatically compared to YS-6
($\tauwls = 1.47$~ns).}
\label{fig:profiles}
\end{figure}

From analytical formulas derived in section~\ref{sec:model}, several important questions can be studied: how much information is present in the leading edge of the detected photon-time distribution, how that information scales with $\Npe$ and $\tauwls$, how geometry redistributes light along the detector, and when electronics or SiPM timing become limiting.\\
The basic timing object is the leading edge of the full single-photon detection-time PDF, $\fdet(t)$. Figure~\ref{fig:profiles} constructs this PDF from scintillation emission, WLS re-emission, the scintillator+WLS convolution, and the final SPTR-smeared detection profile as derived in section ~\ref{sec:model}. The dominant broadening occurs at the WLS step. Slow fibers such as Y-11 broaden the distribution and suppress the early photons that carry the most timing information, whereas faster fibers such as YS-6 preserve more of the scintillator leading edge. The SPTR convolution is a small perturbation on the scale of the scintillator and WLS time constants; the curves in figure~\ref{fig:profiles}(d), spanning 80--300~ps FWHM, are nearly indistinguishable compared with the change induced by the WLS fiber.\\
An important feature of the final profile in figure~\ref{fig:profiles}(d) is the finite rise time and the fact that the PDF maximum is not at $t=0$ but is shifted to the right. These features play a crucial role in time resolution's relation to $\Npe$. 

\subsection{Timing resolution versus photoelectron yield and WLS re-emission time}
\label{sec:sigma_vs_Npe}

\begin{figure}[htbp]
\centering
\includegraphics[width=\textwidth]{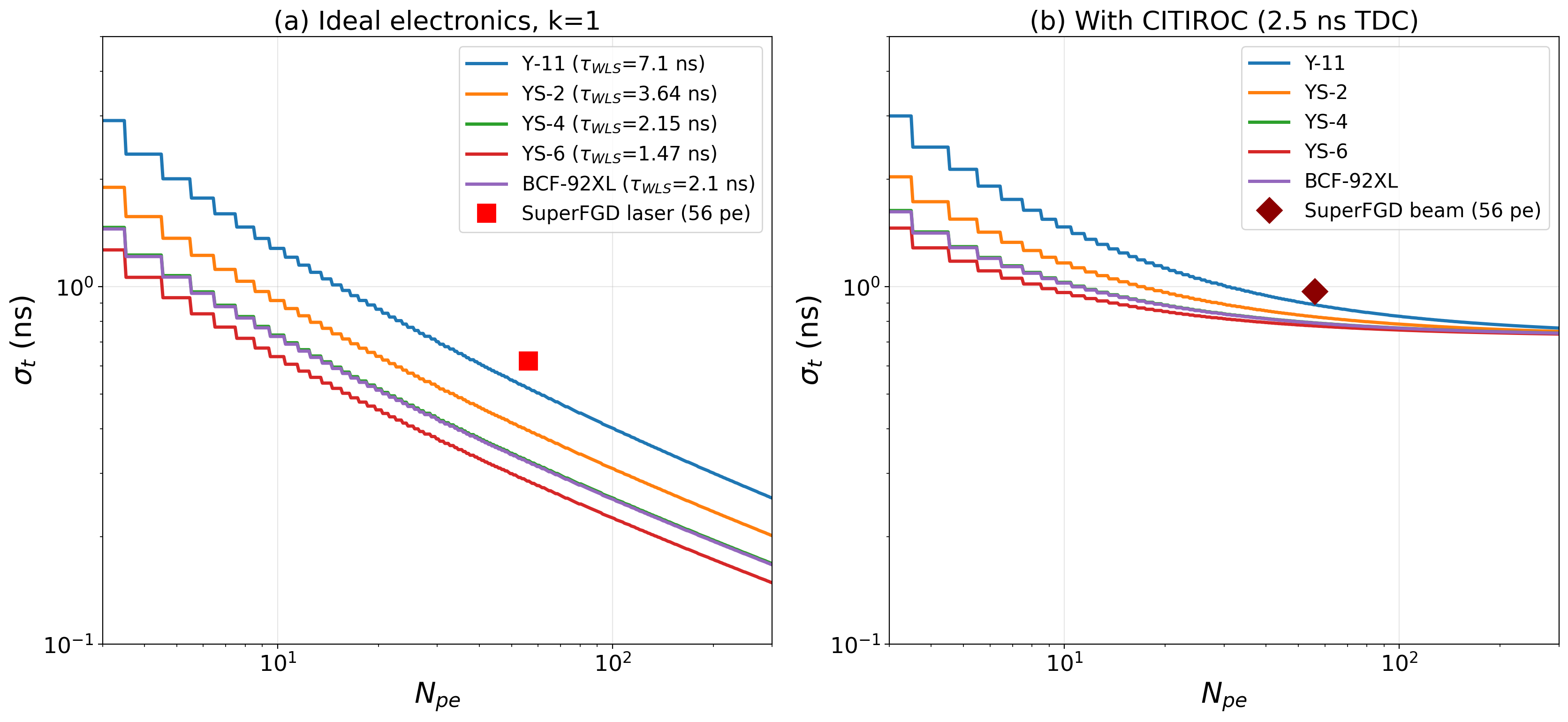}
\caption{Timing resolution $\sigt$ versus photoelectron yield $\Npe$ for leading-edge
discrimination with $k = 1$. (a) Five WLS fiber types with EJ-200 scintillator and
ideal electronics. SuperFGD measurements (laser at $\Npe \approx 56$~\cite{SuperFGD_timing}) are shown as data points.
(b) The same fiber comparison with CITIROC-like 2.5~ns TDC quantization included. The electronics contribution flattens the high-$\Npe$ behavior.}
\label{fig:sigma_Npe}
\end{figure}

At fixed readout topology and electronics, the dominant variables are the detected photoelectron yield and the WLS re-emission time. Figure~\ref{fig:sigma_Npe} shows the most fundamental relationship in this study:
$\sigt$ as a function of $\Npe$ for different WLS fiber types. All curves follow
approximate power laws $\sigt \approx A \cdot \Npe^{-\alpha}$, but with exponents
$\alpha \approx 0.44$--$0.49$ that are significantly smaller than the $\alpha = 1$
expected for a naive PDF with just one decaying exponential. This reduced resolution improvement arises because the convolved PDF $\fdet$ has a finite rise time, preventing the earliest photons
from arriving at $t = 0$ and thus fundamentally limiting the timing information
content per photon.

The fiber-dependent spread at fixed $\Npe$ quantifies the impact of WLS re-emission
time. At $\Npe = 50$, $\sigt$ ranges from 0.300~ns (YS-6) to 0.548~ns (Y-11), a
factor of 1.83. Published SuperFGD measurements are overlaid: the laser calibration
point ($\Npe \approx 56$, $\sigt = 0.62$~ns) and beam test data ($\Npe \approx 56$,
$\sigt = 0.97$~ns, including electronics contribution) lie near the ideal-electronics Y-11 curve of panel~(a) and the Y-11+CITIROC curve of panel~(b), respectively,
consistent with the Y-11 fiber used in SuperFGD~\cite{SuperFGD_2020,SuperFGD_timing}.

Figure~\ref{fig:sigma_Npe}(b) shows the same configurations after adding the CITIROC 2.5~ns TDC quantization contribution. In this electronics-limited case, the high-$\Npe$ improvement flattens because the electronics term is added in quadrature. The SuperFGD beam point lies close to the Y-11+CITIROC curve, while the laser point is closer to the ideal-electronics Y-11 prediction, consistent with the different contributions included in the two measurements.

\begin{figure}[htbp]
\centering
\includegraphics[width=\textwidth]{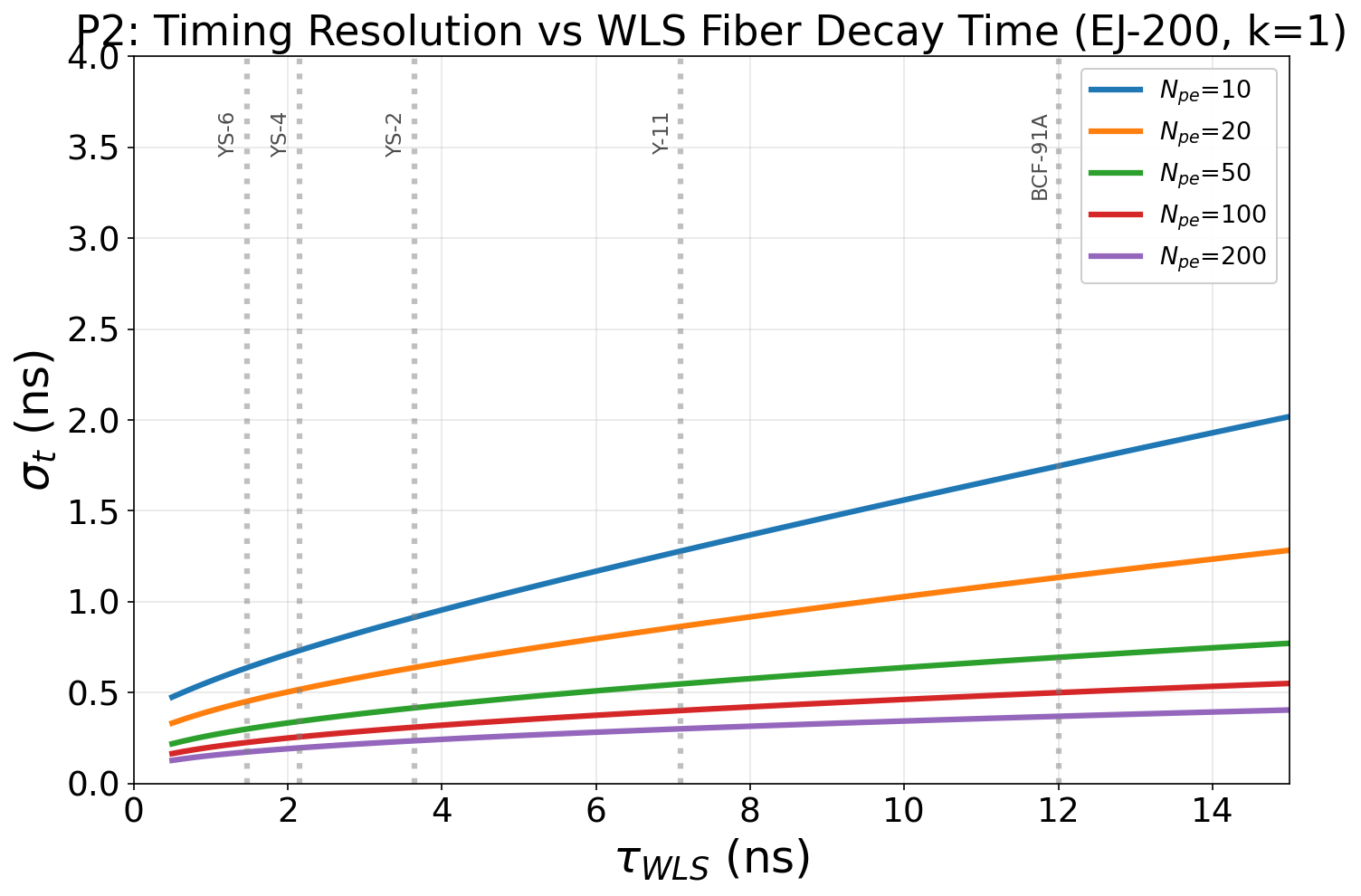}
\caption{Timing resolution as a function of WLS fiber re-emission time $\tauwls$ for
different $\Npe$ values. Vertical lines indicate commercial fiber types. Diminishing
returns are visible below $\tauwls \approx 2$~ns where the scintillator decay time
becomes dominant.}
\label{fig:fiber}
\end{figure}

The continuous $\tauwls$ scan in figure~\ref{fig:fiber} shows where fiber upgrades are most effective. Reducing $\tauwls$ from Y-11 values to the YS-2 range gives a large gain, but the improvement begins to saturate below $\tauwls \approx 2$~ns because the EJ-200 scintillator decay time ($\taud=2.1$~ns) then becomes comparable to or slower than the WLS response. In this regime the WLS fiber no longer dominates the convolution, so further reduction of $\tauwls$ yields smaller gains.

\begin{figure}[htbp]
\centering
\includegraphics[width=\textwidth]{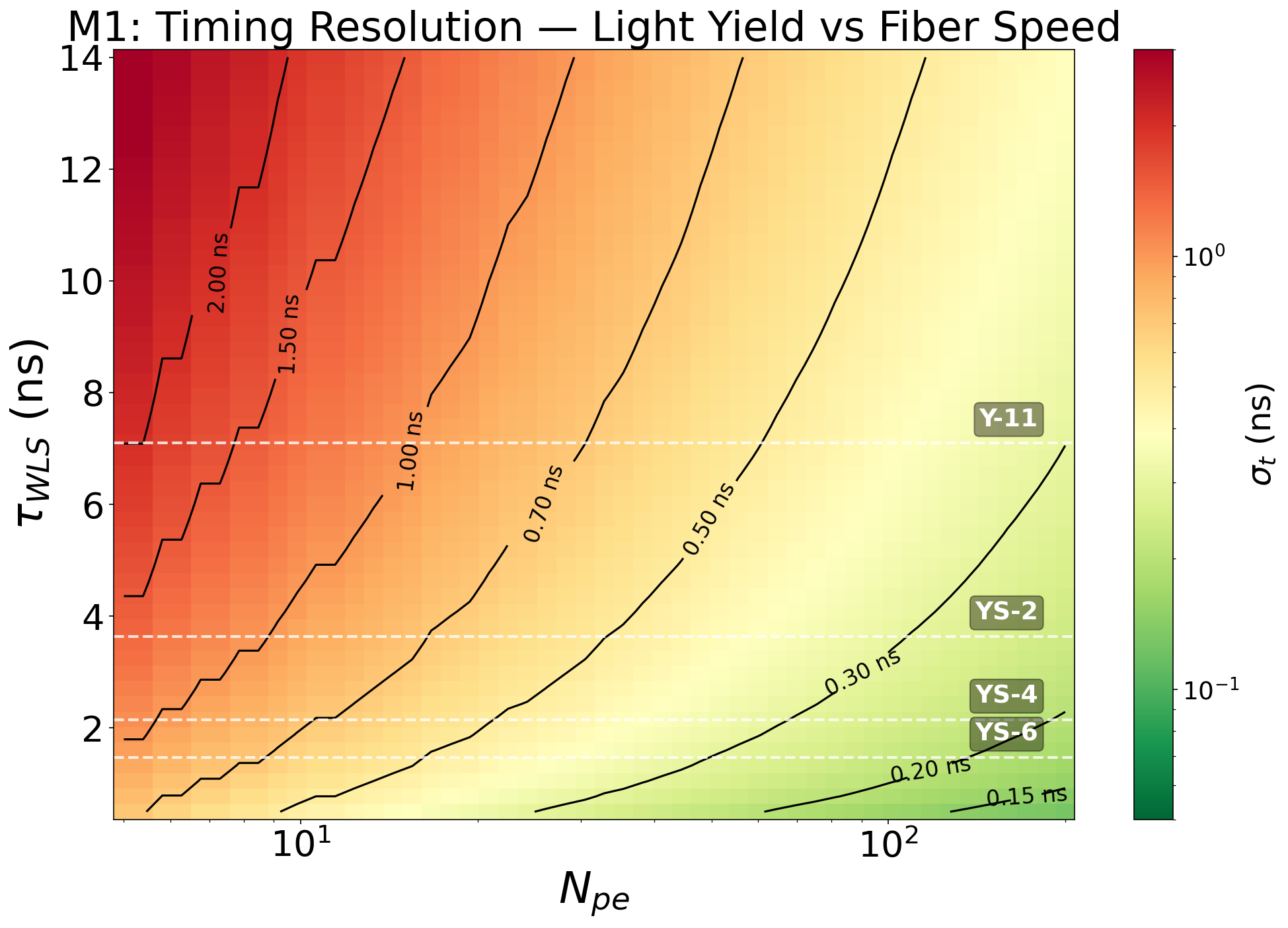}
\caption{Master tradeoff map: $\sigt$ in the $(\Npe,\tauwls)$ plane. Iso-resolution contours are drawn at 0.15~ns, 0.2~ns, 0.3~ns, 0.5~ns, and longer times. Horizontal dashed lines mark commercial fiber types. Following a contour from high to low $\tauwls$ gives the $\Npe$ reduction achievable by switching to a faster fiber.}
\label{fig:map_Npe_tau}
\end{figure}

The same tradeoff is summarized in the two-dimensional map of figure~\ref{fig:map_Npe_tau}. Moving left along an iso-resolution contour quantifies the amount of light yield that can be exchanged for a faster fiber. For the 0.5~ns contour, the required $\Npe$ decreases from roughly 60 for Y-11 to below 20 for YS-6. This is the most important design implication of the analytical model: a fiber with a shorter re-emission time can substitute for a large increase in detected light yield.

\subsection{Geometry, boundary conditions, and readout topology}
\label{sec:position}

\begin{figure}[htbp]
\centering
\includegraphics[width=\textwidth]{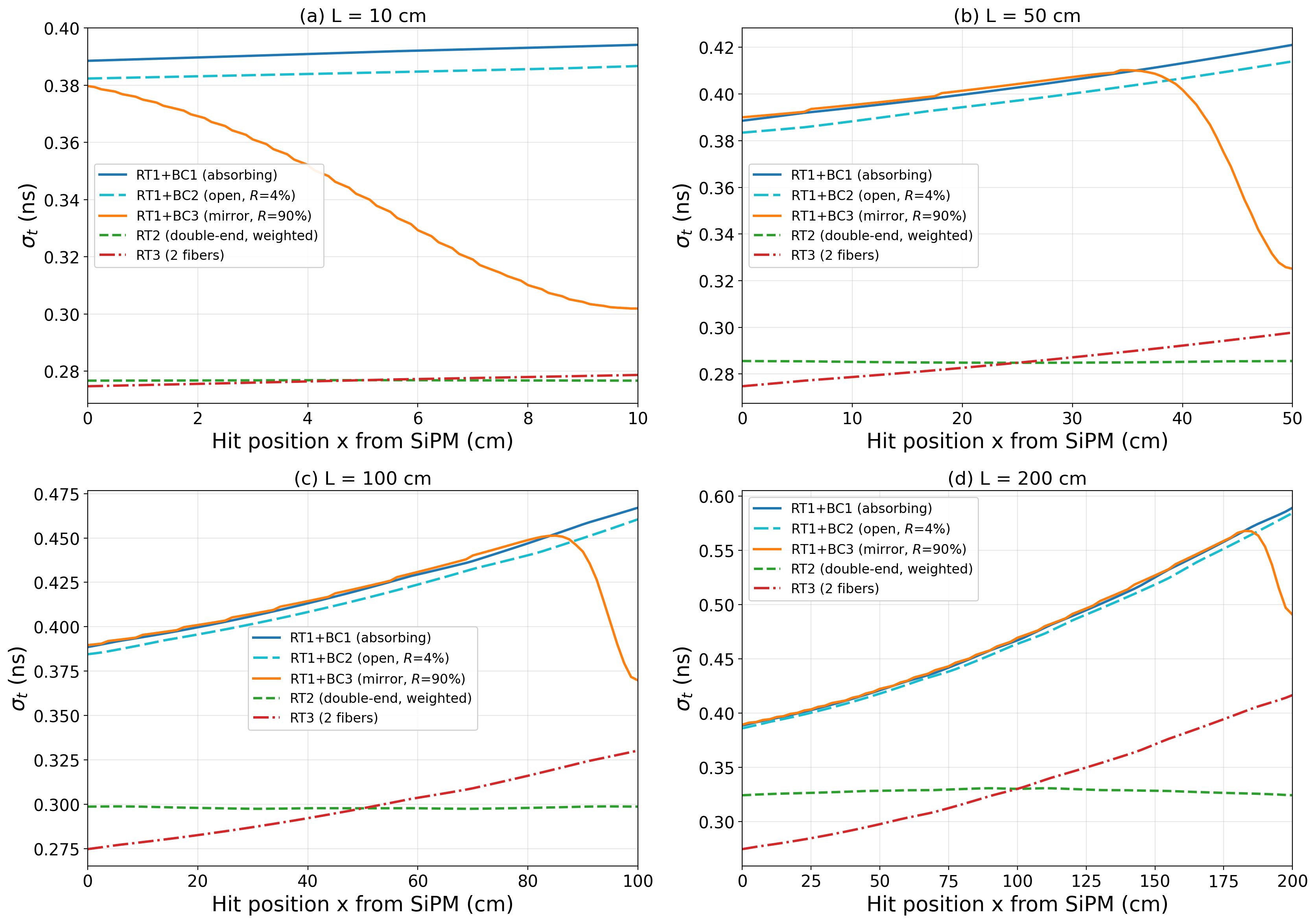}
\caption{Position dependence of $\sigt$ for bar
lengths of (a) 10~cm, (b) 50~cm, (c) 100~cm, and (d) 200~cm.
Multiple boundary conditions (BC1, BC2, BC3) and readout topologies (RT1, RT2, RT3 with two
fibers) are
compared (Y-11 fiber, EJ-200, $\Npe = 108$ at the SiPM end, ideal electronics). Double-ended readout provides nearly flat response regardless of position.}
\label{fig:position}
\end{figure}

Figure~\ref{fig:position} shows the dependence of $\sigt$ on the hit position $x$
along the detector for different boundary conditions and readout topologies at bar
lengths of 10, 50, 100, and 200~cm. For a
100~cm bar [panel~(c)], single-ended readout with BC1 (absorbing far end) or BC2
(open, $R \approx 4\%$) shows gradual degradation from the near end ($x \approx 0$)
to the far end ($x \approx L$) due to the exponential attenuation of $\Npe$ in the
WLS fiber. Double-ended readout (RT2) with weighted time averaging provides a nearly position-independent
resolution, as the loss in one channel is compensated by the gain in the other. This
``self-correcting'' property makes RT2 the preferred topology for long detectors.
Multi-fiber readout (RT3, two fibers) gives the same $\sqrt{2}$ statistical gain as RT2
but, with all fibers read out at the same end, retains the position dependence of RT1.
For a short detector [panel~(a), 10~cm], BC1 and BC2 conditions give
similar performance because transit times are negligible.
For the BC3 (mirror, $R = 90\%$) configuration, photons reaching the far end are reflected back toward the SiPM. The detected light is therefore a mixture of two populations: direct photons with path length $x$ and reflected photons with path length $2L - x$. Each population has its own transit time. The reflected photons arrive later by $\Delta t_\mathrm{reflect} = (2L - 2x) \cdot n_\mathrm{core} / c$. The analytical model handles this by constructing a mixture PDF: $f_\mathrm{det}^\mathrm{BC3}(t) = w_\mathrm{direct} f_\mathrm{det}(t) + w_\mathrm{reflect} f_\mathrm{det}(t - \Delta t_\mathrm{reflect})$, where the weights are determined by the attenuation and mirror reflectivity. Crucially, for a $k = 1$ (first-photon) trigger, the timestamp is $T_{(1)} = \min$ over all detected photons. Because the mirror only adds reflected photons (it does not remove any direct photons), it can only help or be neutral: $T_{(1)}^{\mathrm{BC3}} \leq T_{(1)}^{\mathrm{BC1}}$. For a long bar ($L = 100$~cm), the reflected photons arrive $\sim$$2L \cdot n_\mathrm{core}/c \approx 10$~ns after the direct photons, far too late to compete for the first arrival. In this regime BC3 is effectively equivalent to BC1. For a short bar ($L \lesssim 20$~cm), the mirror delay is only $\sim$1~ns, comparable to the WLS re-emission time, so the reflected photons arrive early enough to augment the leading-edge photon population and modestly improve the first-photon timing.

\begin{figure}[htbp]
\centering
\includegraphics[width=\textwidth]{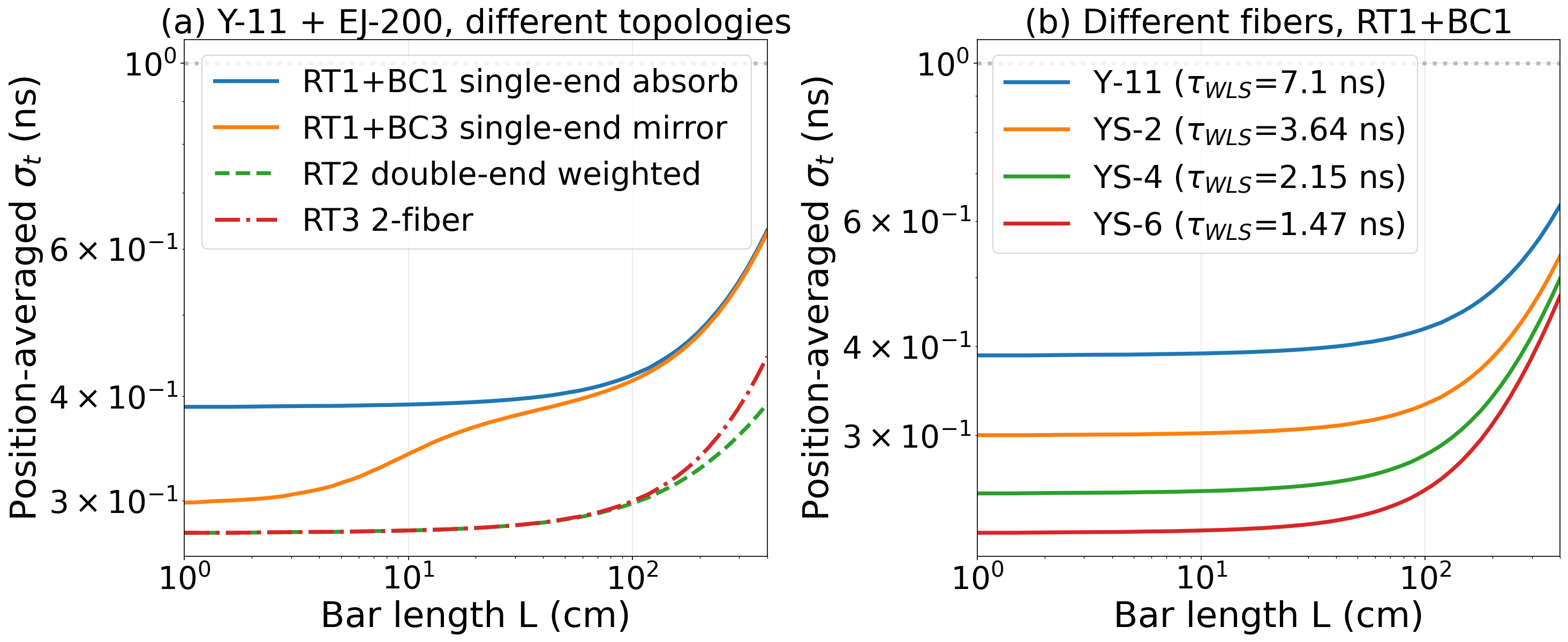}
\caption{Bar length scaling. (a) Position-averaged $\sigt$ versus bar length for
different readout topologies with Y-11 fiber. (b) Length dependence for different
fiber types with RT1+BC1. BC3 (mirror) helps at short lengths and converges to BC1
for long bars where the reflected photons arrive too late to affect the first-photon
trigger.}
\label{fig:length}
\end{figure}

\begin{figure}[htbp]
\centering
\includegraphics[width=\textwidth]{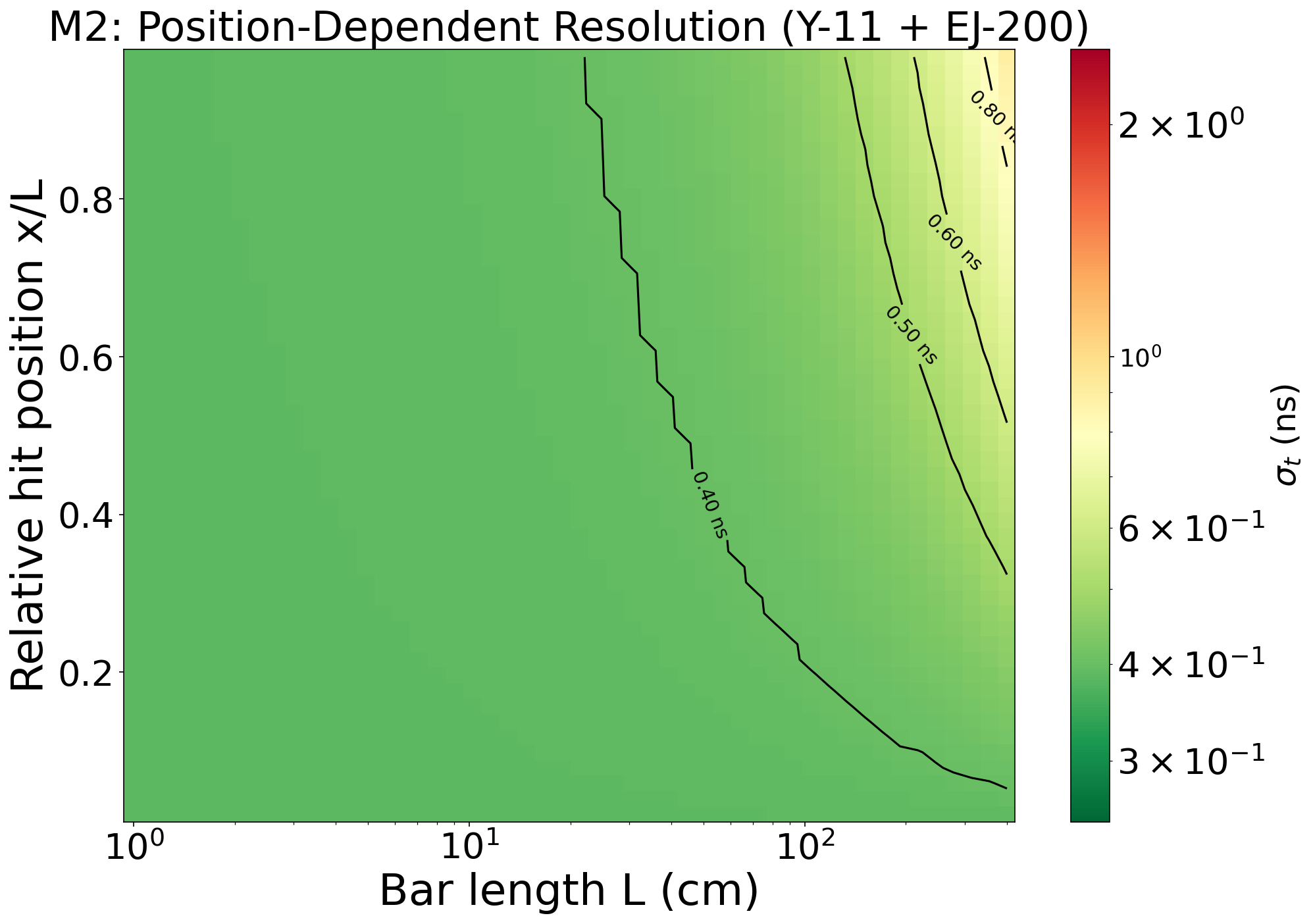}
\caption{Position-length map showing $\sigt$ as a function of bar length $L$ and
normalized hit position $x/L$ for RT1+BC1 readout with Y-11 fiber.}
\label{fig:map_position}
\end{figure}

The same behavior appears in the length scan of figure~\ref{fig:length}. Single-ended
readout degrades gradually with bar length as fiber attenuation reduces $\Npe$ at the midpoint. Mirroring the far end improves short detectors but converges to the absorbing case
for long detectors. Double-ended readout and
multi-fiber readout provide a $\sqrt{2}$ improvement and remain nearly length-independent
up to $L \approx 100$~cm. Beyond 200~cm, even Double-ended readout degrades due to severe attenuation
in both channels. 
A two-dimensional plot can be formulated to visualize the full landscape between the position and length parameters. Figure~\ref{fig:map_position} shows $\sigt$ in the $(L, x/L)$ plane for single-ended
readout with Y-11 fiber. The 1.0~ns contour curves sharply at $L > 100$~cm near the far
end, indicating that long detectors require double-ended readout to maintain uniform
timing performance.

\subsection{Electronics, SiPM timing, and timing pickoff threshold}
\label{sec:electronics}

\begin{figure}[htbp]
\centering
\includegraphics[width=\textwidth]{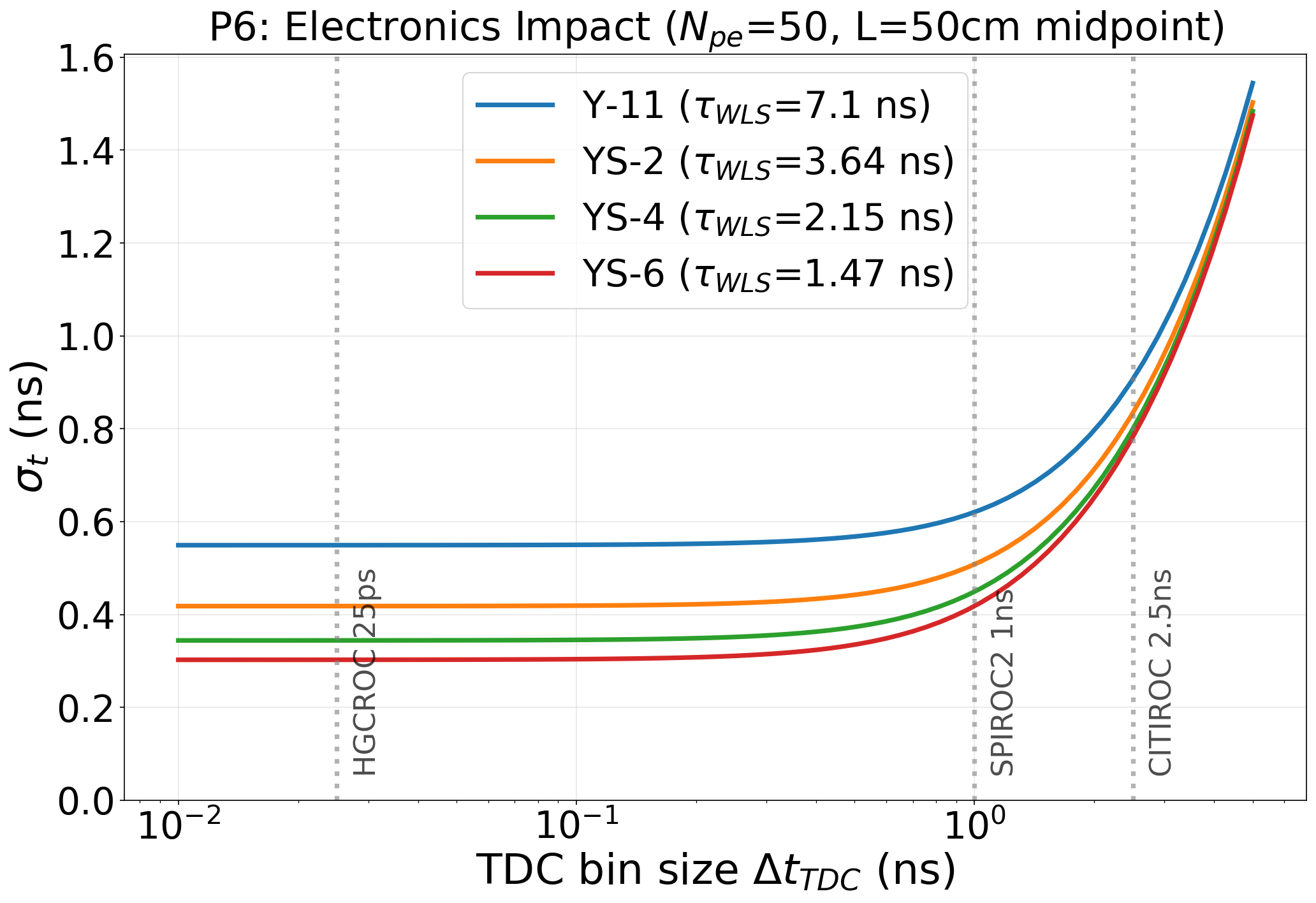}
\caption{Impact of TDC bin size $\Delta t_\mathrm{TDC}$ on timing resolution. The
transition from photon-statistics-limited to electronics-limited operation occurs
around $\Delta t_\mathrm{TDC}\approx 2$~ns for typical Y-11 configurations with
$\Npe=50$. Different curves correspond to different WLS fibers.}
\label{fig:electronics}
\end{figure}

Electronics matter when the digitization contribution is comparable to the photon
statistics term. The TDC bin size contributes
$\sigelec = \Delta t_\mathrm{TDC}/\sqrt{12}$, which adds in quadrature with the
photon-statistics and transit-time terms. Figure~\ref{fig:electronics} shows the
transition from the photon-statistics-limited regime to the electronics-limited
regime. For Y-11 fiber with $\Npe=50$, the crossover occurs at
$\Delta t_\mathrm{TDC}\approx 2$~ns. CITIROC, with a 2.5~ns TDC bin, is therefore
not transparent for this configuration, while HGCROC and DRS4-like timing are
negligible on this scale.

\begin{figure}[htbp]
\centering
\includegraphics[width=\textwidth]{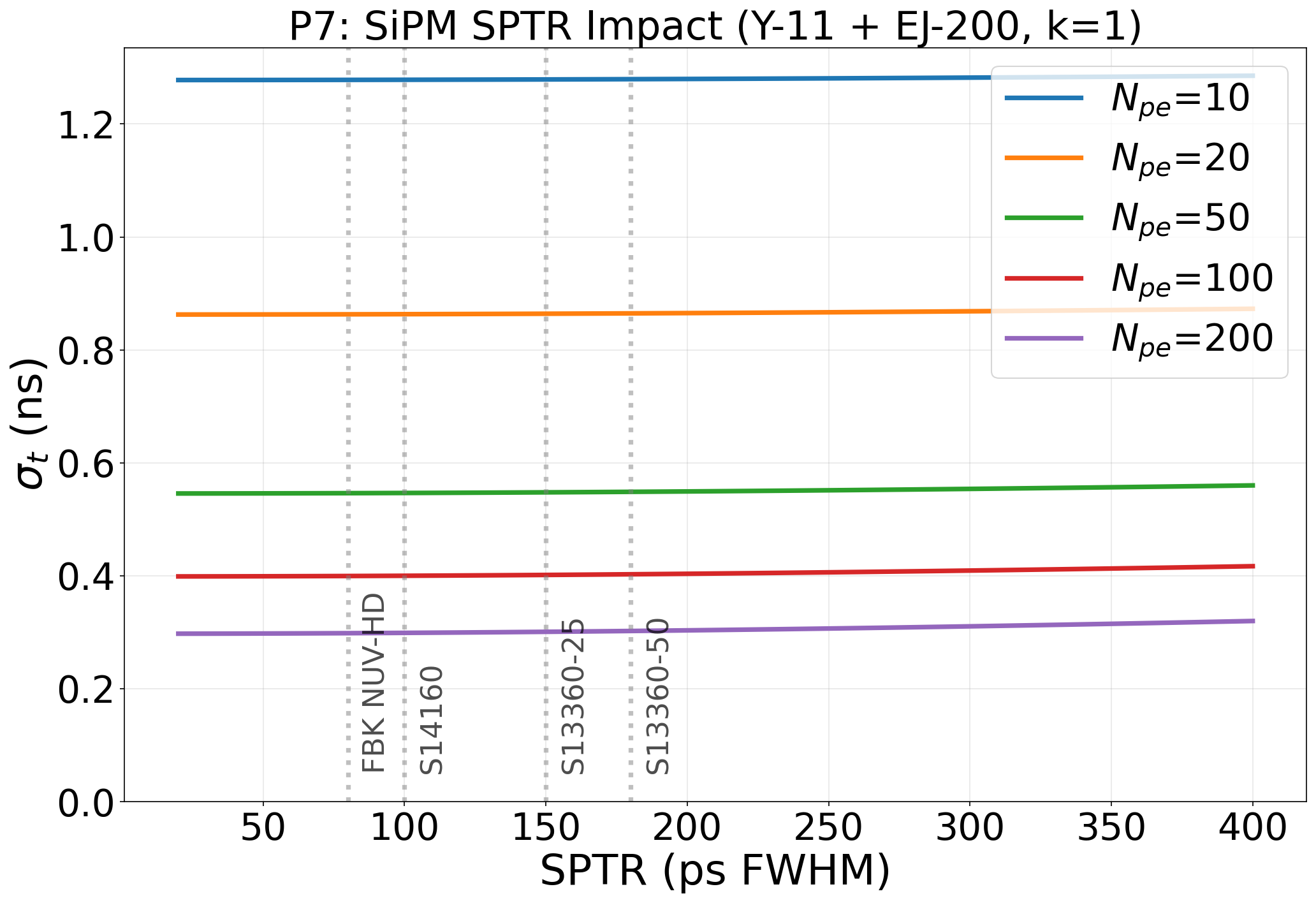}
\caption{SiPM SPTR impact for Y-11 fiber with EJ-200 scintillator at several $\Npe$
values. The dependence on SPTR is weak over the typical 80--300~ps FWHM range,
showing that SPTR is subdominant to photon statistics and WLS timing for most
configurations considered here. SPTR only matters for high-$\Npe$ ($> 100$) configurations with slow fibers.
For typical setups, FBK NUV-HD (80~ps) provides no measurable advantage over standard
Hamamatsu devices (150~ps).}
\label{fig:sptr}
\end{figure}

By contrast, the SiPM SPTR is usually a secondary effect. Figure~\ref{fig:sptr}
shows the Y-11+EJ-200 case at several values of $\Npe$. Varying the SPTR over a
large range changes the total resolution only weakly because the dominant timing
spread is set by the scintillator+WLS photon-time distribution and by photon
statistics. For SiPM choice, PDE is usually more important than SPTR because it
changes $\Npe$ directly; the corresponding two-dimensional map is given in
appendix~\ref{app:analytical_extra}.

\begin{figure}[htbp]
\centering
\includegraphics[width=\textwidth]{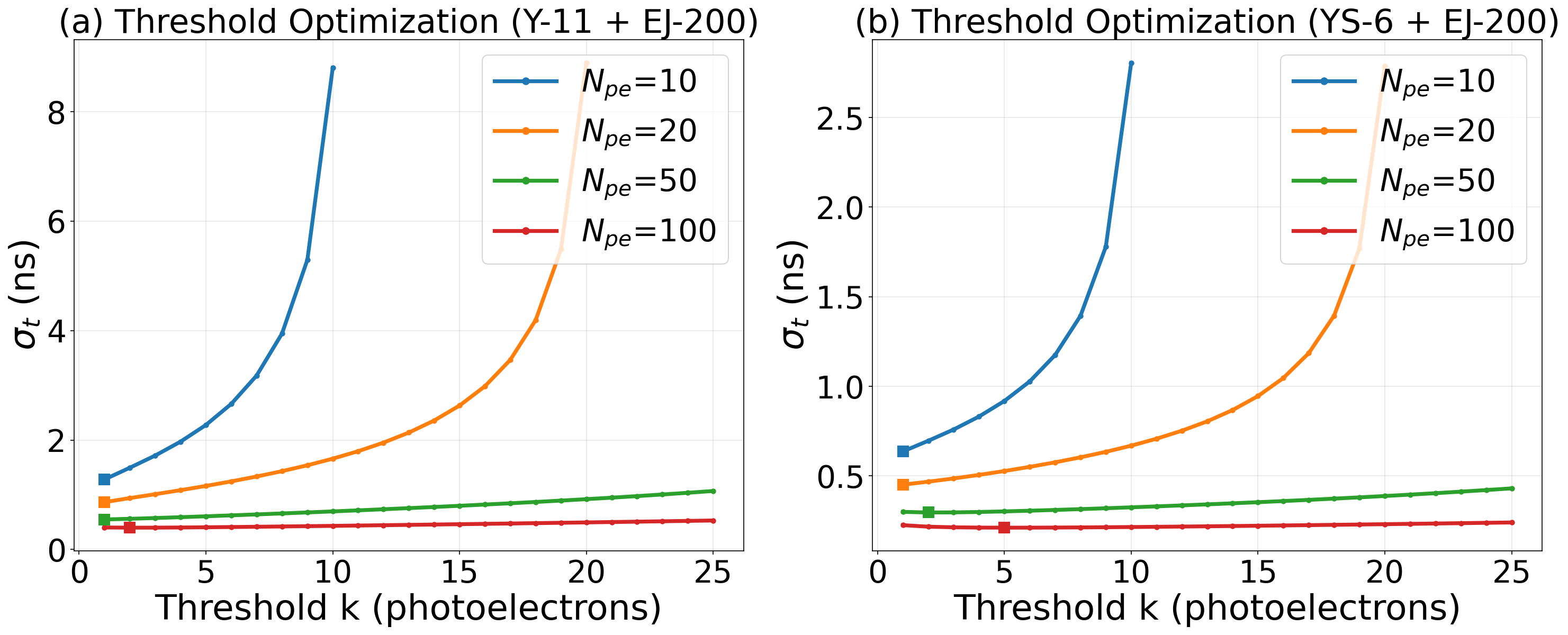}
\caption{Optimal leading-edge threshold $k$. The timing resolution $\sigt$ is shown versus $k$ for different $\Npe$ values.
For the convolved scintillator+WLS detection PDF, $k=1$ is optimal across the
practical range shown.}
\label{fig:threshold}
\end{figure}

For leading-edge timing, the analytical order-statistic calculation predicts that
$k=1$ is optimal across the practical $\Npe$ range considered here. Figure~\ref{fig:threshold}
shows this explicitly: moving to later photoelectrons reduces sensitivity to
single-photon jitter but loses more information by triggering on a slower part of
the photon-time distribution. More detailed timing-pickoff comparisons, including
constant-fraction discrimination in the Toy Monte Carlo, are presented in
section~\ref{sec:mc_pickoff}.

\subsection{Variance budget and sensitivity}
\label{sec:variance}

\begin{figure}[htbp]
\centering
\includegraphics[width=\textwidth]{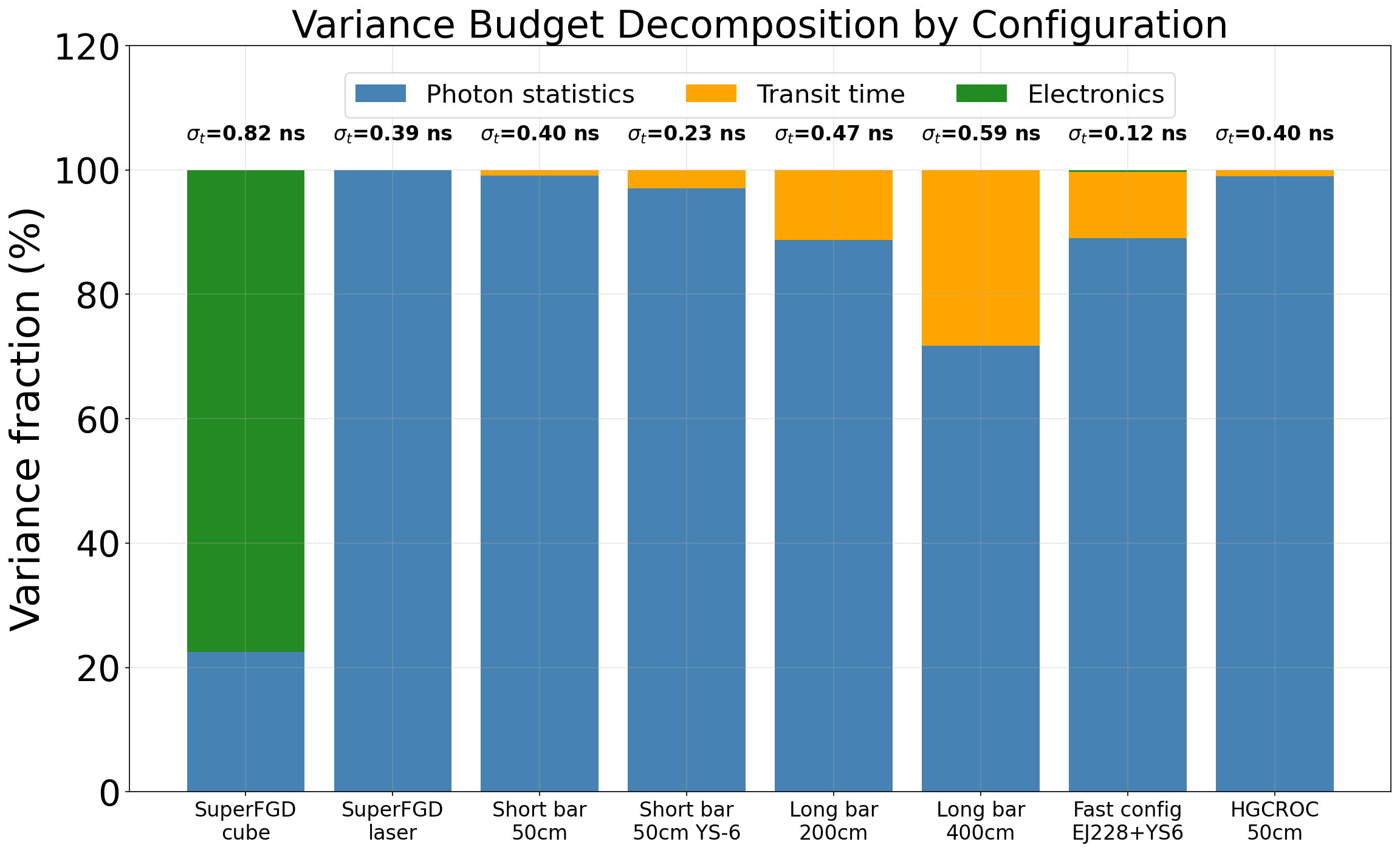}
\caption{Variance budget decomposition for eight representative detector configurations.
The components are photon statistics, fiber transit time, and electronics. Photon
statistics dominate in all configurations except those with coarse TDC quantization.
The total $\sigt$ value is annotated above each bar.}
\label{fig:variance}
\end{figure}

The variance decomposition in figure~\ref{fig:variance} shows the resulting design
hierarchy. Photon statistics dominate most configurations, making $\Npe$ and $\tauwls$ the primary
parameters for optimization. The main exception is a
configuration with coarse timing electronics, such as SuperFGD cube with CITIROC, where the TDC term
can become the largest contribution. Transit-time spread becomes visible for long
bars, but it remains a secondary contribution compared with photon statistics for
most lengths considered here.

\begin{figure}[htbp]
\centering
\includegraphics[width=\textwidth]{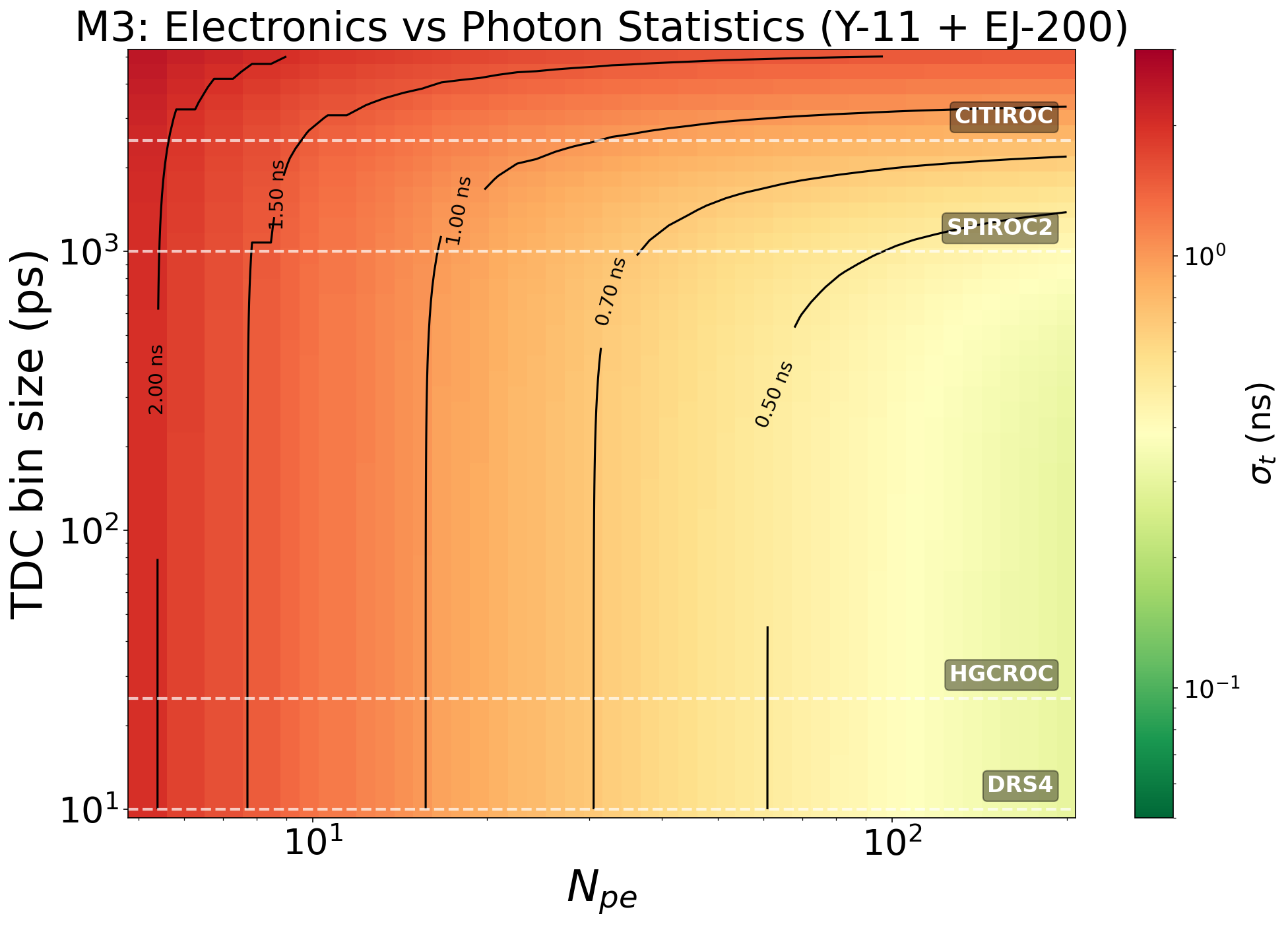}
\caption{Electronics regime diagram: $\sigt$ in the $(\Npe, \Delta t_\mathrm{TDC})$
plane. The diagonal boundary separates photon-statistics-dominated (lower right) from
electronics-dominated (upper left) operation. Specific ASIC positions are marked.}
\label{fig:map_elec}
\end{figure}

Figure~\ref{fig:map_elec} shows $\sigt$ in the $(\Npe, \Delta t_\mathrm{TDC})$ plane,
creating a ``regime diagram'' that delineates the boundary between photon-statistics-dominated
and electronics-dominated operation. Below and to the right of the diagonal, photon
statistics dominate. Above and to the left, electronics compete or dominate. The CITIROC
ASIC at $\Delta t_\mathrm{TDC} = 2.5$~ns enters the electronics-limited regime for
$\Npe > 30$.

\begin{figure}[htbp]
\centering
\includegraphics[width=\textwidth]{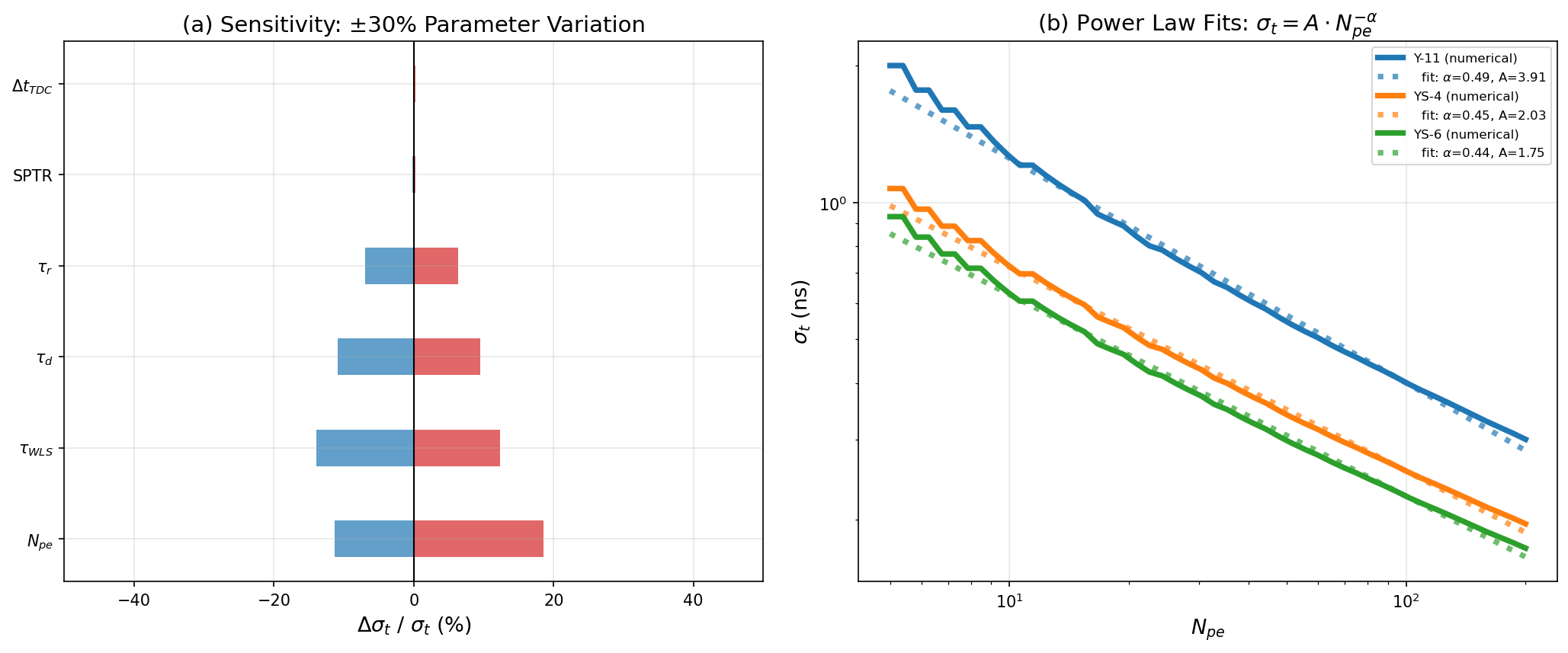}
\caption{Sensitivity analysis. (a) Tornado chart showing the effect of $\pm30\%$
parameter variations around the EJ-200+Y-11, $\Npe=50$ baseline. (b) Power-law
fits $\sigt=A\Npe^{-\alpha}$ for three fiber types, with fitted $\alpha$ values shown.}
\label{fig:sensitivity}
\end{figure}

Figure~\ref{fig:sensitivity} reaches the same conclusion from a parameter-variation
perspective. Around the EJ-200+Y-11, $\Npe=50$ baseline, $\Npe$ and $\tauwls$ give
the largest changes in timing resolution. The scintillator decay time is the next
most important material parameter, while SPTR and fine TDC binning are weak levers
once the system is photon-statistics limited.
Figure~\ref{fig:sensitivity}(b) fits the time resolution with power laws
$\sigt = A\,\Npe^{-\alpha}$ over $5 \leq \Npe \leq 200$, with exponents
$\alpha \approx 0.44--0.49$. The exponent is close to $1/2$ for all fibers, so doubling the light yield gains
only about 27\% because of the finite rise time of the detection PDF. More importantly, this shows changing fibers have only scaling factor effect over the entire resolution curve that stays the same for $\Npe = 10 \text{or} 100$. Additionally, the fitted lines cross the
numerical curves, whose local slope is steeper at low $\Npe$, a point relevant to the Geant4
comparison in section~\ref{sec:g4_parameter_scans}.

\section{Toy Monte Carlo validation}
\label{sec:mc}
The predictions above are done deterministically with fixed analytical parameters and $\Npe$ values. To validate these calculations, we construct a Toy Monte Carlo model with a high number of independent events. For a given set of parameters, each event is processed by having its $\Npe$ drawn from Poisson $(\langle\Npe\rangle)$. For each photoelectron, scintillation emission time is sampled from $\fscint$ (equation \ref{eq:fscint}), as are the WLS re-emission, fiber transit time, and SiPM single photon time resolution jitter from their respective PDF curves. For the mirror boundary condition (BC3), reflected photons with the additional path length and the corresponding extra fiber transit time are included. Finally, the $k$-th photon trigger is applied to record the event timestamp. 
The MC uses no analytical approximations beyond the parameterized
time PDFs described in section~\ref{sec:model}.
Unless stated otherwise, $2\times10^5$ events are generated per parameter point. The statistical uncertainty is calculated as $\delta s = s\sqrt{(\kappa-1)/(4n)}$, where $s$ is the standard deviation of $n$ recorded events, or $\sigt$, and $\kappa$ is the sample kurtosis
of the timestamp distribution. For
$2\times10^5$ events, $\delta s/s \approx 0.1$--$0.3\%$, so the error bars are often smaller
than the markers.

\subsection{Convolution and order statistic validation}

\begin{figure}[htbp]
\centering
\includegraphics[width=\textwidth]{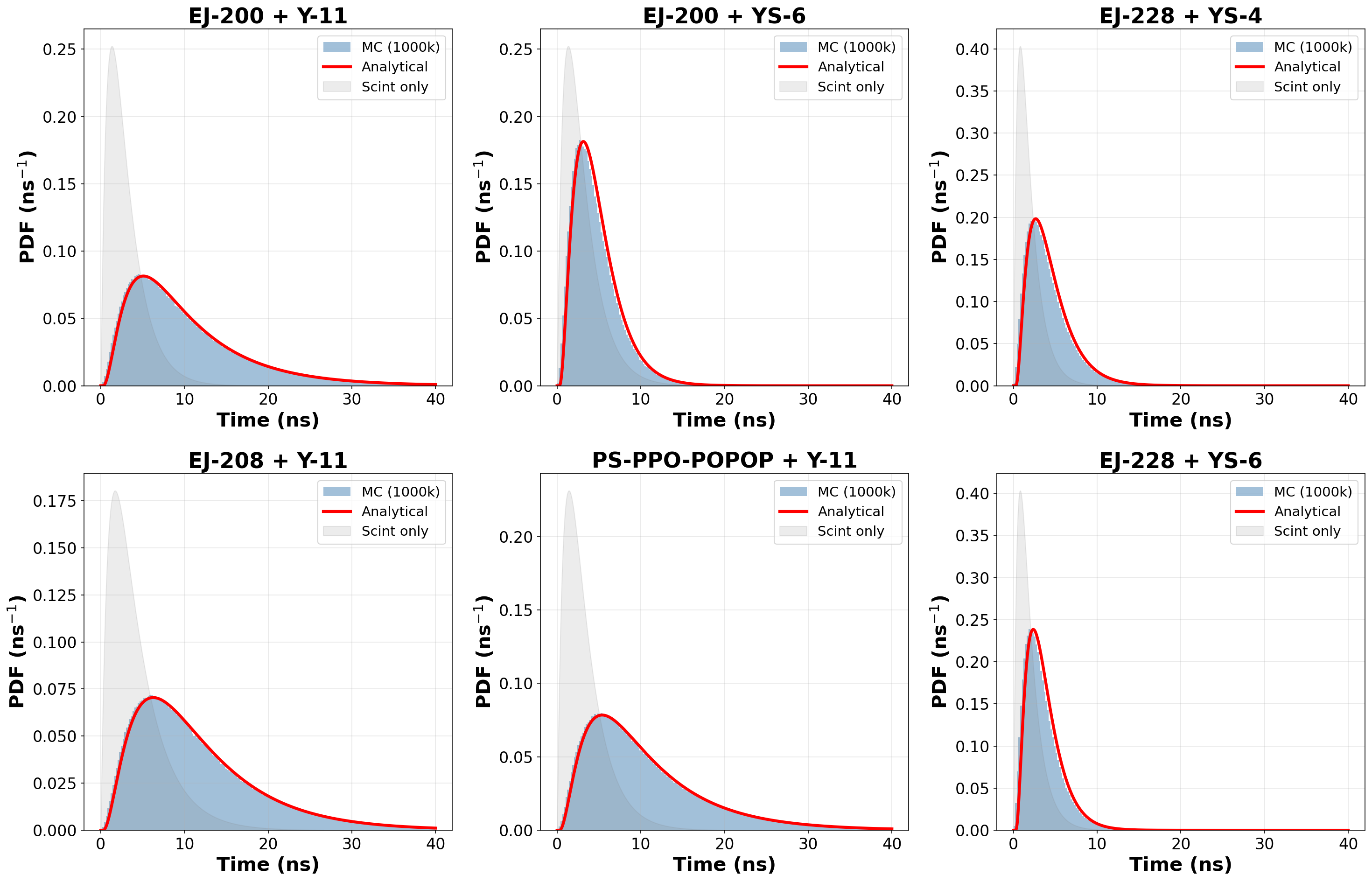}
\caption{Convolution validation: MC-sampled histograms (blue, $10^6$ samples) versus
analytical PDFs (red lines) for six scintillator+WLS fiber combinations. The gray
shaded region shows the scintillator-only PDF for comparison.}
\label{fig:mc_conv}
\end{figure}

The first two checks isolate the two most dominant parts identified in section~\ref{sec:analytical}: effects of $f_\mathrm{scint+WLS}(t)$ and the order statistics.\\
Figure~\ref{fig:mc_conv} compares Toy MC samples of $f_\mathrm{scint+WLS}(t)$ with the
closed-form convolution for several scintillator+fiber combinations. The sampled histograms
reproduce the analytical PDFs, validating the sum-of-exponentials sampling approach. Kolmogorov-Smirnov tests confirm agreement ($p > 0.05$ for all six combinations),
validating that the sum-of-exponentials sampling correctly reproduces the
analytical convolution.

\begin{figure}[htbp]
\centering
\includegraphics[width=\textwidth]{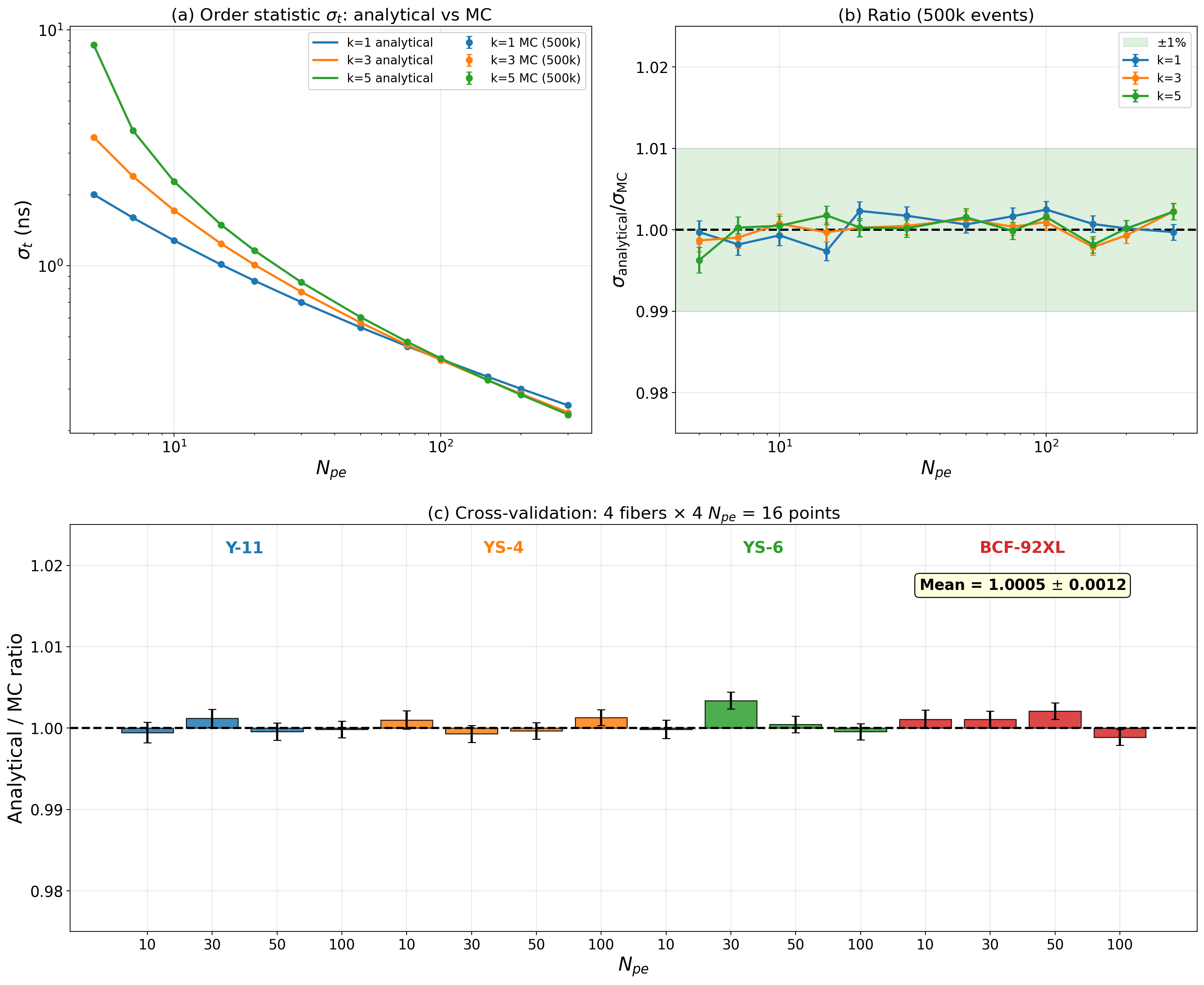}
\caption{Order statistic validation ($5 \times 10^5$ events per point). (a) Analytical
(lines) vs MC (symbols) for $k = 1, 3, 5$. (b) Ratio plot: mean $= 1.0002 \pm 0.0014$.
(c) Cross-validation across 16 independent scintillator+fiber+$\Npe$ combinations.}
\label{fig:mc_order}
\end{figure}

Figure~\ref{fig:mc_order} then compares the analytical order-statistic integration with the
Toy MC timestamp distribution. The ratio is centered at unity at the $10^{-3}$ level. The
cross-validation panel uses four fiber choices and four $\Npe$ values, giving 16 independent
points, all within the plotted sub-percent band.

\subsection{Full fixed-$\Npe$ parameter scan}
\label{sec:mc_scan}

\begin{figure}[htbp]
\centering
\includegraphics[width=\textwidth]{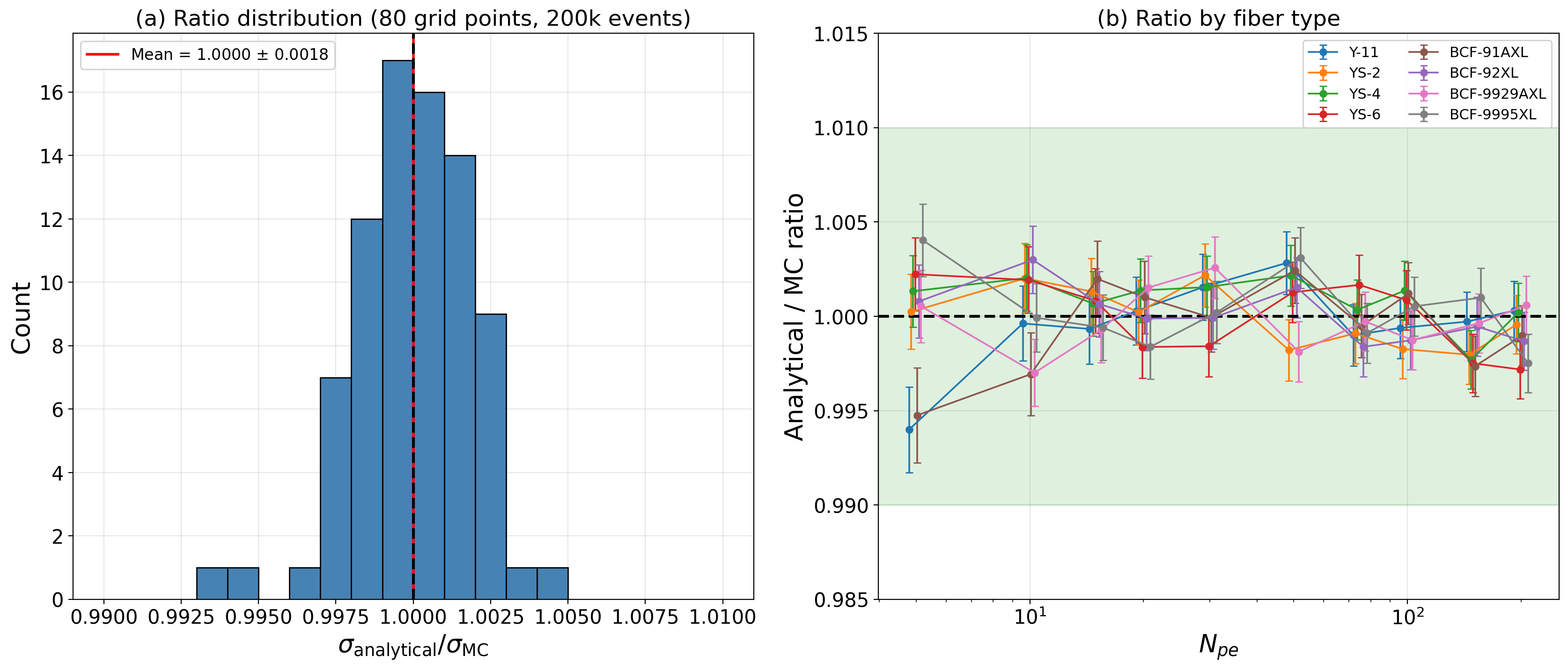}
\caption{Full parameter scan validation: analytical versus Toy MC ratio across 80 fixed-$\Npe$
grid points. (a) Ratio distribution with mean $1.0000\pm0.0018$. (b) Ratio versus $\Npe$ by
fiber type, showing no systematic fiber-dependent bias; error bars show the MC statistical
uncertainty.}
\label{fig:mc_scan}
\end{figure}

Figure~\ref{fig:mc_scan} gives the full fixed-$\Npe$ validation over 80 grid points
(8 fibers $\times$ 10 $\Npe$ values). The ratio
$\sigt^\mathrm{analytical}/\sigt^\mathrm{MC}$ has mean $1.0000\pm0.0018$
(rms spread of the 80 ratios, equal to the statistical uncertainty of a single point), with all
points within $\pm0.6\%$. This demonstrates that the analytical order-statistic calculation
and the event-by-event Toy MC are numerically consistent when the assumptions are matched.

\subsection{Effects beyond the fixed-$\Npe$ analytical calculation}
\label{sec:mc_beyond_fixed}

\begin{figure}[htbp]
\centering
\includegraphics[width=\textwidth]{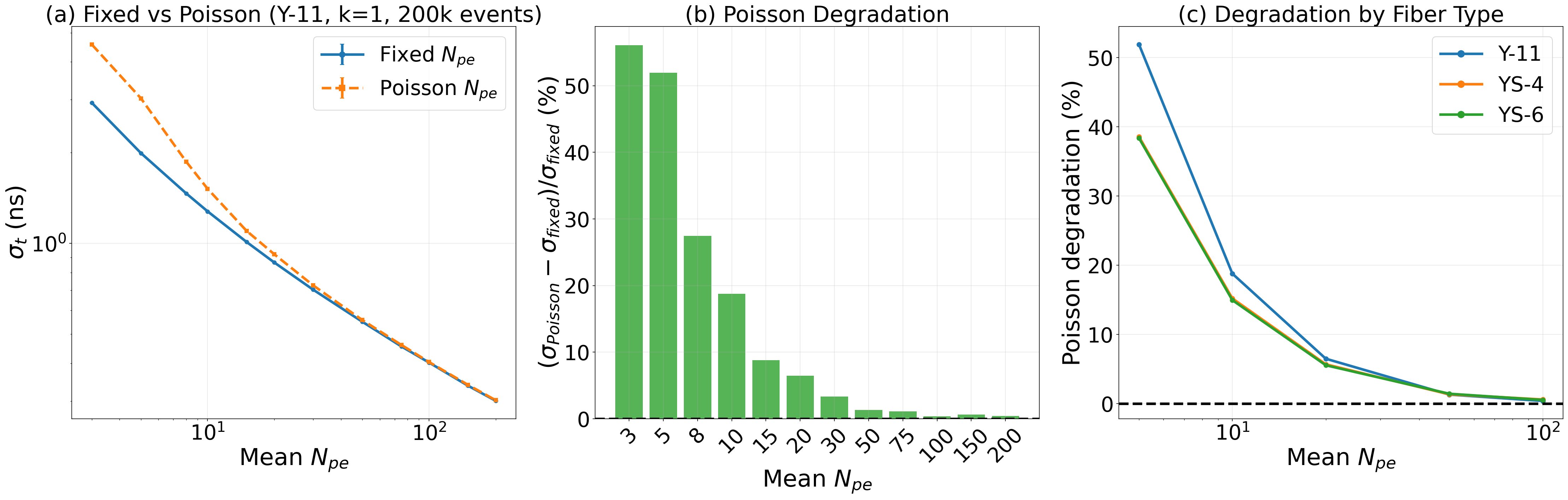}
\caption{Impact of Poisson fluctuations in $\Npe$ on timing resolution. Fixed-$N$ and
Poisson-$N$ Toy MC results are compared for three fiber types across $\Npe$ from 3 to 200.}
\label{fig:mc_poisson}
\end{figure}

The baseline analytical calculation evaluates the order statistic at fixed $\Npe$. Real events,
however, have light-yield fluctuations. Figure~\ref{fig:mc_poisson} compares fixed-$N$ and
Poisson-distributed photon counts. The degradation is large at very low light yield, where a
substantial fraction of events contain only one or two detected photons, but it falls below about
8\% for $\Npe>20$. This quantifies the slight inaccuracy in the analytical model for low-$\Npe$ applications, but also indicates that the analytical predictions are reliable for $\Npe > 20$ systems, which covers most practical configurations.

The Toy MC also checks detector-level effects that are cumbersome to express in a single closed
form. The results validate the boundary-condition picture: a mirror helps for short
bars, where reflected photons arrive early enough to contribute to the first-photon population,
but becomes nearly neutral for long bars. The MC with a BC3 configuration model confirms the double-ended
readout behavior and its hovering around the expected $\sqrt{2}$ improvement for nearly symmetric
channels. The ideal $1/\sqrt{N_\mathrm{ch}}$ scaling for independent multi-channel was also verified with the Toy MC results averaging to within about $\pm1\%$ of the analytical expectation.  
The detailed boundary-condition, double-ended-readout, and multi-channel comparison plots are presented in appendix~\ref{app:mc_detector_effects}.

\subsection{Timing discrimination methods}
\label{sec:mc_pickoff}

\begin{figure}[htbp]
\centering
\includegraphics[width=\textwidth]{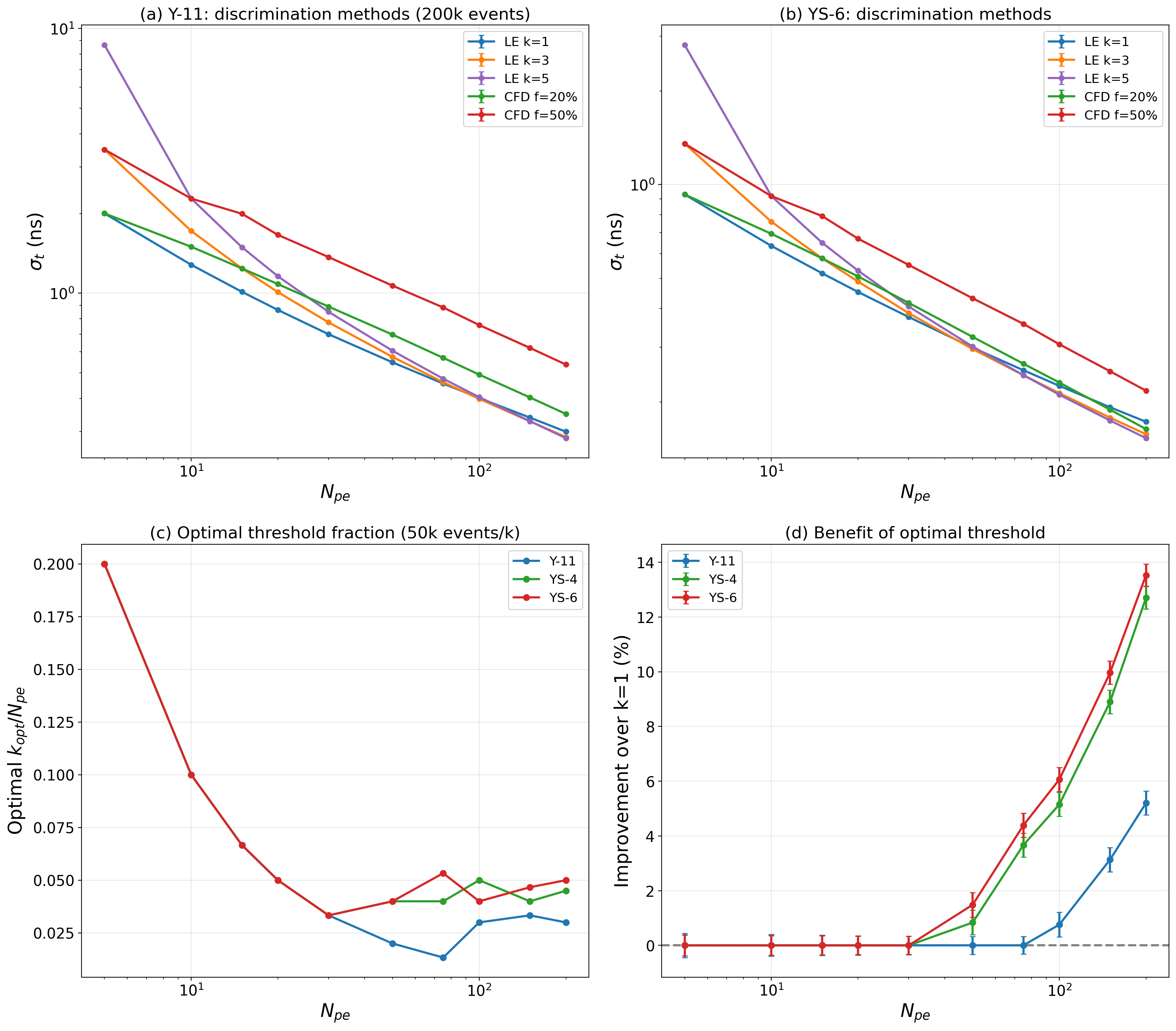}
\caption{Timing discrimination method comparison ($2 \times 10^5$ events per point).
(a) Y-11: LE ($k = 1, 3, 5$) and CFD (20\%, 50\%). (b) Same for YS-6. (c) Optimal
threshold fraction $k_\mathrm{opt}/\Npe$ ($5\times10^4$ events per point). (d) Improvement from optimal $k$ over $k = 1$.}
\label{fig:mc_cfd}
\end{figure}

As shown in section \ref{sec:analytical}, $k$=1 seems to be the best overall triggering threshold for the detector. This result is further verified by Figure \ref{fig:mc_cfd}. For both Y-11 and YS-6 fibers, $k$=1 outperforms constant-fraction discrimination (CFD) methods throughout all ranges and is only barely surpassed by higher $k$ thresholds at very high $\Npe$ regimes. This $k$=1-comparable performance by higher $k$ thresholds could result from the sharper rising edge of the detection-time PDF at high $\Npe$. 

\subsection{Decision matrix}
\label{sec:decision}

\begin{figure}[htbp]
\centering
\includegraphics[width=\textwidth]{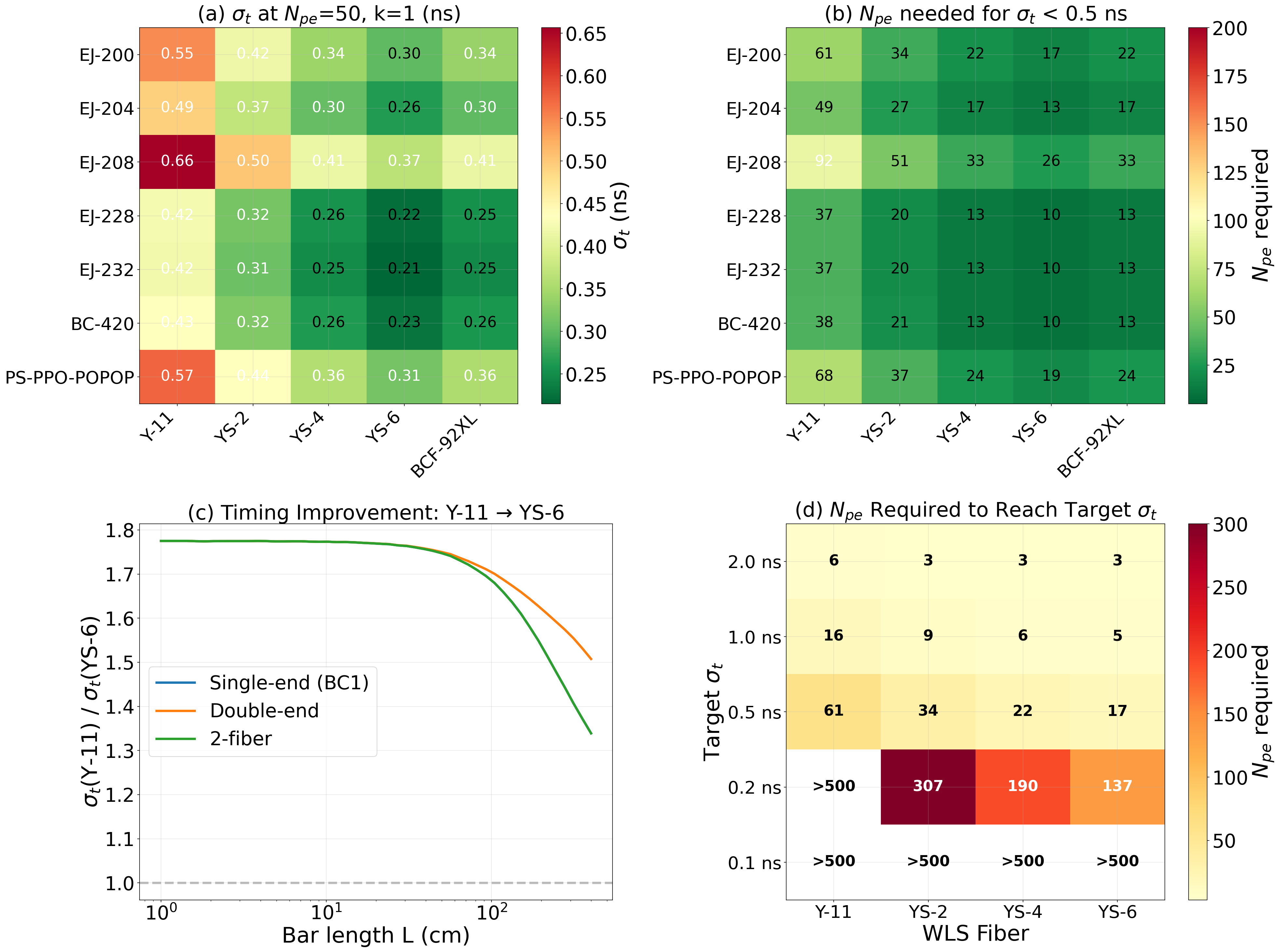}
\caption{Decision matrix for detector design. (a) $\sigt$ at $\Npe = 50$ for all
material combinations. (b) $\Npe$ required for $\sigt < 0.5$~ns. (c) Y-11 to YS-6
improvement ratio versus bar length. (d) $\Npe$ required for target $\sigt$ values.}
\label{fig:decision}
\end{figure}

Figure~\ref{fig:decision} consolidates the study results into actionable design guidance.
Panel~(a) shows $\sigt$ at $\Npe = 50$ for all scintillator$\times$fiber combinations as
a heatmap. Panel~(b) shows the $\Npe$ required to achieve $\sigt < 0.5$~ns for each fiber.
Panel~(c) quantifies the improvement from switching Y-11 to YS-6 as a function of bar
length and readout topology. Panel~(d) presents the $\Npe$ required to reach various target
$\sigt$ values, providing a direct lookup for detector design. 
At fixed $\Npe$, fast scintillator+fiber combinations give the best timing. For a target resolution,
faster fibers substantially reduce the required light yield. The gain from switching Y-11 to YS-6
is largest in the regime where WLS timing dominates and becomes less dramatic once other effects
or attenuation dominate. These trends lead to a practical hierarchy: first establish the achievable
$\Npe$, then choose the fastest viable WLS fiber, then select a readout topology that controls
position dependence, and finally ensure the electronics contribution is below the photon-statistics
floor.

\section{Geant4 optical simulation}
\label{sec:geant4}
The Toy MC validates the mathematical implementation of the analytical model, but it uses the
same parameterized photon-time distributions and simplified transport assumptions. A more
rigorous test is provided by a full Geant4 optical simulation that tracks scintillation photons,
WLS re-emission, fiber propagation, boundary interactions, and SiPM detection in a physics-first, three-dimensional geometry.\\

\subsection{Simulation setup}
\label{sec:g4_setup}
The simulation is implemented in Geant4 version 11.4.01~\cite{Geant4_2003,Geant4_2016}. The modeled detector contains a PVT
scintillator bar with configurable light yield, bi-exponential scintillation timing, wavelength-dependent
bulk absorption, and Birks quenching (Geant4 material \texttt{G4\_PLASTIC\_SC\_\allowbreak VINYLTOLUENE},
$n = 1.58$, bulk absorption length 380~cm, Birks constant $k_B = 0.126$~mm/MeV, $\taur = 0.9$~ns,
$\taud = 2.1$~ns). The WLS fiber is modeled as a double-clad structure with
core, inner-cladding, and outer-cladding refractive indices, wavelength-dependent absorption,
green re-emission, configurable WLS decay time, and bulk attenuation (350~cm) for trapped photons. The
scintillator wrapping is represented by a diffuse reflective optical surface, and the SiPM is modeled
as a silicon absorber with random PDE acceptance and Gaussian SPTR smearing. The far end can
be absorbing (BC1), open/Fresnel-like (BC2), or mirrored (BC3).

The primary particle is a 1~GeV $\mu^-$ incident perpendicular to the bar at a configurable
position along the fiber axis. Each parameter point is simulated with 1000 muons (10~cm bar)
or 800 muons (100~cm bar); the light yield is set per scan to reach the $\Npe$ regime of
interest (250 and 1000~photons/MeV for the 10~cm bar, 150--3000~photons/MeV for the $\Npe$ scan,
and 1500~photons/MeV for the 100~cm bar). For each event, the simulation records the detected photoelectron
yield and the first-photon timestamp at each instrumented end. The Geant4 $\sigt$ is the
standard deviation of the first-photon timestamps. The statistical uncertainty, shown as error
bars in all Geant4 figures, is obtained from 1000 bootstrap resamples of the events and amounts
to 2.5--5.5\% per point for $\langle\Npe\rangle \geq 10$ and 5.5--15\% for
$\langle\Npe\rangle < 10$. In all comparisons below, the
analytical prediction is evaluated using the Geant4-measured $\Npe$. This isolates the timing
model from the separate problem of predicting absolute light collection. The analytical
prediction is evaluated at fixed $N = \langle\Npe\rangle$, whereas the Geant4 $\sigt$ includes
the event-by-event fluctuations of the number of detected photoelectrons (Poisson statistics
and the Landau distribution of the energy deposit); the effect of such fluctuations was
quantified with the Toy MC in section~\ref{sec:mc_beyond_fixed} and is referred to below.

\subsection{WLS, position, boundary, and readout-topology checks}
\label{sec:g4_position_topology}

\begin{figure}[htbp]
\centering
\includegraphics[width=\textwidth]{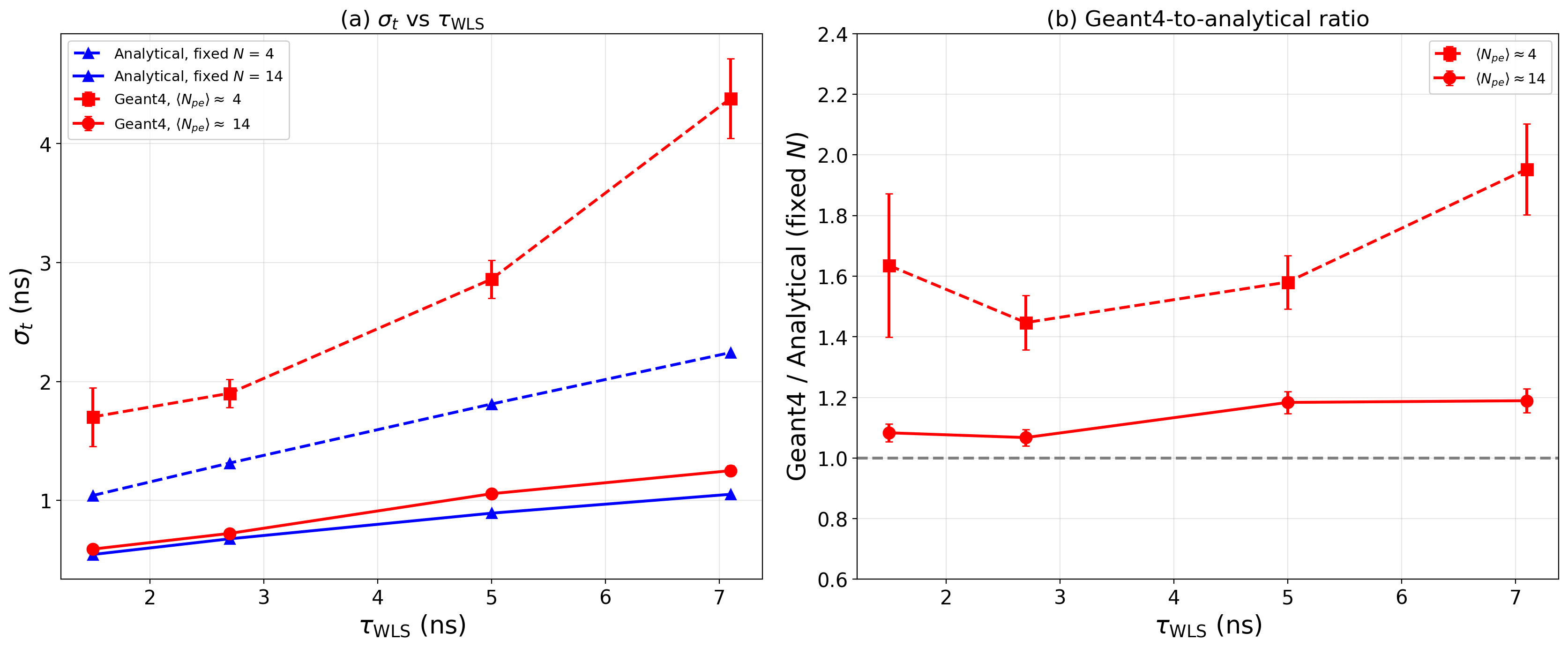}
\caption{WLS re-emission time scan at two $\Npe$ regimes.
(a) Dashed: original low-$\Npe$ ($\approx 4$) comparison showing Poisson-dominated
discrepancy. Solid: high-$\Npe$ ($\approx 14$) comparison.
Blue: analytical prediction at fixed $N = \langle \Npe \rangle$.
(b) G4/analytical ratio. Error bars
show the statistical uncertainty (bootstrap; 800 events per point).}
\label{fig:g4_wls}
\end{figure}

The first test verifies if the strong dependence on $\tauwls$ identified by the analytical analysis can be recovered. 
We vary $\tauwls$ from 1.5 to 7~ns
while keeping all other parameters fixed. Figure~\ref{fig:g4_wls} compares the Geant4
results with the analytical prediction. 

The default Geant4 configuration ($\langle \Npe \rangle \approx 4$) sits in a regime
where Poisson fluctuations dominate. To make the comparison meaningful, we run an
additional set of simulations at 4$\times$ the default light yield, raising
$\langle \Npe \rangle$ to $\approx 14$ (dashed lines show the original low-$\Npe$
results for reference). At $\Npe \approx 14$, the G4/analytical ratio is
1.07--1.19 ($\pm 0.03$--$0.04$) across the $\tauwls$ range, confirming that the analytical model
correctly captures the $\tauwls$ dependence. The offset has
the sign and size expected from the event-by-event fluctuations of $\Npe$, which the
fixed-$\Npe$ calculation does not include: the Toy MC of section~\ref{sec:mc_beyond_fixed}
(figure~\ref{fig:mc_poisson}) shows that Poisson fluctuations alone increase $\sigt$ by about
10\% at $\Npe \approx 14$ and by about 50\% at $\Npe \approx 4$, where the Geant4-to-analytical
ratio is 1.4--2.0. The fluctuations of $\Npe$ in the Geant4 simulation are somewhat wider than
Poisson because the energy deposit itself fluctuates (Landau distribution), which is consistent
with the ratios lying slightly above the Poisson expectation.

\begin{figure}[htbp]
\centering
\includegraphics[width=\textwidth]{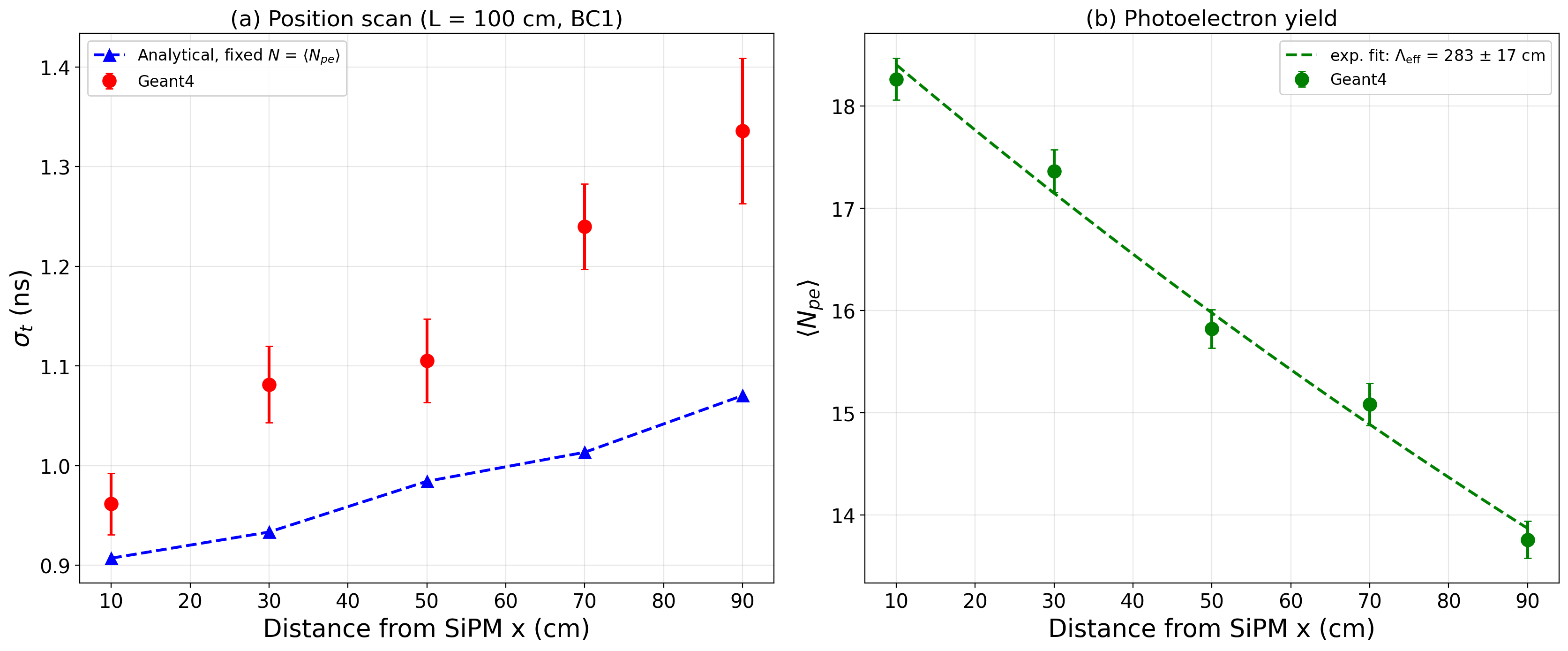}
\caption{Geant4 position scan for a 100~cm scintillator bar with single-ended readout (BC1).
Left: timing resolution from Geant4 (red circles; 800 events per point) compared with the analytical prediction
evaluated at the Geant4-measured $\Npe$ (blue triangles). Right: mean detected photoelectron
yield versus position, showing the expected exponential attenuation (fit:
$\Lambda_\mathrm{eff} = 283 \pm 17$~cm). Hit positions are given as the distance from the SiPM.}
\label{fig:g4_position}
\end{figure}

The second validation tests whether the analytical model correctly predicts the
position dependence of $\sigt$ for a 100~cm bar with single-ended BC1 readout.
As the hit position moves away from the SiPM, $\Npe$ decreases exponentially
due to fiber attenuation, and $\sigt$ increases correspondingly.
Figure~\ref{fig:g4_position} compares the Geant4 results with the analytical
prediction evaluated at the Geant4-measured $\Npe$. The prediction
reproduces the rising trend; the Geant4 points lie 6--25\% above it, an offset of the same size
as in the $\tauwls$ scan at a similar $\langle \Npe \rangle$ of 14--18.

\begin{figure}[htbp]
\centering
\includegraphics[width=0.65\textwidth]{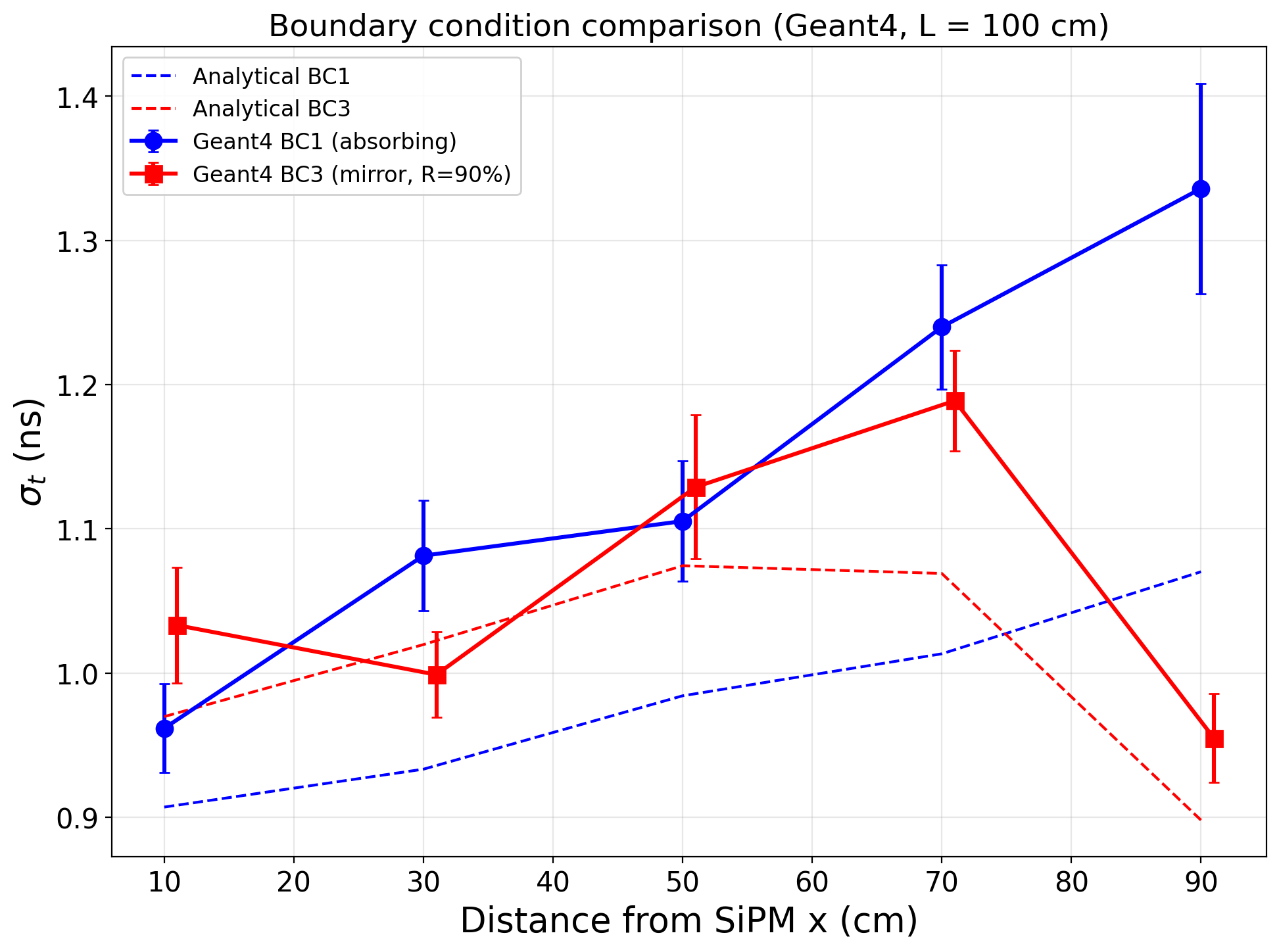}
\caption{Geant4 boundary condition comparison: absorbing (BC1, blue) versus mirror
(BC3, red) far end for a 100~cm bar (dashed lines: analytical predictions; 800 events per
point). The mirror provides a significant improvement only for hits
close to the mirrored end.}
\label{fig:g4_boundary}
\end{figure}

The third test validates the mirror boundary prediction. For a $k = 1$ trigger, the
analytical model predicts that BC3 should be effectively neutral for a 100~cm bar because
reflected photons arrive $\sim$$2L \cdot n_\mathrm{core}/c \approx 10$~ns after the direct
photons, far too late to affect the first-photon arrival, except for hits close to the
mirrored end, where the reflected photons arrive within about 1~ns
(figure~\ref{fig:position}(c)). Figure~\ref{fig:g4_boundary}
confirms this: the BC1 and BC3 curves are consistent
within the statistical uncertainties for hits 10--70~cm from the SiPM, while at 90~cm (close, 10~cm,
from the mirror) BC3 improves the resolution from $1.34 \pm 0.07$ to $0.96 \pm 0.03$~ns.
This is consistent with the analytical expectation and the toy MC boundary condition
study (appendix~\ref{app:mc_detector_effects}): the mirror only improves the resolution when the hit position is close to the mirror enough that the reflected photons can boost $\Npe$.

\begin{figure}[htbp]
\centering
\includegraphics[width=0.65\textwidth]{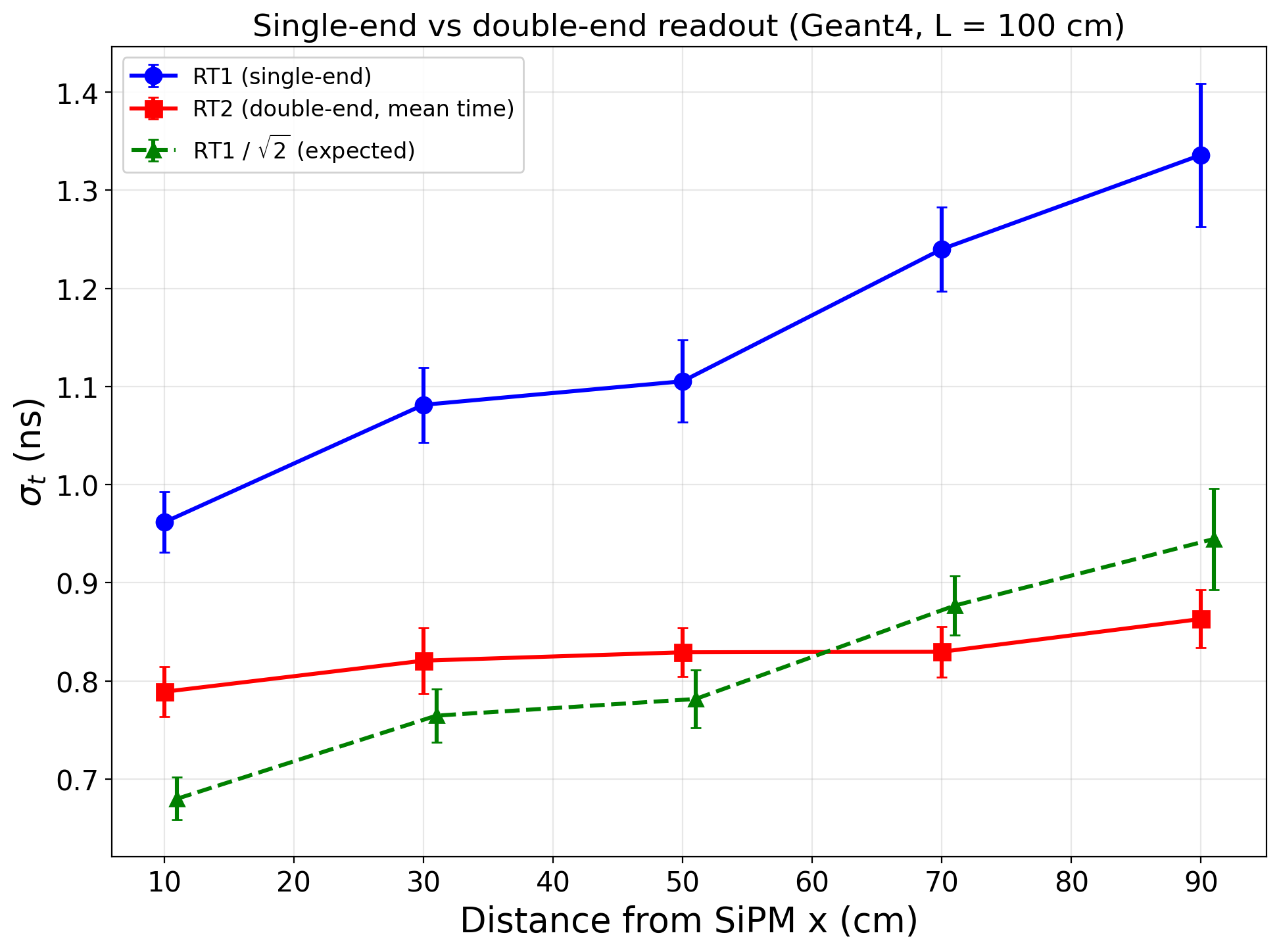}
\caption{Geant4 readout topology comparison: single-ended RT1 (blue) versus double-ended
RT2 mean-time (red). The green dashed line shows the expected $\mathrm{RT1}/\sqrt{2}$
scaling (800 events per point; error bars are statistical). The double-ended readout provides both improved resolution and position uniformity.}
\label{fig:g4_readout}
\end{figure}

The fourth test validates the $\sqrt{2}$ improvement and positional invariance prediction for double-ended readout.
In the analytical model, the weighted combination of two independent single-ended
measurements reduces $\sigt$ by $\sqrt{2}$ for symmetric detectors. The Geant4
simulation tests this with realistic optical correlations between the two ends.
Figure~\ref{fig:g4_readout} confirms the $\sqrt{2}$ improvement (at the bar center, the
ratio of the single-end to the double-end resolution is $1.39 \pm 0.06$) and shows that the
double-ended mean-time largely eliminates position dependence. This is consistent
with analytical results in section \ref{sec:position} and toy MC results of section~\ref{sec:mc_beyond_fixed}.

\subsection{SPTR and $\Npe$ scans} 
\label{sec:g4_parameter_scans}

\begin{figure}[htbp]
\centering
\includegraphics[width=\textwidth]{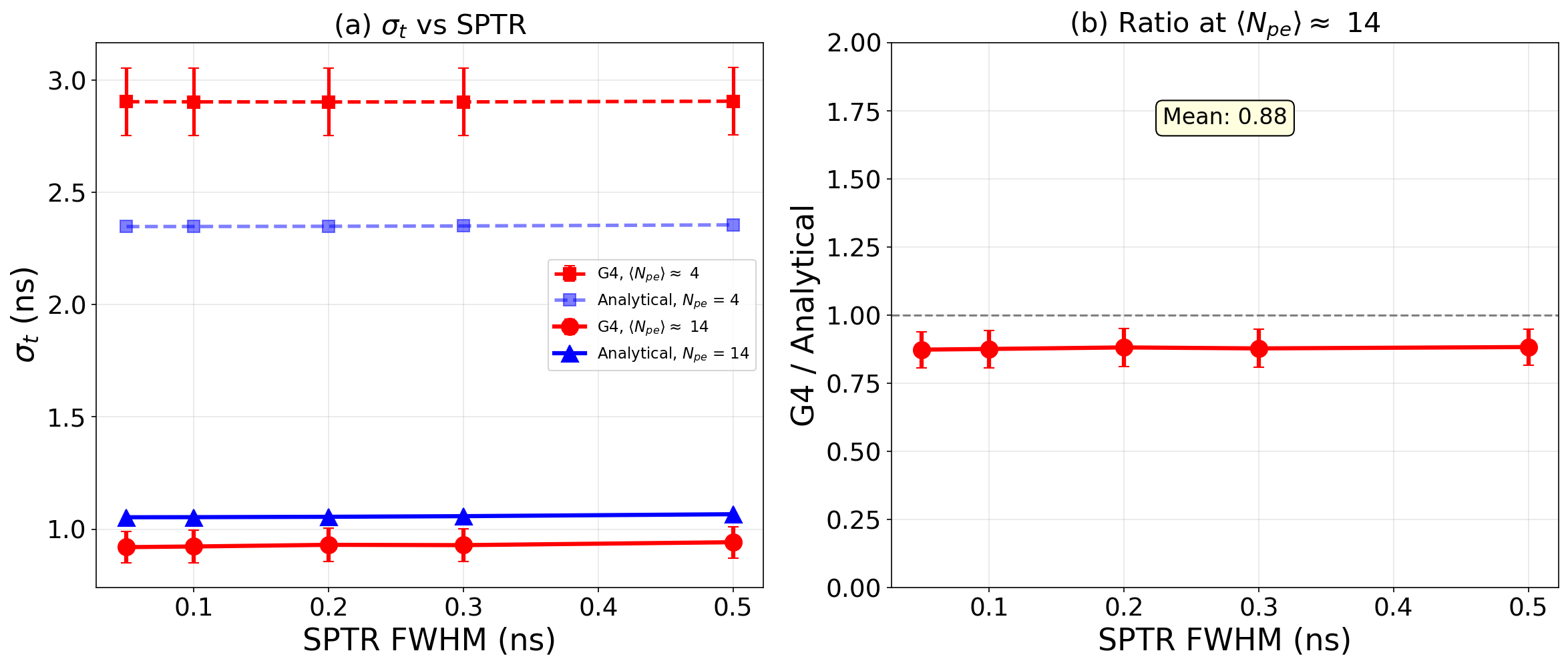}
\caption{SPTR scan at two $\Npe$ regimes.
(a) Dashed: $\Npe \approx 4$. Solid: $\Npe \approx 14$.
The weak SPTR dependence is confirmed at both levels.
(b) G4/analytical ratio at $\Npe \approx 14$: constant at $\approx 0.88$. Error bars show the statistical uncertainty, obtained from 1000 bootstrap resamples of the events (3000 muons per point at $\Npe \approx 4$ and about 1900 at $\Npe \approx 14$). The same simulated events are used for all SPTR values, so the points of a series are statistically correlated.}
\label{fig:g4_sptr}
\end{figure}

The SPTR scan in figure~\ref{fig:g4_sptr} confirms the weak dependence already seen in the
analytical model. At both low and higher light yield, changing the SPTR produces little change in
the total timing resolution compared with the effect of photon statistics and WLS timing.
At $\Npe \approx 14$, the G4/analytical ratio is $0.88$ (constant across SPTR),  lower than in the $\tauwls$ scan of section~\ref{sec:g4_position_topology}, a difference that is examined in section~\ref{sec:g4_effects}. 

\begin{figure}[htbp]
\centering
\includegraphics[width=0.65\textwidth]{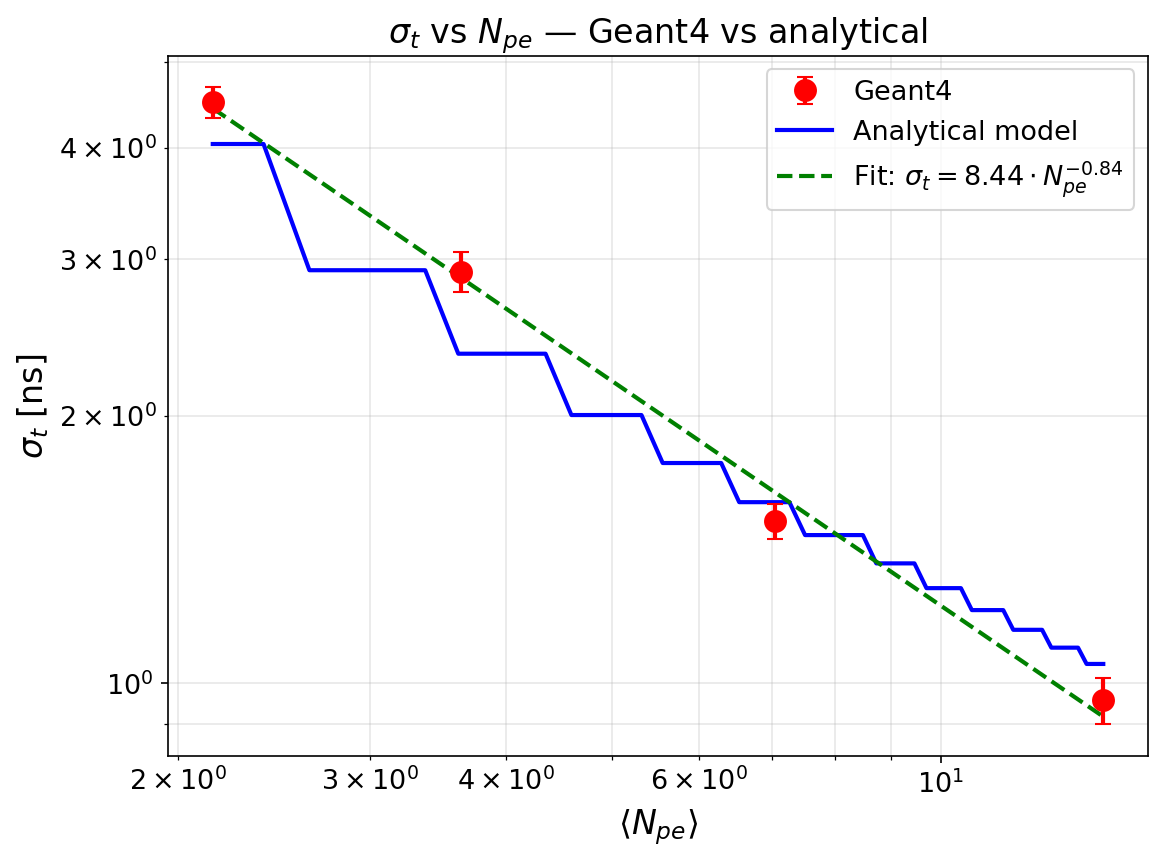}
\caption{Timing resolution versus mean $\Npe$ from Geant4 (red circles), obtained
by varying the scintillator light yield. The solid blue line shows the analytical
prediction. The dashed green line is a power-law fit to the Geant4 data.}
\label{fig:g4_npe}
\end{figure}

A slight discrepancy is identified when we scan through $\Npe$ with G4 simulations in figure \ref{fig:g4_npe}. While the trend of a descending power-law scaling for increased $\Npe$ is consistent, the fitted curve for the Geant4 data yields $\alpha\approx0.84$$\pm$ 0.04, steeper than the $0.44-0.49$ identified from analytical models.  Part of the difference between the two curves stem from the the fact at lower $\Npe$ the analytical curve is by itself steeper (its exponent is 0.70 for the four points of figure 26). The rest of the difference is explained in section~\ref{sec:g4_effects}. 

\subsection{Origin of the differences between Geant4 and the analytical model}
\label{sec:g4_effects}

The ratio between the Geant4 resolution and the analytical prediction is different throughout different tests above. At $\langle\Npe\rangle \approx 14$ the Geant4 resolution lies 7--19\% above the prediction in the $\tauwls$ scan of section~\ref{sec:g4_position_topology} but $12 \pm 7\%$ below it in the SPTR scan of figure~\ref{fig:g4_sptr}, and the exponent fitted in figure~\ref{fig:g4_npe} is larger than the analytical one. We need to trace these differences to the optical configuration of the simulation and to the fluctuations of $\Npe$, using a dedicated set of runs in which individual effects are switched on and off.

The scans of figures~\ref{fig:g4_sptr} and~\ref{fig:g4_npe} were made with assumptions that every wavelength-shifted photon is delayed by exactly $\tauwls$ (the default time profile of Geant4) instead of by an exponentially distributed time, and light that reaches the surface of the bar is absorbed, so that only direct light is collected. The WLS decay then does not broaden the photon arrival times. The time spread comes from the scintillation emission and from a small number of Cherenkov photons that reach the SiPM without being wavelength shifted, one $\tauwls$ ahead of the others. This gives a resolution below the analytical value at $\langle\Npe\rangle \approx 14$ and a steeper dependence on $\Npe$. Figures~\ref{fig:g4_sptr} and~\ref{fig:g4_npe} test the weak dependence on the SPTR and the general trend with $\Npe$, but not the absolute scale of the prediction.

The dedicated runs use Geant4 version 11.2.2 with a $1 \times 1 \times 100$~cm$^3$ bar and a 1~mm Y-11 fiber on its axis, hit at the bar center and read out at one end. The WLS time profile is exponential, and the Geant4 rise-time parameter is chosen such that the simulated scintillation pulse coincides with equation~(\ref{eq:fscint})\footnote{Geant4 parameterizes the pulse as $e^{-t/\taud}\,(1 - e^{-t/\tau_1})$, which equals equation~(\ref{eq:fscint}) for $\tau_1 = \taur\taud/(\taud - \taur) = 1.575$~ns.}. Each configuration has 4000 muons, and the uncertainties are obtained from 1000 bootstrap resamples of the events.

Two effects separate the Geant4 resolution from the fixed-$\Npe$ prediction. The first is the event-by-event fluctuation of $\Npe$. The fixed-$\Npe$ calculation can be extended to a fluctuating number of photoelectrons without further approximation. If $P(N)$ is the probability of detecting $N$ photoelectrons, and $\mu_N$ and $V_N$ are the mean and the variance of $T_{(k:N)}$ obtained from equation~(\ref{eq:orderpdf}), the law of total variance gives
\begin{equation}
\sigphoton^2 = \sum_{N \geq k} \tilde{P}(N)\left[ V_N + \mu_N^2 \right] - \Big[ \sum_{N \geq k} \tilde{P}(N)\,\mu_N \Big]^2 ,
\qquad \tilde{P}(N) = \frac{P(N)}{\sum_{N' \geq k} P(N')} ,
\label{eq:compound}
\end{equation}
where events with $N<k$ give no timestamp. For a Poisson $P(N)$, equation~(\ref{eq:compound}) agrees with the Poisson-$N$ Toy MC of figure~\ref{fig:mc_poisson} to within 0.4\%, and for Y-11 it exceeds the fixed-$\Npe$ value by 59\% at $\langle\Npe\rangle = 4$, 18\% at 10, 10\% at 14, and 6\% at 20. In the Geant4 runs, the distribution of $\Npe$ is wider than Poisson because the energy deposit also fluctuates, and equation~(\ref{eq:compound}) evaluated with the simulated distribution exceeds the fixed-$\Npe$ value by about 17\% at $\langle\Npe\rangle \approx 14$.

The second effect is that the photons do not reach the SiPM with the idealized time distribution $\fdet$. Table~\ref{tab:g4_effects} compares four configurations. In configuration~A the assumptions of the analytical model are reproduced as closely as possible. The surface of the bar is absorbing, so that only direct light is collected, and the WLS emission spectrum is restricted to wavelengths at which the dye does not absorb. The Geant4 resolution is then $(6 \pm 2)\%$ above equation~(\ref{eq:compound}). This residual is attributed to the propagation in the fiber, where the simulation tracks skew rays whose paths are longer than assumed in equation~(\ref{eq:transit}). Restoring the overlap between the emission and absorption spectra of the fiber (configuration~B), which allows re-emitted photons to be absorbed and re-emitted again, or a diffuse reflective coating of the bar (configuration~C), which adds the time needed to collect the light, raises the offset to 16\% in either case. With both (configuration~D) it is $(22 \pm 2)\%$. These effects delay part of the photons and deplete the leading edge of the arrival-time distribution. The fraction of photons that arrive within 2~ns of the earliest possible time is 4.5\% for $\fdet$, 2.8\% in configuration~A, and 1.2\% in configuration~D. When $\fdet$ in equation~(\ref{eq:compound}) is replaced by the arrival-time distribution of the same simulation, the calculation reproduces the Geant4 resolution to within 3\% in all four configurations (last column of table~\ref{tab:g4_effects}). The order-statistic treatment is therefore accurate, and the offset comes from the single-photon time distribution.

\begin{figure}[htbp]
\centering
\includegraphics[width=0.65\textwidth]{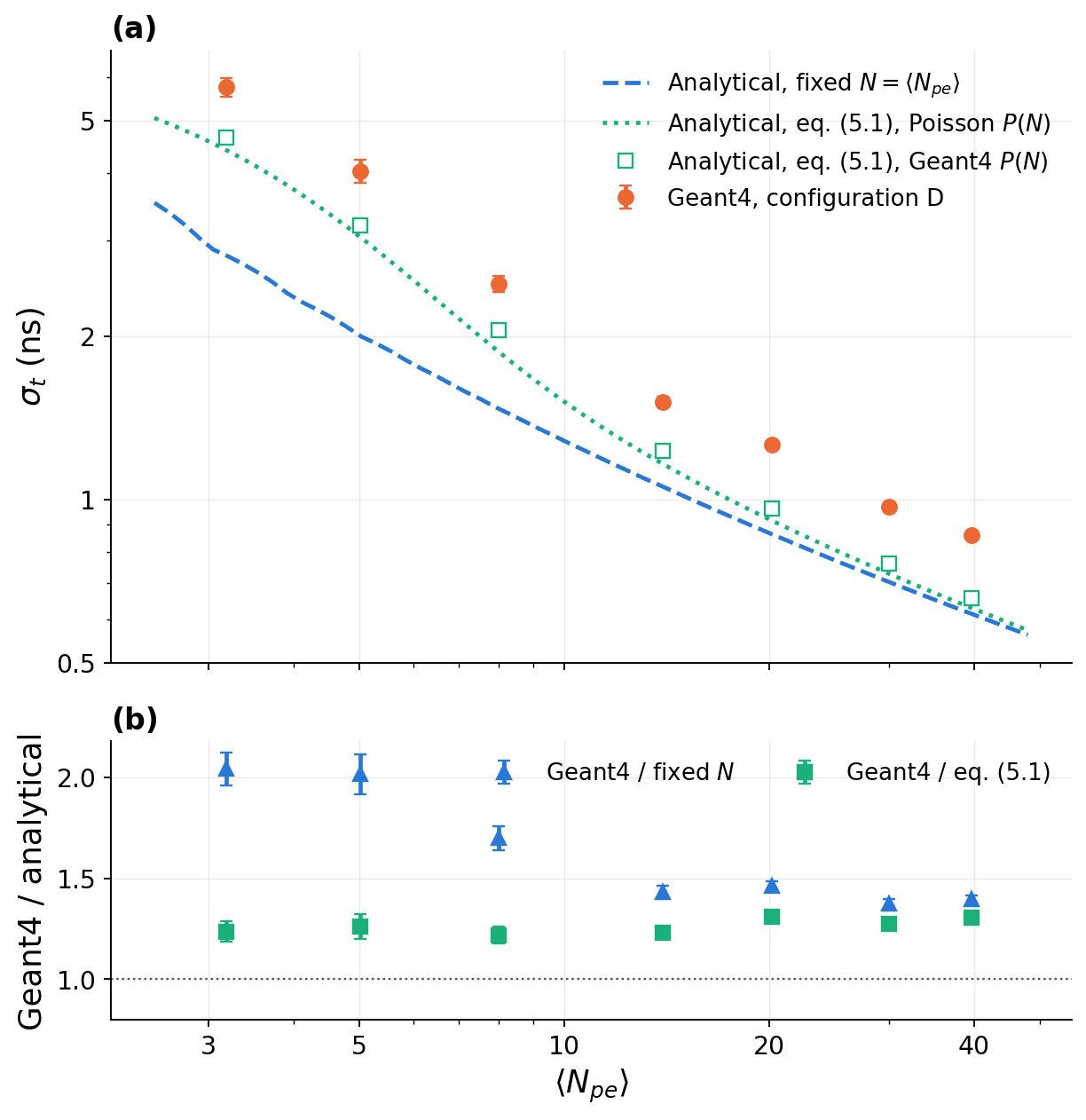}
\caption{(a) Timing resolution versus mean $\Npe$ in configuration D of table~\ref{tab:g4_effects} (circles, 3000 events per point), compared with the fixed-$\Npe$ prediction (dashed line) and with equation~(\ref{eq:compound}) evaluated with the Geant4 distribution of $\Npe$ (open squares) and with a Poisson distribution (dotted line). (b) Ratio of the Geant4 resolution to the fixed-$\Npe$ prediction (triangles) and to equation~(\ref{eq:compound}) (squares). Error bars show the statistical uncertainty (bootstrap).}
\label{fig:g4_npe_ext}
\end{figure}

In Figure~\ref{fig:g4_npe_ext}, we repeat the $\Npe$ scan in configuration~D with 3000 independent events per point. A power-law fit over $3 \leq \langle\Npe\rangle \leq 40$ gives $\alpha = 0.77 \pm 0.01$ for Geant4, 0.60 for the fixed-$\Npe$ calculation at the same points, and 0.79 for equation~(\ref{eq:compound}). The steeper slope of the simulation is therefore the effect of the fluctuations of $\Npe$ at low light yield and not an additional discrepancy. The ratio of the Geant4 resolution to equation~(\ref{eq:compound}) is nearly constant over the scan, between 1.22 and 1.31. In this configuration, the resolution changes by less than 1\% when the SPTR is varied between 50 and 500~ps, which confirms the weak dependence found in figure~\ref{fig:g4_sptr}.

In summary, the analytical model reproduces the dependences seen in the simulation, while its absolute scale rests on two idealizations, a fixed $\Npe$ and the time distribution $\fdet$. In configurations A to D the Geant4 resolution exceeds the fixed-$\Npe$ prediction by 24\% to 43\% at $\langle\Npe\rangle \approx 14$, of which about 17\% is due to the fluctuations of $\Npe$. The smaller offsets of section~\ref{sec:g4_position_topology} indicate that the transport contribution depends on the optical properties assumed for the bar and the fiber. For an absolute prediction in a specific geometry, the distribution of $\Npe$ and the single-photon arrival-time distribution should both be taken from a simulation or a measurement and inserted in equation~(\ref{eq:compound}).

\begin{table}[htbp]
\centering
\caption{Geant4 resolution at $\langle\Npe\rangle \approx 14$ for a hit at the center of a 100~cm bar with Y-11 fiber, compared with the fixed-$\Npe$ prediction, with equation~(\ref{eq:compound}) using $\fdet$, and with equation~(\ref{eq:compound}) using the photon arrival-time distribution of the same simulation in place of $\fdet$ (column marked $^{\ast}$). Uncertainties are statistical.}
\label{tab:g4_effects}
\smallskip
{\small
\begin{tabular}{lcccc}
\toprule
 & & \multicolumn{3}{c}{Geant4 / analytical} \\
Configuration & $\sigt^\mathrm{G4}$ (ns) & fixed $\Npe$ & eq.~(\ref{eq:compound}) & eq.~(\ref{eq:compound})$^{\ast}$ \\
\midrule
Configuration (figures~\ref{fig:g4_sptr} and~\ref{fig:g4_npe}) & $0.92 \pm 0.07$ & $0.88 \pm 0.07$ & $0.75 \pm 0.06$ & n/a \\
\midrule
A, direct light only, no spectral overlap & $1.31 \pm 0.03$ & $1.24 \pm 0.03$ & $1.06 \pm 0.02$ & $1.00 \pm 0.02$ \\
B, as A with spectral overlap & $1.44 \pm 0.03$ & $1.37 \pm 0.03$ & $1.16 \pm 0.03$ & $1.00 \pm 0.02$ \\
C, as A with reflective coating & $1.42 \pm 0.03$ & $1.35 \pm 0.03$ & $1.16 \pm 0.02$ & $1.02 \pm 0.02$ \\
D, as A with both & $1.51 \pm 0.03$ & $1.43 \pm 0.03$ & $1.22 \pm 0.02$ & $0.98 \pm 0.02$ \\
\bottomrule
\end{tabular}}
\end{table}

\subsection{Aggregate comparison and role of $\Npe$ calibration} 

\begin{figure}[htbp]
\centering
\includegraphics[width=\textwidth]{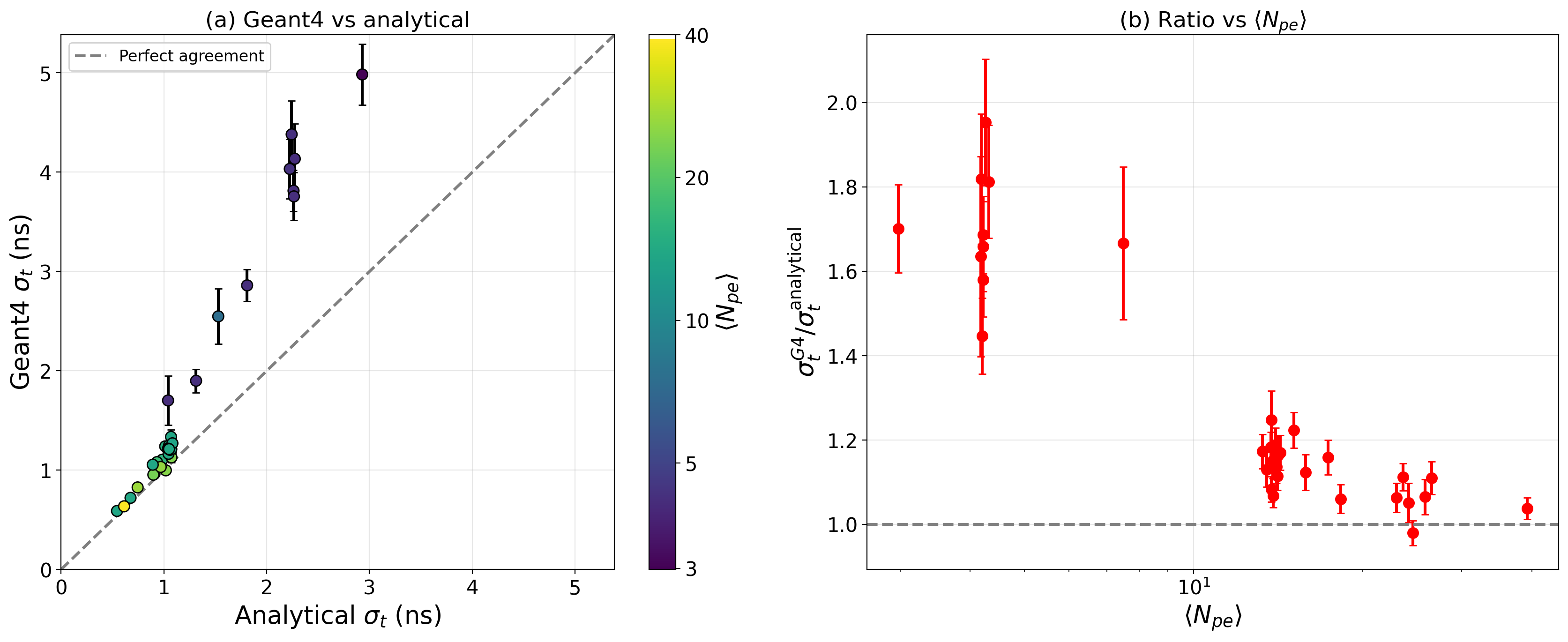}
\caption{Aggregate Geant4 validation. (a) Geant4 versus analytical timing resolution for all RT1 runs, colored by $\langle \Npe \rangle$. (b) G4 resolution versus analytical resolution ratio at different $\langle \Npe \rangle$.}
\label{fig:g4_master}
\end{figure}

Figure~\ref{fig:g4_master} combines results from all 34 RT1 Geant4 configurations and compares them against analytical predictions with the same configuration parameters. The G4/analytical ratio is 1.12 (rms spread 0.06) for the 24
points with $\langle \Npe \rangle \geq 10$ and 1.70 (rms 0.13) for the 10 points below. The large $\sigt$ points are the main source of scatter, and small $\sigt$ points agree much more consistently between G4 and analytical predictions. When taken in the context of figure~\ref{fig:g4_npe}, we see this trend can be explained by the fact that larger $\sigt$ points are associated with lower $\Npe$ that are much more susceptible to Poisson fluctuations of $\Npe$, which inflate the inclusive timing width (section~\ref{sec:mc_beyond_fixed}), and whose statistical uncertainties are also largest. At higher $\Npe$, smaller $\sigt$, where those fluctuations are reduced, the agreement is tighter, although the Geant4 values remain systematically above the fixed-$\Npe$ prediction by about 10\%.

We note that the Geant4 validation spans $\Npe \approx 3$--39, while the
lookup tables emphasize $\Npe \approx 50$--200. The effective attenuation length in the simulation
($\Lambda_\mathrm{eff} \approx 283 \pm 17$~cm from an exponential fit to the $\Npe$ profile) is shorter
than the tabulated $\lambda_\mathrm{att} = 350$~cm, likely due to cladding mode
losses and wavelength-dependent absorption not captured by the single-exponential
parameterization. This does not invalidate the timing comparison (since the analytical
model is evaluated at the Geant4 $\Npe$), but it underscores the importance of
calibrating $\Npe$ from simulation or measurement for maximum accuracy of resolution predictions. 

The practical implication of the Geant4 validation is twofold. First, the analytical
timing model is reliable for comparative studies (e.g., fiber selection, readout topology
optimization) where the relative predictions matter more than absolute values.
Second, for absolute $\sigt$ predictions in specific detector geometries, we recommend using Geant4 to determine $\Npe$ for the specific geometry then apply the analytical timing model with the Geant4-derived
$\Npe$, keeping in mind that the fixed-$\Npe$ calculation underestimates $\sigt$ when
$\Npe \lesssim 20$ (section~\ref{sec:mc_beyond_fixed}). This two-step approach leverages the strengths of both methods and will provide the most robust timing resolution predictions.

Sections~\ref{sec:analytical} through~\ref{sec:geant4} showed
that the analytical model reproduces the Toy MC at the sub-percent level and follows the Geant4
simulation to within 10--20\% for $\langle \Npe \rangle \gtrsim 10$ for configurations in section~\ref{sec:g4_position_topology} when evaluated at the correct
$\Npe$, the remaining difference being due to the event-by-event
fluctuations of $\Npe$ and to photon-transport effects that are not part of the idealized time profile. Section~\ref{sec:g4_effects} shows that these two contributions can raise the Geant4 resolution to 43\% above the fixed-$\Npe$ value at $\langle \Npe \rangle \approx 14$, so that the analytical resolution should be read as a lower bound
The remaining practical question is: for a given detector, what $\Npe$ one should
use. It depends on details that the analytical model abstracts
away. The number of WLS fibers per scintillator element, the surface treatment, the
fiber-to-scintillator coupling, and the overall geometry all affect how many
scintillation photons ultimately reach the SiPM. A recent Geant4 optical study of
the SuperFGD unit cell~\cite{SuperFGD_optical} illustrates this complexity well. Of
the $\sim$18,000 scintillation photons produced by a MIP in a 1~cm cube, 53\% are
absorbed in the scintillator bulk, 17\% escape as optical crosstalk to neighboring
cubes, and the remaining 30\% are captured and wavelength-shifted by the three orthogonal Y-11 fibers. The resulting
$\Npe \approx 56$ per fiber~\cite{SuperFGD_timing} cannot be predicted from
material properties alone. It must be measured in a prototype or simulated with a
geometry-specific Geant4 model.

The analytical timing model, however, does not need to solve the light collection
problem. It takes $\Npe$ as a given and predicts $\sigt$. This separation is by
design. The light collection depends on engineering choices that vary from one
detector to the next. The timing physics, once $\Npe$ photons have been detected, is
universal. This section uses the model to map out the timing landscape and identify
the most effective paths to better resolution.

\section{Comparison with measurements and detector-design guidance}
\label{sec:design_guidance}
After the analytical model and its two validation layers have been established, we now compare
the predictions with published detector measurements and collect the results into practical design
guidance. \\

\subsection{Benchmark comparison with published measurements}
\label{sec:benchmark}

\begin{figure}[htbp]
\centering
\includegraphics[width=\textwidth]{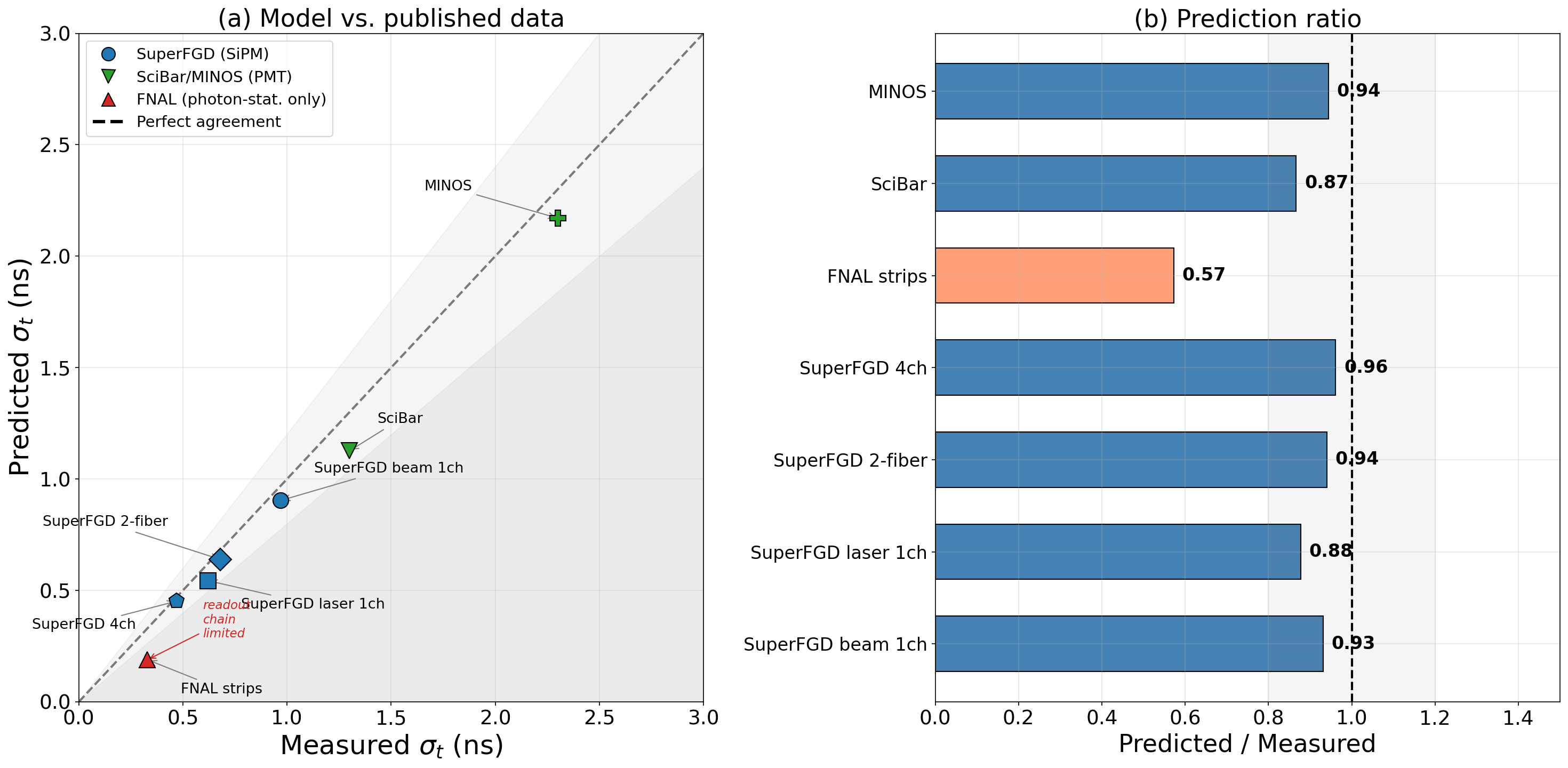}
\caption{Benchmark validation against published measurements from SuperFGD ($\Npe = 56$, SiPM)~\cite{SuperFGD_timing}, SciBar ($\Npe \sim 14$ avg, PMT)~\cite{SciBar_timing}, MINOS ($\Npe \sim 3$/end, PMT)~\cite{MINOS_timing}, and FNAL ($\Npe \sim 69$/end, SiPM)~\cite{FNAL_strips}. (a) Predicted versus measured $\sigt$. Six of seven benchmarks fall within $\pm$20\% of perfect agreement (dashed line). The FNAL point (triangle) shows only the photon-statistics floor, as the CAMAC-era readout chain dominates the measured resolution. (b) Prediction ratios.}
\label{fig:benchmark}
\end{figure}

To assess the predictive accuracy of the analytical model, we compare its predictions against published timing measurements from four experiments spanning a wide range of detector sizes and photoelectron yields: SuperFGD~\cite{SuperFGD_timing}, FNAL test beam strips~\cite{FNAL_strips}, SciBar~\cite{SciBar_timing}, and MINOS~\cite{MINOS_timing}. For each experiment, the model is evaluated using the detector parameters reported in the original publication. The photosensor single-photon timing jitter enters through the transit time spread (TTS) for multi-anode PMTs (SciBar H8804, MINOS R5900-M16, both 300~ps FWHM~\cite{Lang_M16,Barker_M64}) or the SPTR for SiPMs (SuperFGD S13360-1325CS, 150~ps FWHM; FNAL S13360-3050CS, 150~ps FWHM~\cite{Gundacker_SPTR}). Figure~\ref{fig:benchmark} shows the comparison. 
The four SuperFGD configurations use $\Npe = 56$ per fiber~\cite{SuperFGD_timing} with Y-11 WLS fiber~\cite{SuperFGD_2020} ($\tauwls = 7.1$~ns) and CITIROC readout ($\sigelec = 0.72$~ns). The model under-predicts all four: beam single-channel ($\sigt^\mathrm{pred} = 0.90$~ns vs.\ $\sigt^\mathrm{meas} = 0.97$~ns, ratio 0.93), laser intrinsic ($\sigt^\mathrm{pred} = 0.54$~ns vs.\ $\sigt^\mathrm{meas} = 0.62$~ns, ratio 0.88; an oscilloscope-based laser measurement without the CITIROC TDC contribution), 2-fiber average ($\sigt^\mathrm{pred} = 0.64$~ns vs.\ $\sigt^\mathrm{meas} = 0.68$~ns, ratio 0.94), and 4-channel average ($\sigt^\mathrm{pred} = 0.45$~ns vs.\ $\sigt^\mathrm{meas} = 0.47$~ns, ratio 0.96; two cubes with two fibers each along a track segment). This systematic under-prediction (ratios 0.88--0.96) is expected for an idealized model that omits real-world degradation from trigger time-walk, dark-count contamination, and optical cross-talk. The under-prediction also confirms the aggregate G4/Analytical result in  figure \ref{fig:g4_master} where G4 consistently over-predicts the analytical on most $\Npe$ values. 

SciBar uses 300~cm polystyrene strips with Y-11 fiber and H8804 multi-anode PMT readout~\cite{SciBar_timing}. The model prediction ($\sigt^\mathrm{pred} = 1.13$~ns) agrees with the measured 1.3~ns (ratio 0.87). The strip-center light yield of $\Npe \approx 14$ assumed here is derived from the measured 16.5~photoelectrons/cm near the PMT end~\cite{SciBar_timing}, attenuated over half the strip length. MINOS uses 800~cm strips with Y-11 fiber and R5900-M16 PMT readout~\cite{MINOS_timing}. At $\sim$3~photoelectrons per strip end (six photoelectrons summed over both ends at strip center) with double-ended mean-time readout, the predicted resolution ($\sigt^\mathrm{pred} = 2.17$~ns) agrees well with the measured 2.3~ns (ratio 0.94). The remaining 6\% discrepancy is consistent with unmodeled ASDLite discriminator time-walk and 640~MHz TDC quantization jitter (1.56~ns bins).

The FNAL strip measurement~\cite{FNAL_strips} uses Bicron 404A scintillator with BCF-92 fiber ($\tauwls \approx 2.1$~ns~\cite{Yamazumi_BCFXL}) and S13360-3050CS SiPM readout, achieving 137~total photoelectrons per strip at the high-gain setting ($\sim$69 per end, taken as half the published total) in a 100~cm double-ended configuration. At this high $\Npe$ with a fast fiber, the photon-statistics contribution predicted by the model ($\sigt^\mathrm{pred} = 0.19$~ns) falls well below the measured mean-time resolution of 0.33~ns. The discrepancy arises because the FNAL readout chain (passive splitter, amplifier, constant-threshold discriminator, CAMAC TDC) contributes jitter that dominates over photon statistics at high $\Npe$. This case illustrates the regime where the model provides a floor on achievable timing, not the full system resolution.

\subsection{Timing optimization}
\label{sec:geom_lc}

\begin{figure}[htbp]
\centering
\includegraphics[width=\textwidth]{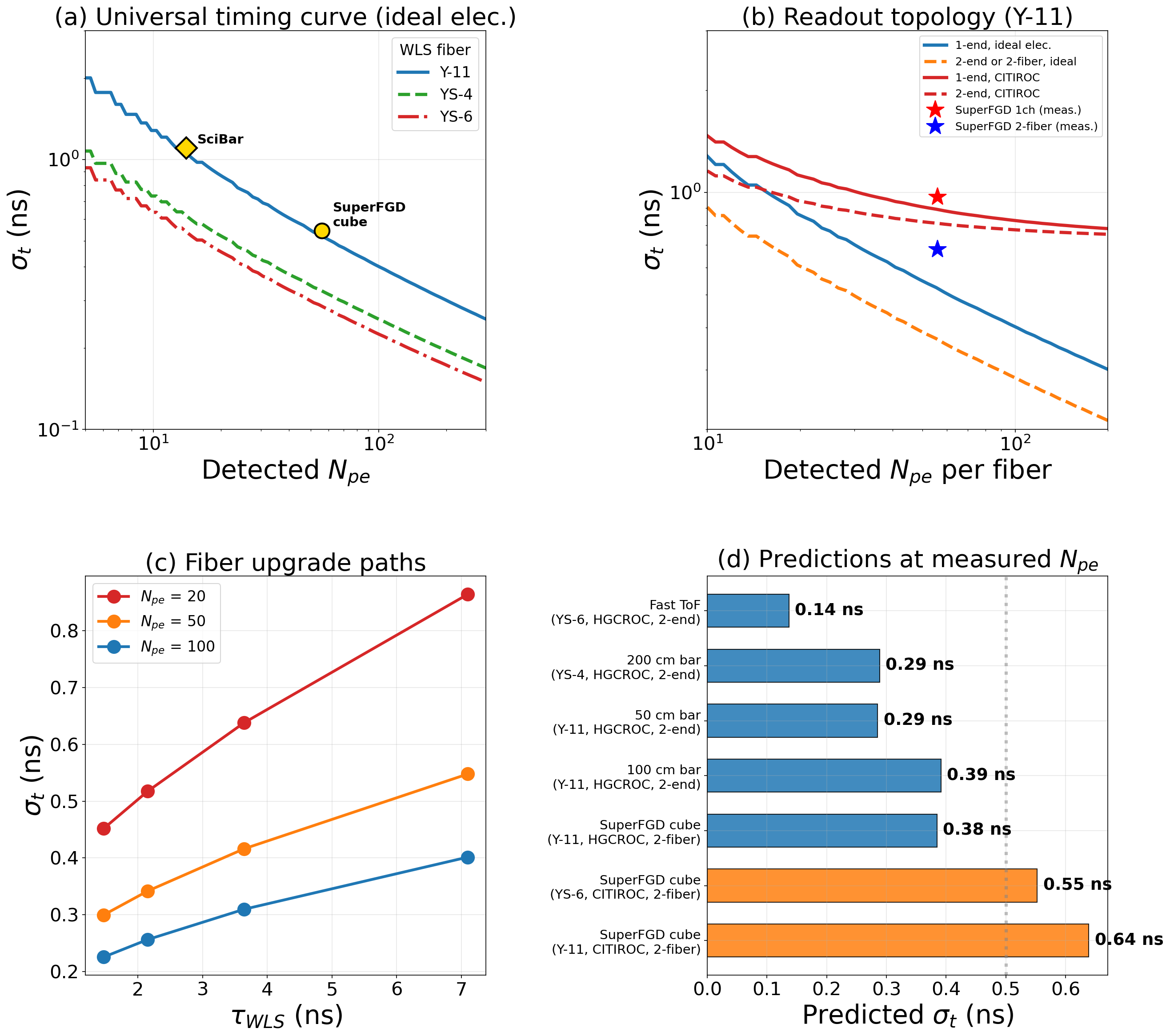}
\caption{Timing predictions for real detectors.
(a)~$\sigt$ vs detected $\Npe$ for three fiber types with ideal electronics.
Stars mark detectors at their measured $\Npe$.
(b)~Effect of readout topology and electronics (Y-11 fiber). SuperFGD measured
values shown as stars at $\Npe = 56$.
(c)~Fiber upgrade paths at $\Npe = 20$, 50, and 100.
(d)~Predicted $\sigt$ for seven practical configurations at published $\Npe$.}
\label{fig:geometry}
\end{figure}

Figure~\ref{fig:geometry}(a) shows $\sigt$ as a function of detected $\Npe$ for
three WLS fiber types. These curves are the central deliverable of the analytical
framework. Given a measured or simulated $\Npe$ for any detector geometry, one reads
off $\sigt$ directly. Two representative detectors are marked at their published
$\Npe$ values on the Y-11 curve. The spread between fiber types at a given $\Npe$
is striking. At $\Npe = 50$, Y-11 gives 0.55~ns while YS-6 gives 0.30~ns. This
factor-of-two difference comes entirely from the WLS re-emission time and is
available as an upgrade path without modifying the detector geometry or light
collection.

The interplay between readout topology and electronics is shown in
figure~\ref{fig:geometry}(b) for Y-11 fiber. Double-ended readout reduces $\sigt$ by
$\sqrt{2}$ in the photon-statistics-limited regime (high $\Npe$, fast electronics).
But when the electronics jitter is large, as with the CITIROC ASIC
($\sigelec = 0.72$~ns), the $\sqrt{2}$ improvement is partially offset because the
electronics noise also combines in quadrature. The SuperFGD measured timing values
(0.97~ns single-channel, 0.68~ns two-fiber) at $\Npe = 56$ sit between the CITIROC
single-end and double-end model curves, confirming that the model captures the
correct behavior even when electronics jitter is significant.

Figure~\ref{fig:geometry}(c) shows the timing improvement from switching fibers at
three representative $\Npe$ values. The improvement from Y-11 to YS-6 is
1.8$\times$ at $\Npe = 100$ and grows to 1.9$\times$ at $\Npe = 20$, because at
lower $\Npe$ the broader WLS tail has a larger relative effect on the first-photon
order statistic. For any existing detector that currently uses Y-11, a fiber swap to
YS-6 is the single most impactful timing upgrade. It requires no changes to the
scintillator, SiPM, or electronics.

Finally, figure~\ref{fig:geometry}(d) collects predictions for seven concrete
configurations spanning the range from small cubes to fast time-of-flight counters.
All predictions use measured or published $\Npe$ values, not the naive formula. The
SuperFGD cube with its current Y-11 fiber and CITIROC readout is predicted at 0.64~ns per channel (2-fiber average). Replacing Y-11 with YS-6 brings this to
0.55~ns. Replacing CITIROC with HGCROC brings it to 0.38~ns. For a dedicated fast
ToF counter with YS-6, $\Npe = 150$, HGCROC electronics, and double-ended readout,
the model predicts 0.14~ns.

\subsection{Comprehensive lookup tables}
\label{sec:lookup}
\begin{figure}[htbp]
\centering
\includegraphics[width=\textwidth]{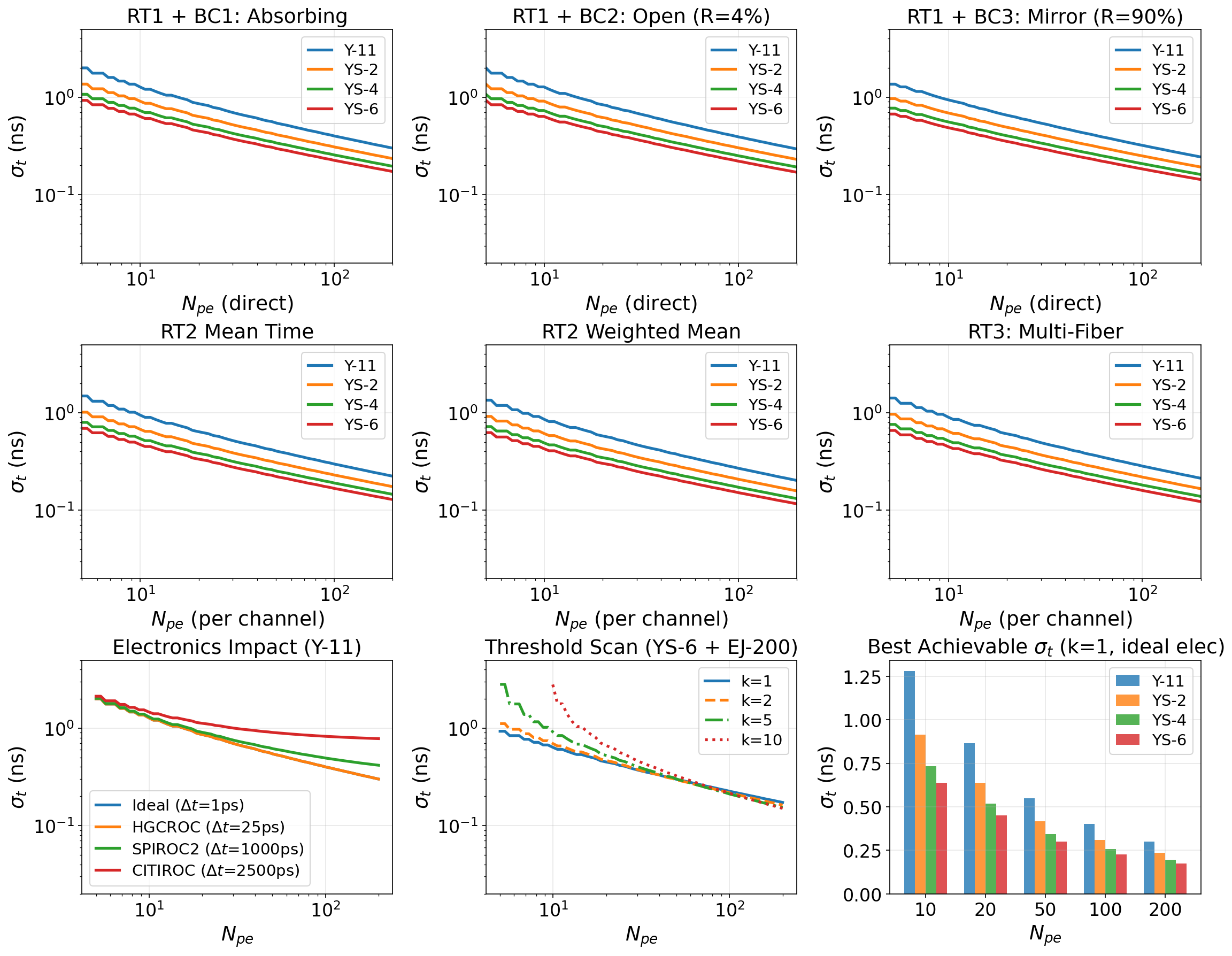}
\caption{Nine-panel comprehensive lookup figure covering all readout topologies, boundary
conditions, electronics configurations, and threshold options. Each panel shows $\sigt$
versus $\Npe$ for four WLS fiber types.}
\label{fig:lookup}
\end{figure}

Figure~\ref{fig:lookup} provides a nine-panel compendium covering the full parameter
space. The top row shows $\sigt$ versus $\Npe$ for single-ended readout with the three
boundary conditions (BC1, BC2, BC3). The middle row covers double-ended readout (mean
time, weighted mean, multi-fiber). The bottom row shows the electronics impact, threshold
scan, and a bar chart of the best achievable $\sigt$ at key $\Npe$ values.

This figure serves as a ``one-stop shop'' for a detector designer who needs to quickly
assess the expected timing performance of a configuration.

More complete lookup material, including $\Npe$--$\sigt$ tables, scintillator--fiber matrices, topology tables, and electronics tables, is provided in appendix~\ref{app:lookup}.
\section{Discussion}
\label{sec:discussion}

\subsection{Key findings}

The comprehensive parameter scan reveals several findings of practical importance for
detector design:

\begin{enumerate}[label=(\alph*)]
  \item Photon statistics domination. In more than 90\% of the configurations studied, photon statistics contribute $> 90\%$ of the total timing variance. The two exceptions are systems with coarse TDC (CITIROC at 2.5~ns), where electronics contributes up to 63\% of the variance, and very long bars ($L > 200$~cm) where transit time contributes a non-negligible but still minority share of the position-averaged variance ($\sim$2--11\%). This finding means that for most applications, the optimization strategy reduces to maximizing $\Npe$ and minimizing $\tauwls$. 

  \item WLS fiber as the primary lever. The WLS fiber re-emission time $\tauwls$ is the single most impactful hardware parameter. Switching from Y-11 ($\tauwls = 7.1$~ns) to YS-6 ($\tauwls = 1.47$~ns) improves timing by a factor of 1.7--2.2 across the studied $\Npe$ range (table~\ref{tab:lookup1}), equivalent to increasing $\Npe$ by a factor of $\sim$4. This ``exchange rate'' makes fiber upgrading the most cost-effective path to better timing in existing detector designs. 

  \item Power law scaling. The timing resolution follows $\sigt \approx A \cdot \Npe^{-\alpha}$ with $\alpha \approx 0.44$--$0.49$, significantly slower than the $\Npe^{-1}$ scaling expected for a pure exponential emitter. This departure arises because the convolved detection PDF $\fdet$ has a finite rise time (determined by the interplay of $\taur$, $\taud$, and $\tauwls$), which suppresses the probability of detecting photons near $t = 0$. The practical consequence is that doubling $\Npe$ improves timing by only 26--29\%, not 50\% as naive scaling would suggest.

  \item Mirror far-end behavior. For a $k = 1$ (first-photon) trigger, a far-end mirror (BC3) can only add photons to the detection pool (it never removes direct photons), so it can only improve or leave unchanged the first-photon arrival time: $T_{(1)}^{\mathrm{BC3}} \leq T_{(1)}^{\mathrm{BC1}}$. In practice, the mirror provides a modest net benefit for short detectors ($L \lesssim 20$~cm), where the reflected photons arrive within $\sim$1~ns of the direct photons and augment the leading-edge photon population. For long bars ($L \gtrsim 100$~cm), the reflected photons travel an
extra $\sim$$2L \cdot n_\mathrm{core}/c$ and arrive $\sim$10--20~ns after the first direct
photon, far too late to affect the first-photon trigger. In this regime, BC3 is effectively
neutral (BC3 $\approx$ BC1) except close to the mirrored end, as confirmed by the Geant4 simulation (figure~\ref{fig:g4_boundary}).

  \item Double-ended readout. The weighted mean time from double-ended readout achieves the theoretically expected
$\sqrt{2}$ improvement, verified by MC to within 2\%. Beyond resolution improvement,
RT2 provides crucial position uniformity: the non-uniform response of single-ended readout
(which can vary by a factor of 2 across a 200~cm bar) is replaced by nearly flat response.

\end{enumerate}

\subsection{Comparison with published results}

The benchmark validation of section~\ref{sec:benchmark} compares the model against seven published measurements spanning four experiments. The SuperFGD configurations ($\Npe = 56$, Y-11 fiber, SiPM readout) show systematic under-prediction with ratios of 0.88--0.96, the expected behavior for an idealized model that omits real-world degradation effects such as trigger time-walk, dark-count contamination, optical cross-talk, and fiber-coupling losses. The SciBar long-bar benchmark (300~cm, Y-11 fiber, H8804 PMT readout, $\sigt = 1.3$~ns) agrees at ratio 0.87. The MINOS benchmark (800~cm strips, R5900-M16 PMT readout, $\sigt = 2.3$~ns, $\sim$3~photoelectrons per strip end) agrees at ratio 0.94, with the small remaining discrepancy attributable to ASDLite discriminator time-walk and TDC quantization jitter. The FNAL strip measurement ($\Npe \approx 69$ per end, BCF-92 fiber, SiPM readout, $\sigt = 0.33$~ns) is the sole significant outlier: the model predicts only the photon-statistics floor (0.19~ns), well below the measured value, because the CAMAC-era analog readout chain contributes jitter that dominates at high $\Npe$ with fast fiber. This case illustrates that the model provides a lower bound on achievable timing, and that readout electronics design becomes the limiting factor when photon statistics alone do not dominate.

\subsection{Limitations and outlook}

The Geant4 validation of section~\ref{sec:geant4} demonstrates that the analytical timing
model is most accurate when evaluated at the correct $\Npe$, but the prediction of $\Npe$ itself
from first principles remains challenging. The geometry correction factors of
section~\ref{sec:geom_lc} provide approximate treatments. For precise $\Npe$ predictions in
specific detector geometries, a Geant4 optical simulation calibrated to prototype data is
recommended before applying the analytical model to the Geant4-derived $\Npe$.

The Geant4 comparison itself is limited to $\langle \Npe \rangle \lesssim 40$ and to a single bar read out through one fiber. Of the two contributions separated in section~\ref{sec:g4_effects}, the fluctuation of $\Npe$ is handled by
equation~(\ref{eq:compound}), whereas the photon-transport contribution is available only as a simulated arrival-time distribution. A parameterization in terms of the fiber and bar properties, which would allow $\fdet$ to be corrected without a dedicated simulation, has not
been attempted. Extending the comparison to higher light yields, where this contribution should dominate the residual offset, and to multi-element geometries such as segmented cube detectors, in which optical cross-talk between neighboring elements alters both $\Npe$ and the arrival-time distribution, is left to future work.

Some additional effects are intentionally overlooked in this study and merit future investigations. Pulse pile-up at high event rates could artificially increase the $\Npe$ of a detector, and the corresponding SiPM saturation from the high $\Npe$ may also have some unexpected effects on timing resolutions. The same artificial bloating of $\Npe$ can also happen from the afterpulse or crosstalk of the sensors. On the physical design side, scintillator and WLS parameters fluctuate with temperature in a real detector, and the slight bending of fibers in a detector could also introduce variations in the actual light transport efficiency.

\section{Conclusions}
\label{sec:conclusions}

We developed a semi-analytical framework for predicting the timing resolution of plastic
scintillator detectors with WLS fiber and SiPM readout. The model constructs the detected
single-photon time distribution from scintillation emission, WLS re-emission, and SiPM response;
uses order statistics to convert that distribution into an event timestamp; and adds detector-level
terms from fiber transport, electronics, and readout topology.

The analytical calculation is internally validated by a high-statistics Toy Monte Carlo. At fixed
$\Npe$, the analytical and Toy MC predictions agree at the sub-percent level across 80 grid points.
The Toy MC also quantifies effects beyond the fixed-$\Npe$ calculation, including Poisson light-yield
fluctuations, boundary conditions, multi-channel averaging, and timing-pickoff choices. A full
Geant4 optical simulation provides a second validation layer and shows that the analytical timing
model remains predictive when evaluated at the Geant4-measured $\Npe$.

The main design conclusions are as follows. Photon statistics dominate most practical configurations,
so the first priority is to maximize the detected $\Npe$. The WLS re-emission time is the strongest
hardware lever at fixed $\Npe$, with fast fibers providing nearly a factor-of-two timing improvement
relative to Y-11 in representative cases. The scaling with $\Npe$ is slower than $1/\Npe$ because the
scintillator+WLS detection-time PDF has a finite leading edge. Double-ended readout provides the
expected statistical improvement and, more importantly, strong position uniformity for long bars.
Mirrors help mainly for short detectors. Electronics should be made fine enough that its contribution
lies below the photon-statistics floor; once that condition is met, further TDC improvement gives
little gain.

The resulting lookup tables, design maps, and validation comparisons provide a quantitative basis
for optimizing scintillator timing detectors. The recommended workflow for a specific detector is to
measure or simulate the detected $\Npe$, then use the analytical timing model and design maps to
select the fiber, readout topology, and electronics configuration needed to reach the target timing
resolution.

\appendix

\section{Tabulated results}
\label{app:lookup}
Table~\ref{tab:lookup1} provides the complete $\sigt$ versus $\Npe$ lookup for each
fiber type with EJ-200 scintillator, $k = 1$ threshold, and ideal electronics. These
values represent the fundamental photon-statistics-limited resolution achievable with
each fiber.

\begin{table}[htbp]
\centering
\caption{Timing resolution $\sigt$ (ns) versus $\Npe$ for five WLS fiber types.
EJ-200 scintillator, $k = 1$, ideal electronics.}
\label{tab:lookup1}
\smallskip
\begin{tabular}{rrrrrr}
\toprule
$\Npe$ & Y-11 & YS-2 & YS-4 & YS-6 & BCF-92XL \\
\midrule
5   & 2.003 & 1.365 & 1.073 & 0.931 & 1.063 \\
10  & 1.279 & 0.914 & 0.732 & 0.637 & 0.725 \\
15  & 1.011 & 0.737 & 0.595 & 0.519 & 0.590 \\
20  & 0.864 & 0.638 & 0.517 & 0.452 & 0.513 \\
30  & 0.701 & 0.525 & 0.428 & 0.375 & 0.425 \\
50  & 0.548 & 0.416 & 0.342 & 0.300 & 0.339 \\
75  & 0.455 & 0.349 & 0.288 & 0.253 & 0.286 \\
100 & 0.401 & 0.310 & 0.256 & 0.226 & 0.254 \\
150 & 0.338 & 0.263 & 0.219 & 0.193 & 0.217 \\
200 & 0.301 & 0.235 & 0.196 & 0.173 & 0.194 \\
\bottomrule
\end{tabular}
\end{table}

Table~\ref{tab:lookup2} shows $\sigt$ for all scintillator $\times$ fiber combinations at
$\Npe = 50$, providing a complete material selection guide.

\begin{table}[htbp]
\centering
\caption{Timing resolution $\sigt$ (ns) for all scintillator $\times$ WLS fiber
combinations at $\Npe = 50$, $k = 1$, ideal electronics.}
\label{tab:lookup2}
\smallskip
\begin{tabular}{lrrrrr}
\toprule
Scintillator & Y-11 & YS-2 & YS-4 & YS-6 & BCF-92XL \\
\midrule
EJ-200      & 0.548 & 0.416 & 0.342 & 0.300 & 0.339 \\
EJ-204      & 0.493 & 0.371 & 0.303 & 0.265 & 0.300 \\
EJ-208      & 0.657 & 0.502 & 0.415 & 0.366 & 0.412 \\
EJ-228      & 0.425 & 0.316 & 0.256 & 0.222 & 0.253 \\
EJ-232      & 0.421 & 0.309 & 0.249 & 0.215 & 0.246 \\
BC-420      & 0.435 & 0.323 & 0.262 & 0.228 & 0.259 \\
PS+PPO+POPOP & 0.573 & 0.436 & 0.359 & 0.315 & 0.356 \\
\bottomrule
\end{tabular}
\end{table}

Table~\ref{tab:lookup3} shows the position-averaged $\sigt$ for all readout topologies as
a function of bar length.

\begin{table}[htbp]
\centering
\caption{Position-averaged $\sigt$ (ns) for all readout topologies. Y-11 fiber, EJ-200
scintillator, $\Npe \approx 108$ at SiPM end (reference light yield assumed for this table),
ideal electronics.}
\label{tab:lookup3}
\smallskip
\begin{tabular}{lrrrrr}
\toprule
Configuration & $L = 1$ & $L = 10$ & $L = 50$ & $L = 100$ & $L = 200$~cm \\
\midrule
RT1+BC1 (absorbing) & 0.388 & 0.391 & 0.403 & 0.423 & 0.474 \\
RT1+BC2 (open)      & 0.382 & 0.385 & 0.397 & 0.417 & 0.469 \\
RT1+BC3 (mirror) & 0.298 & 0.340 & 0.395 & 0.420 & 0.474 \\
RT2 (double-end)    & 0.275 & 0.276 & 0.285 & 0.298 & 0.328 \\
RT3 2-fiber         & 0.275 & 0.276 & 0.285 & 0.299 & 0.335 \\
RT3 3-fiber         & 0.224 & 0.226 & 0.233 & 0.244 & 0.273 \\
\bottomrule
\end{tabular}
\end{table}

Table~\ref{tab:lookup4} quantifies the electronics impact.

\begin{table}[htbp]
\centering
\caption{Electronics impact on timing resolution. Y-11 fiber, EJ-200, $\Npe = 50$.}
\label{tab:lookup4}
\smallskip
\begin{tabular}{lrrrr}
\toprule
Electronics & $\Delta t_\mathrm{TDC}$ (ps) & $\sigelec$ (ns) & $\sigt$ (ns) & $f_\mathrm{elec}$ (\%) \\
\midrule
CITIROC  & 2500 & 0.722 & 0.906 & 63.4 \\
SPIROC2  & 1000 & 0.289 & 0.619 & 21.7 \\
HGCROC   &   25 & 0.007 & 0.548 &  0.0 \\
DRS4     &   10 & 0.003 & 0.548 &  0.0 \\
Ideal    &    1 & 0.000 & 0.548 &  0.0 \\
\bottomrule
\end{tabular}
\end{table}

\subsection{Complete summary}
The complete set of lookup tables, design maps, and parametric scans presented in this paper provides comprehensive coverage of the parameter space relevant to scintillator timing detector design. The tabulated results (tables~\ref{tab:lookup1}--\ref{tab:lookup4}) together with the nine-panel lookup figure (figure~\ref{fig:lookup}) serve as the primary reference for detector optimization.

\section{Additional analytical and simulation plots}
\label{app:analytical_extra}

This appendix collects analytical maps and simulation plots that are useful for detailed detector design.

\subsection{Extra 2D maps}
\label{app:analytical_2D}
\begin{figure}[H]
\centering
\includegraphics[width=\textwidth]{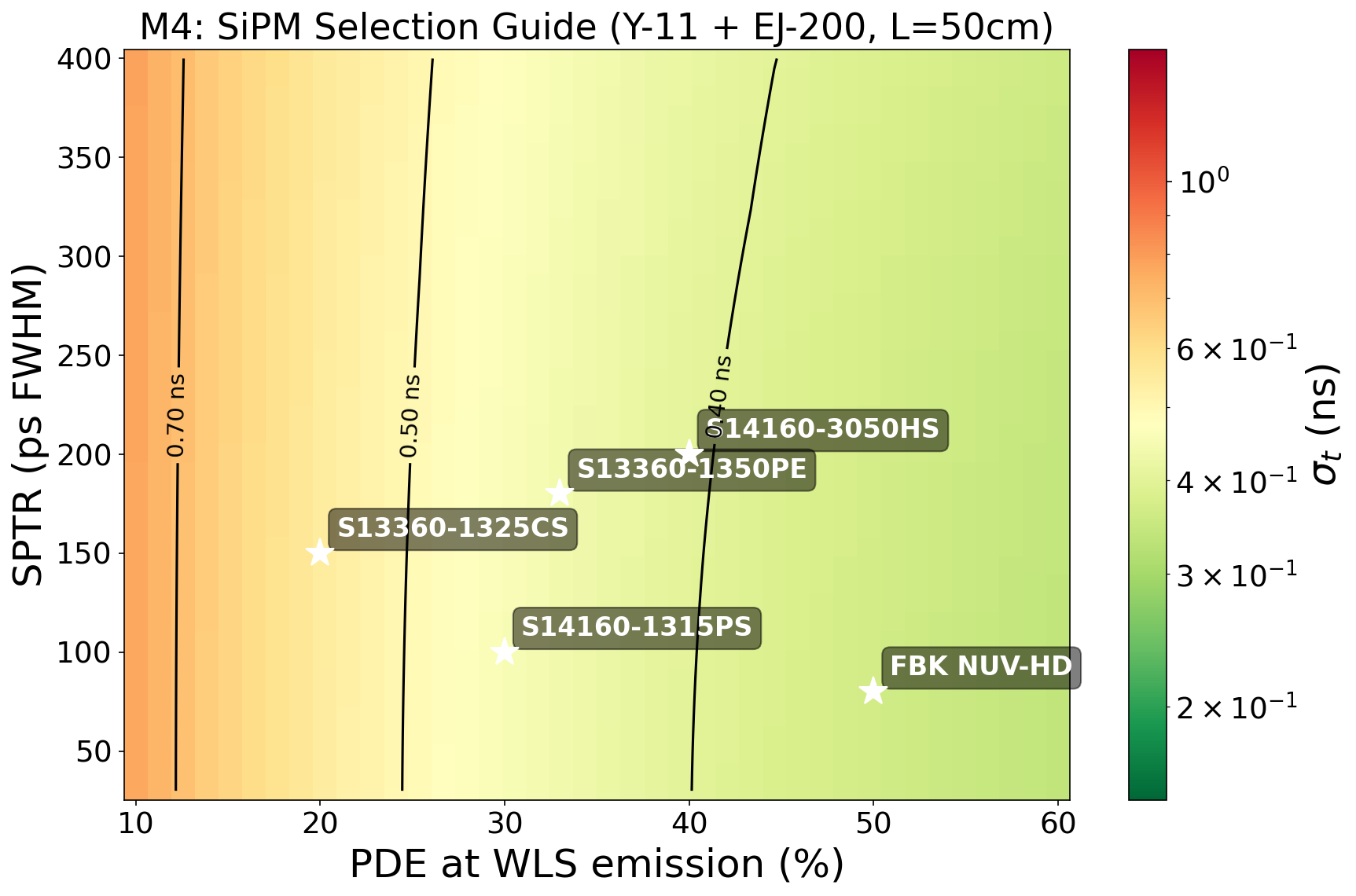}
\caption{SiPM selection guide: $\sigt$ in the PDE--SPTR plane. }
\label{fig:map_sipm}
\end{figure}

\begin{figure}[H]
\centering
\includegraphics[width=\textwidth]{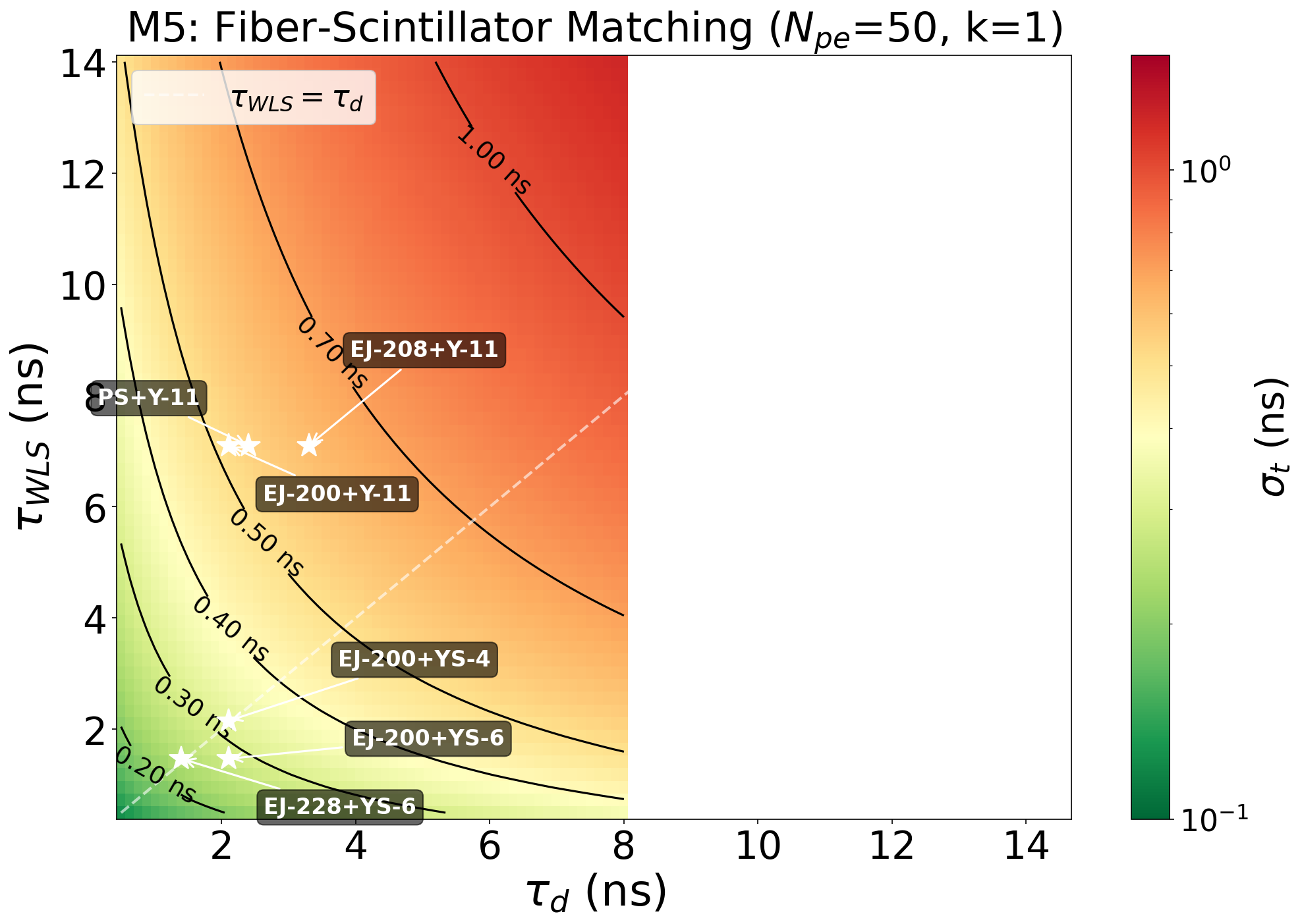}
\caption{Fiber-scintillator matching map: $\sigt$ in the $(\taud,\tauwls)$ plane at $\Npe=50$.}
\label{fig:map_matching}
\end{figure}

In figure \ref{fig:map_sipm} we present the relationship between PDE\&SPTR and $\sigt$ to guide SiPM selection. In all ranges PDE effects are much more prominent than SPTR effects since it directly affects detected $\Npe$. This isolates photon detection efficiency as the more important parameter for improving the timing resolution. 
In figure \ref{fig:map_matching} we present the relationship between $\tauwls$ \& $\taud$ and $\sigt$ to study their relative importance. Here we can see both parameters are important with the diagonal separating the WLS-dominated and scintillator-dominated regimes. 

\subsection{Detector effects}
\label{app:mc_detector_effects} 

\begin{figure}[H]
\centering
\includegraphics[width=0.95\textwidth]{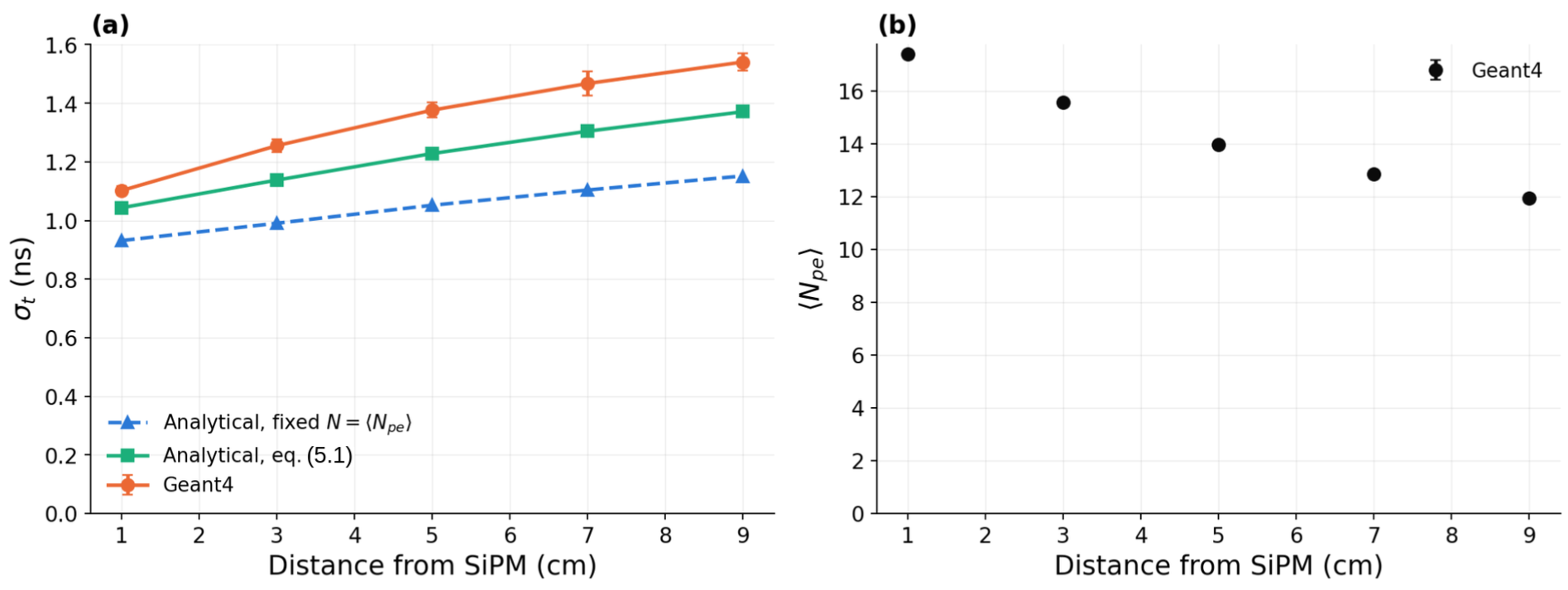}
\caption{Left: Geant4 timing resolution versus hit position for a 10~cm short bar (red circles)
compared to the analytical prediction at fixed matched $\Npe$ (blue triangles) and analytical prediction with $\Npe$ correction following eq.\ref{eq:compound}.
Right: photoelectron yield profile showing $\Npe$ decreasing from $\approx 16$ to $12$ across the bar.}
\label{fig:g4_short}
\end{figure}

\begin{figure}[htbp]
\centering
\includegraphics[width=\textwidth]{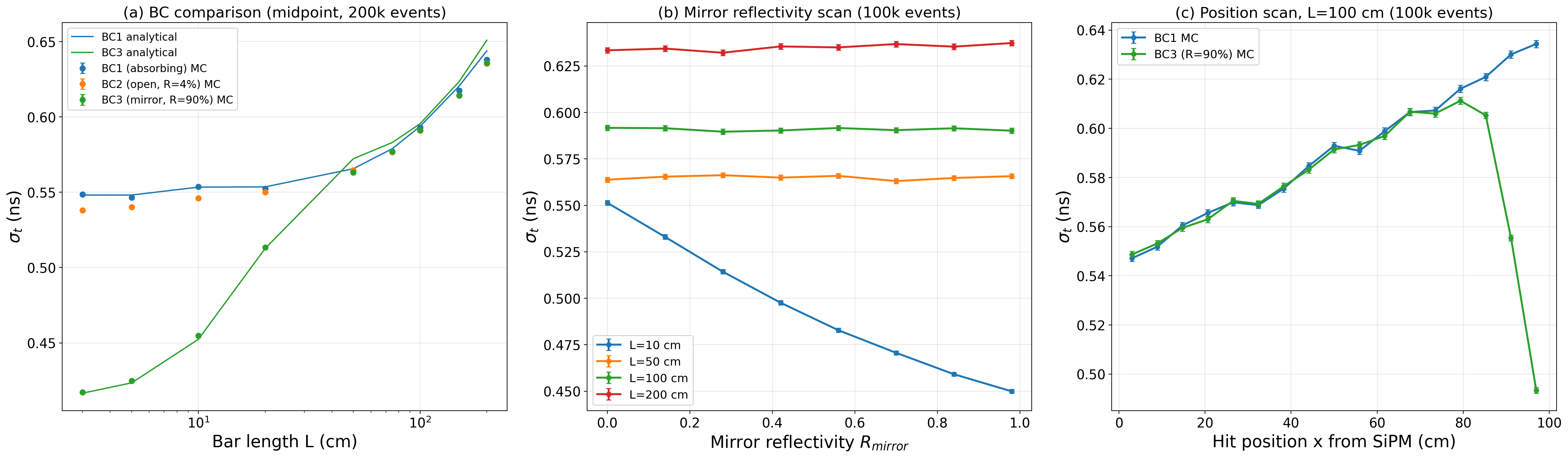}
\caption{Boundary-condition Toy MC study (Y-11, EJ-200, $\Npe = 50$ at the SiPM end). (a) Comparison of absorbing, open, and mirrored
far ends versus bar length at the midpoint, with the analytical BC1 and BC3 predictions as
lines. (b) Mirror reflectivity scan for different lengths.
(c) Position scan at $L=100$~cm comparing BC1 and BC3. Panels (b) and (c) use $10^5$ events
per point; error bars show the MC statistical uncertainty.}
\label{fig:mc_bc}
\end{figure}

\begin{figure}[htbp]
\centering
\includegraphics[width=\textwidth]{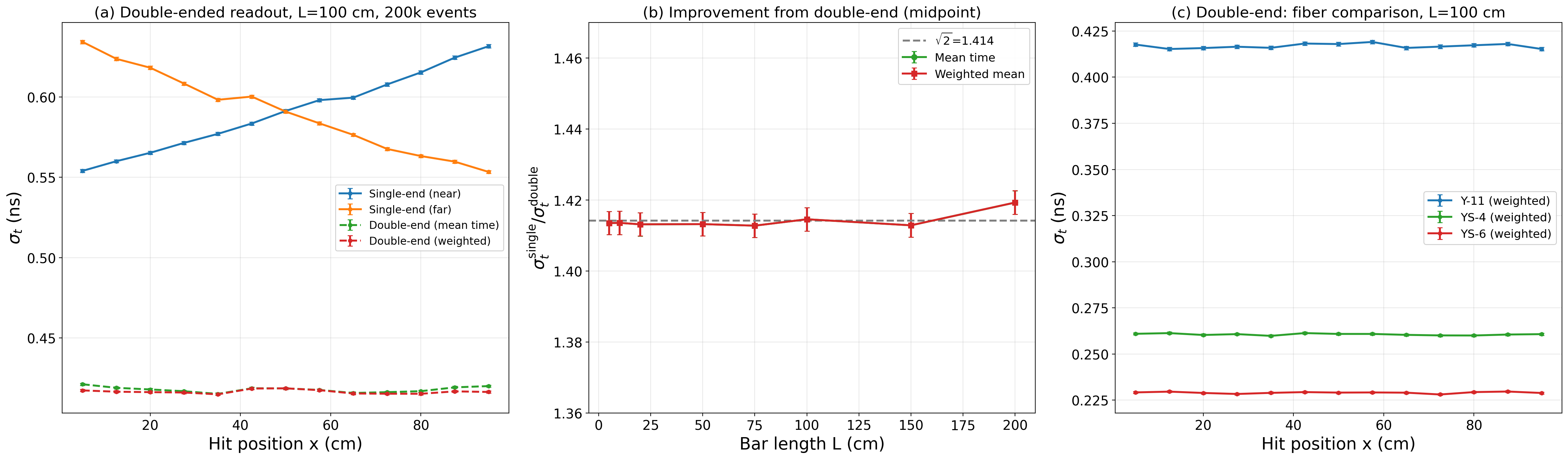}
\caption{Double-ended readout Toy MC study (EJ-200, $\Npe = 50$ at each SiPM end). (a) Position scan at $L=100$~cm for the two
single-ended measurements and their weighted combination. (b) Improvement ratio versus bar
length, hovering around $\sqrt{2}$. Mean and weighted mean curves sit exactly on top of each other.  (c) Fiber comparison for double-ended readout at $L=100$~cm
($10^5$ events per point). Error bars show the MC statistical uncertainty.}
\label{fig:mc_double}
\end{figure}

The following figures from G4 and toy MC simulations paint a detailed story of under what conditions does the model fit with simulations. Figure~\ref{fig:g4_short} presents the timing resolution for a bar
with single-ended BC1 readout. The Geant4 results follow the analytical prediction
at matched $\Npe$, lying 13--19\% above it at all positions as in
section~\ref{sec:g4_position_topology}, with a nearly uniform $\Npe \approx 14$ profile across the bar.
The position-independent ratio at this length scale is
consistent with the model in a regime where
transit time contributions are negligible. Figure \ref{fig:mc_bc} and  \ref{fig:mc_double} plot the Toy MC simulation of $\sigt$ for different far-end boundary conditions and readout topologies. A few signatures can be identified from the plot that match our predictions: the $\sigt$ associated with BC3 is much better than other conditions for short bars where reflected transit time is negligible but falls back to the baseline for longer configurations, BC3 performance improves for very far positions, the double-ended condition provides the best and most consistent timing resolution, improvements of the double-ended topology over single-ended readout hovers around $\sqrt{2}$, and the single-ended timing resolution improves monotonically as the hit position comes closer to the sensor. 

\begin{figure}[htbp]
\centering
\includegraphics[width=\textwidth]{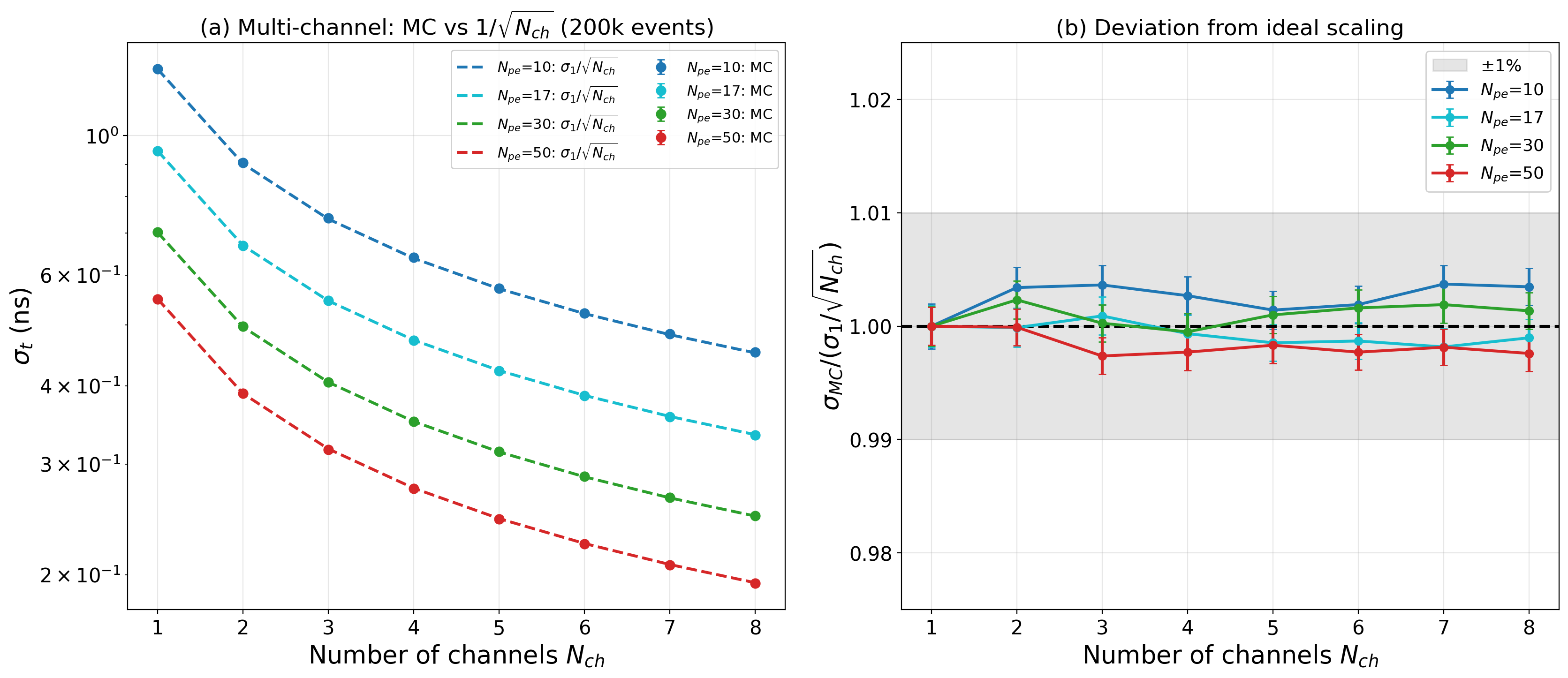}
\caption{Multi-channel scaling Toy MC study (Y-11, EJ-200). Left: $\sigt$ versus number of independent
channels for four $\Npe$ values. Dashed lines show ideal $1/\sqrt{N_\mathrm{ch}}$ scaling.
Right: Ratio of Toy MC to ideal scaling, all within $\pm1\%$. Error bars show the MC
statistical uncertainty}
\label{fig:mc_multi}
\end{figure}

\clearpage
\section{Material properties and detector components}
\label{app:materials}

The study covers a comprehensive set of commercially available detector components.
Table~\ref{tab:scintillators} lists the scintillator materials,
table~\ref{tab:fibers} the WLS fibers,
table~\ref{tab:sipms} the SiPMs, and
table~\ref{tab:electronics} the electronics options.

\begin{table}[htbp]
\centering
\caption{Scintillator material properties used in this study. All materials are
polyvinyltoluene (PVT) based except PS+PPO+POPOP (polystyrene base). Light yields are
for unquenched scintillation.}
\label{tab:scintillators}
\smallskip
\begin{tabular}{lccccc}
\toprule
Material & $\taur$ (ns) & $\taud$ (ns) & Light yield (ph/MeV) & $\lambda_\mathrm{peak}$ (nm) & $n$ \\
\midrule
EJ-200   & 0.9  & 2.1 & 10\,000 & 425 & 1.58 \\
EJ-204   & 0.7  & 1.8 & 10\,400 & 408 & 1.58 \\
EJ-208   & 1.0  & 3.3 &  9\,200 & 435 & 1.58 \\
EJ-228   & 0.5  & 1.4 & 10\,200 & 391 & 1.58 \\
EJ-232   & 0.35 & 1.6 &  8\,400 & 370 & 1.58 \\
BC-420   & 0.5  & 1.5 & 10\,000 & 391 & 1.58 \\
PS+PPO+POPOP & 0.9 & 2.4 & 8\,000 & 420 & 1.59 \\
\bottomrule
\end{tabular}
\end{table}

\begin{table}[htbp]
\centering
\caption{WLS fiber properties. Kuraray fibers (Y-series) and Saint-Gobain/Luxium fibers
(BCF-series) are listed. All fibers are 1~mm diameter; the Kuraray fibers are multi-clad, while the BCF-XL series is single-clad}. Decay times $\tauwls$ for Y-series fibers are measured values from Kodama \textit{et al.}~\cite{Koshio_fiber} at 10~cm injection distance. Those for BCF-92XL, BCF-9929AXL, and BCF-9995XL are measured values from Yamazumi \textit{et al.}~\cite{Yamazumi_BCFXL} (the manufacturer's nominal decay time is 2.7~ns for all three). NA and trapping-efficiency values are
manufacturer specifications~\cite{Kuraray_fiber,SaintGobain_fiber}. $\varepsilon_\mathrm{cap}$ is the trapping efficiency per propagation direction (multi-clad Kuraray: NA $= 0.72$, $\varepsilon_\mathrm{cap} = 5.4\%$; single-clad BCF: NA $= 0.58$, $\varepsilon_\mathrm{cap} = 3.4\%$). Attenuation lengths are conservative design values; dedicated measurements report long-component attenuation lengths of 4.4--5.6~m for the Y-series~\cite{Koshio_fiber}.

\label{tab:fibers}
\smallskip
\begin{tabular}{lcccccc}
\toprule
Fiber & $\tauwls$ (ns) & $\Lambda_\mathrm{att}$ (cm) & NA & $n_\mathrm{core}$ & $\varepsilon_\mathrm{cap}$ & $\lambda_\mathrm{em}$ (nm) \\
\midrule
Y-11(200)  & 7.10  & 350 & 0.72 & 1.59 & 0.054 & 476 \\
YS-2       & 3.64  & 350 & 0.72 & 1.59 & 0.054 & 474 \\
YS-4       & 2.15  & 300 & 0.72 & 1.59 & 0.054 & 470 \\
YS-6       & 1.47  & 300 & 0.72 & 1.59 & 0.054 & 462 \\
BCF-91AXL  & 12.00 & 400 & 0.58 & 1.60 & 0.034 & 494 \\
BCF-92XL   & 2.10  & 400 & 0.58 & 1.60 & 0.034 & 492 \\
BCF-9929AXL & 2.10 & 400 & 0.58 & 1.60 & 0.034 & 492 \\
BCF-9995XL & 2.41  & 400 & 0.58 & 1.60 & 0.034 & 450 \\
\bottomrule
\end{tabular}
\end{table}

\begin{table}[htbp]
\centering
\caption{SiPM properties at the WLS emission wavelength. PDE values are
conservative estimates at the WLS emission peak ($\sim$470--490~nm) and
typical overvoltage. SPTR values are taken from
published device-characterization measurements~\cite{Gundacker_SPTR} and the manufacturer technical note}~\cite{Hamamatsu_MPPC}. The FBK
NUV-HD value of 80~ps represents best-case performance (typical: 90--100~ps). 
\label{tab:sipms}
\smallskip
\begin{tabular}{lcccc}
\toprule
Model & Area (mm$^2$) & Pitch ($\mu$m) & PDE & SPTR (ps FWHM) \\
\midrule
S13360-1325CS  & $1.3\times1.3$ & 25 & 0.20 & 150 \\ 
S14160-1315PS  & $1.3\times1.3$ & 15 & 0.30 & 100 \\ 
S13360-1350PE  & $1.3\times1.3$ & 50 & 0.33 & 180 \\ 
S14160-3050HS  & $3.0\times3.0$ & 50 & 0.40 & 200 \\
FBK NUV-HD     & $3.0\times3.0$ & 35 & 0.50 &  80 \\
\bottomrule
\end{tabular}
\end{table}

\begin{table}[htbp]
\centering
\caption{Front-end electronics configurations. The effective timing contribution is
$\sigelec = \Delta t_\mathrm{TDC}/\sqrt{12}$. The SPIROC2 entry is the TDC design resolution; the timing resolution achieved in test beam was coarser.}
\label{tab:electronics}
\smallskip
\begin{tabular}{lccc}
\toprule
Electronics & $\Delta t_\mathrm{TDC}$ (ps) & $\sigelec$ (ns) & Method \\
\midrule
CITIROC+FPGA & 2500 & 0.722 & LE \\
SPIROC2      & 1000 & 0.289 & LE \\
HGCROC       &   25 & 0.007 & CFD \\
DRS4         &   10 & 0.003 & dCFD \\
Ideal        &    1 & 0.000 & LE \\
\bottomrule
\end{tabular}
\end{table}

\clearpage
%

\end{document}